\documentclass[a4paper,11pt]{article}
\pdfoutput=1
\usepackage{jheppub}
\usepackage{amsmath}
\usepackage{float}
\usepackage{multirow}

\newcommand{\ii}{\mathrm{i}}
\newcommand{\rme}{\mathrm{e}}

\newcommand{\Tr}{\mathrm{Tr}\,}

\newcommand{\diag}{\mathrm{diag}\,}

\newcommand{\cP}{{\mathcal{P}}}
\newcommand{\cQ}{{\mathcal{Q}}}

\newcommand{\one}{{\rm 1\kern -.9mm l}}

\newcommand{\be}{\begin{equation}}
\newcommand{\ee}{\end{equation}}

\newdimen\tableauside\tableauside=1.0ex
\newdimen\tableaurule\tableaurule=0.4pt
\newdimen\tableaustep
\def\phantomhrule#1{\hbox{\vbox to0pt{\hrule height\tableaurule
width#1\vss}}}
\def\phantomvrule#1{\vbox{\hbox to0pt{\vrule width\tableaurule
height#1\hss}}}
\def\sqr{\vbox{%
  \phantomhrule\tableaustep
\hbox{\phantomvrule\tableaustep\kern\tableaustep\phantomvrule\tableaustep}%
  \hbox{\vbox{\phantomhrule\tableauside}\kern-\tableaurule}}}
\def\squares#1{\hbox{\count0=#1\noindent\loop\sqr
  \advance\count0 by-1 \ifnum\count0>0\repeat}}
\def\tableau#1{\vcenter{\offinterlineskip
  \tableaustep=\tableauside\advance\tableaustep by-\tableaurule
  \kern\normallineskip\hbox
    {\kern\normallineskip\vbox
      {\gettableau#1 0 }%
     \kern\normallineskip\kern\tableaurule}%
  \kern\normallineskip\kern\tableaurule}}
\def\gettableau#1 {\ifnum#1=0\let\next=\null\else
  \squares{#1}\let\next=\gettableau\fi\next}
\def\XXint#1#2#3{{\setbox0=\hbox{$#1{#2#3}{\int}$}
     \vcenter{\hbox{$#2#3$}}\kern-.5\wd0}}

\usepackage[usenames,dvipsnames,svgnames,table]{xcolor}

\usepackage{tikz}
\usetikzlibrary{shapes.symbols,shapes.geometric,shapes.misc,circuits.logic.US}
\usetikzlibrary{matrix}
\usetikzlibrary{graphs}
\usetikzlibrary{trees}
\usetikzlibrary{arrows,arrows.spaced}
\usetikzlibrary{decorations.markings,patterns}
\usetikzlibrary{backgrounds,fit}

\tikzstyle{gauge} = [circle, text centered, draw=black, minimum height=1.5cm]
\tikzstyle{squashedflavor} = [rectangle, text centered, draw=black, minimum height=1cm,minimum width=1.5cm]
\tikzstyle{flavor} = [rectangle, text centered, draw=black, minimum height=1.5cm,minimum width=1.5cm]
\tikzstyle{oldgaugedflavor} = [and gate,draw, point up, minimum height=1cm,draw=black]
\tikzstyle{gaugeS} = [circle, text centered, draw=black, minimum height=6ex]
\tikzstyle{flavorS} = [rectangle, text centered, draw=black,minimum height=6ex,minimum width=6ex]

\makeatletter
\tikzset{arc style/.initial={}}
\pgfdeclareshape{dash circled box}{
    \inheritsavedanchors[from={rectangle}]
    \inheritsavedanchors[from={circle}]
    \inheritanchorborder[from={rectangle}] 
    
    \inheritanchor[from={circle}]{center}
    \inheritanchor[from={rectangle}]{center}
    \inheritanchor[from={rectangle}]{south}
    \inheritanchor[from={rectangle}]{west}
    \inheritanchor[from={rectangle}]{north}
    \inheritanchor[from={rectangle}]{east}

    \backgroundpath{
    \southwest \pgf@xa=\pgf@x \pgf@ya=\pgf@y
    \northeast \pgf@xb=\pgf@x \pgf@yb=\pgf@y
    \pgf@xc=\pgf@xb \advance\pgf@xc by-5pt 
    \pgf@yc=\pgf@yb \advance\pgf@yc by-5pt
    
    \pgf@xx=\pgf@xa \advance\pgf@xx by10pt 
    \pgf@yy=\pgf@ya \advance\pgf@yy by10pt

    \pgfpathmoveto{\pgfpoint{\pgf@xa}{\pgf@ya}}
    \pgfpathlineto{\pgfpoint{\pgf@xa}{\pgf@yb}}
    \pgfpathlineto{\pgfpoint{\pgf@xc}{\pgf@yb}}
    \pgfpathlineto{\pgfpoint{\pgf@xb}{\pgf@yc}}
    \pgfpathlineto{\pgfpoint{\pgf@xb}{\pgf@ya}}
    \pgfpathclose
    \pgfpathmoveto{\pgfpoint{\pgf@xc}{\pgf@yb}}
    \pgfpathlineto{\pgfpoint{\pgf@xc}{\pgf@yc}}
    \pgfpathlineto{\pgfpoint{\pgf@xb}{\pgf@yc}}
    \pgfpathlineto{\pgfpoint{\pgf@xc}{\pgf@yc}}
    
    \pgfpathcircle{\pgfpoint{\pgf@xx}{\pgf@yy}}{\pgf@xb} 
 }
}
\makeatother

\makeatletter%
\pgfdeclareshape{barn}{
    \inheritsavedanchors[from={rectangle}]
    \inheritanchorborder[from={rectangle}] 
     \inheritanchor[from={rectangle}]{north}
    \inheritanchor[from={rectangle}]{center}
    \inheritanchor[from={rectangle}]{south}
    \inheritanchor[from={rectangle}]{west}
    \inheritanchor[from={rectangle}]{east}

    \inheritbackgroundpath[from={rectangle}]
   
    \backgroundpath{
    \southwest \pgf@xa=\pgf@x \pgf@ya=\pgf@y
    \northeast \pgf@xb=\pgf@x \pgf@yb=\pgf@y
    \pgf@xc=\pgf@xb \advance\pgf@xc by-\pgf@xa     
    \pgf@yc=\pgf@yb \advance\pgf@yc by-\pgf@ya    
    \advance\pgf@yb by-0.5\pgf@yc
    \pgfpathmoveto{\pgfpoint{\pgf@xa}{\pgf@ya}}
    \pgfpathlineto{\pgfpoint{\pgf@xa}{\pgf@yb}}
    \pgfpatharc{180}{0}{.5\pgf@xc}  
    \pgfpathlineto{\pgfpoint{\pgf@xb}{\pgf@ya}}
    \pgfpathclose
 }
}
\makeatother
\tikzstyle{test1} = [dash circled box, text centered, draw=black, minimum height=1cm,minimum width=1cm]
\tikzstyle{test2} = [barn, text centered, draw=black, minimum height=1cm,minimum width=1cm]
\tikzstyle{gaugedflavor} = [barn,draw, text centered, minimum height=1.5cm,minimum width=1.5cm,draw=black]
\tikzstyle{gaugedflavorS} = [barn,draw, text centered, minimum width=6ex,minimum height=6ex,draw=black]

\pgfdeclareshape{cross}{
  \inheritsavedanchors[from={rectangle}]
    \inheritanchorborder[from={rectangle}] 
     \inheritanchor[from={rectangle}]{north}
    \inheritanchor[from={rectangle}]{center}
    \inheritanchor[from={rectangle}]{south}
    \inheritanchor[from={rectangle}]{west}
    \inheritanchor[from={rectangle}]{east}
\inheritbackgroundpath[from={rectangle}]
    \backgroundpath{
    \draw[solid] (.6ex,-.6ex) -- (-.6ex,.6ex);
    \draw[solid] (.6ex,.6ex) -- (-.6ex,-.6ex);
    }
}

\title{\boldmath Surface operators, dual quivers and contours}
\author[a]{S.~K.~Ashok,}
\affiliation[a]{Institute of Mathematical Sciences \\
Homi Bhabha National Institute (HBNI)\\
IV Cross Road, C.~I.~T.~Campus, \\
  Taramani, Chennai, 600113  Tamil Nadu, India \\}
\emailAdd{sashok@imsc.res.in} 
   
\author[a]{S.~Ballav,}
\emailAdd{sballav@imsc.res.in}
   
\author[b,c]{M.~Bill\`o,}
\affiliation[b]{Universit\`a di Torino, Dipartimento di Fisica}

\affiliation[c]{Arnold-Regge Center and I.\,N.\,F.\,N. - sezione di Torino, \\
Via P. Giuria 1, I-10125 Torino, Italy\\}
\emailAdd{billo@to.infn.it}

\author[b]{E.~Dell'Aquila,}
\emailAdd{edellaquila@gmail.com}

\author[b,c]{M.~Frau,}
\emailAdd{frau@to.infn.it}

\author[a]{V.~Gupta,}
\emailAdd{varungupta@imsc.res.in}

\author[b,c]{R.~R.~John,}
\emailAdd{renjan.rajan@to.infn.it}

\author[d,c]{and A.~Lerda\,}
\affiliation[d]{Universit\`a del Piemonte Orientale, Dipartimento di Scienze e Innovazione Tecnologica\\
Viale T. Michel 11, I-15121 Alessandria, Italy\\}
\emailAdd{lerda@to.infn.it}

\abstract{We study half-BPS surface operators in four dimensional ${\mathcal N}=2$ SU($N$) gauge 
theories, and analyze their low-energy effective action on the four dimensional Coulomb branch using 
equivariant localization. We also study surface operators as coupled 2d/4d quiver gauge theories with an 
SU$(N)$ flavour symmetry. In this description, the same surface operator can be described 
by different quivers that are related to each other by two dimensional Seiberg duality. 
We argue that these dual quivers correspond, on the localization side, to distinct integration contours that 
can be determined by the Fayet-Iliopoulos parameters of the two dimensional gauge nodes. 
We verify the proposal by mapping the solutions of the twisted chiral ring equations of the 2d/4d quivers 
onto individual residues of the localization integrand.
}

\keywords{Supersymmetric gauge theories, ramified instantons, surface operators, duality}

\begin{document}
\maketitle
\flushbottom

\section{Introduction}
\label{intro}
Surface operators in 4d gauge theories are natural two dimensional generalizations 
of Wilson and 't Hooft loops which can provide valuable information about the 
phase structure of the gauge theories 
\cite{Gukov:2014gja}. 
In this paper we study the low-energy effective action of surface operators 
in pure ${\mathcal N}=2$ 4d gauge theories from two distinct points of view, namely
as monodromy defects \cite{Gukov:2006jk,Gukov:2008sn} and
as coupled 2d/4d quiver gauge theories \cite{Gaiotto:2009fs, Gaiotto:2013sma}.
In the first approach, one specifies how the 4d gauge fields are affected by the presence of 
the surface operator by imposing suitable 
boundary conditions in the path-integral. In this framework the non-perturbative 
effects are described in terms of ramified
instantons \cite{Gukov:2006jk} whose partition function can be computed using 
equivariant localization methods \cite{Kanno:2011fw,Gaiotto:2013sma,
Gorsky:2017hro,Ashok:2017odt, Ashok:2017bld,Ashok:2017lko}. 
{From} the ramified instanton partition function one can extract two holomorphic 
functions \cite{Alday:2009fs,Alday:2010vg}:
one is the prepotential $\mathcal{F}$ that governs the low-energy effective action of the 4d
$\mathcal{N}=2$ gauge theory on the Coulomb branch; the other is the twisted 
chiral superpotential $\mathcal{W}$ that describes the 2d dynamics on the defect.

In the second description of the surface operators, one considers coupled 
2d/4d theories that are $(2,2)$ supersymmetric sigma models with an ultraviolet 
description as a gauged linear sigma model (GLSM). The low-energy dynamics 
of such a GLSM is completely determined by a 
twisted chiral superpotential ${\mathcal W}(\sigma)$ that depends on the twisted chiral 
superfields $\sigma$ containing the 2d vector fields \cite{Witten:1993yc}. 
By giving a vacuum expectation value (v.e.v.) to the adjoint scalar 
of the 4d $\mathcal{N}=2$ gauge
theory, one introduces twisted masses in the 2d quiver 
theory \cite{Aharony:1997bx, Hanany:1997vm}.
At a generic point on the 4d Coulomb branch, the 2d theory is therefore massive in the
infrared and the 2d/4d coupling mechanism is determined via the resolvent of the 4d gauge 
theory \cite{Gaiotto:2013sma}. The resulting massive vacua of the GLSM 
are solutions to the twisted chiral ring equations, which are obtained by 
extremizing  ${\mathcal W}(\sigma)$ with respect to the twisted chiral superfields. 

The main goal of this work is to clarify the precise relationship between the above two
descriptions of the surface operators and provide a dictionary to map calculable quantities 
on one side to the other.
In our previous works \cite{Ashok:2017lko, Ashok:2017bld} the first steps in 
this direction were already taken by showing that there is a precise correspondence between 
the massive vacua of the 2d/4d gauge theory and the monodromy 
defects in the ${\mathcal N}=2$ gauge theory.
In fact, the effective twisted chiral superpotential of the 2d/4d quiver 
gauge theory evaluated in a given massive vacuum exactly coincides with the one 
computed from the 4d ramified instanton partition 
function \cite{Ashok:2017lko, Ashok:2017bld}. This equality was shown 
in a specific class of models that are described by oriented quiver diagrams. 
Recently, this result has been proven in full generality
in \cite{Nekrasov:2017rqy,Jeong:2018qpc}.

An important feature of the $(2,2)$ quiver theories that was not fully discussed in
our previous papers is Seiberg duality \cite{Seiberg:1994pq,Benini:2014mia}. This is
an infrared equivalence between two gauge theories that have different 
ultraviolet realizations. 
In this work we fill this gap and consider all possible quivers obtained from the oriented ones 
by applying 2d Seiberg duality. While all such quivers have different gauge groups and 
matter content, once the 4d Coulomb v.e.v.'s  are turned on, it is possible to find 
a one-to-one map between their massive vacua. Therefore it becomes clear that 
they must describe the same surface operator from the point of view of the 4d gauge 
theory; indeed, the different twisted chiral superpotentials, evaluated in the respective 
vacua, all give the same result.
This equality of superpotentials gives a strong hint that the choice of a Seiberg duality 
frame might have an interpretation as distinct contours of integration on the localization 
side: the equality of the superpotentials would then be a simple consequence of multi-dimensional residue theorems.   

In this work we show that this expectation is correct and provide a detailed map between 
a given quiver realization of the surface operator and a particular choice of contour in the 
localization integrals. This contour prescription can be conveniently encoded in a Jeffrey-Kirwan (JK) 
reference vector \cite{JK1995}, whose coefficients turn out to be related to the Fayet-Iliopoulos
(FI) parameters of the 2d/4d quiver. 
While the twisted superpotentials are equal irrespective of the choice of contour, the map relates the individual residues on the localization side to the individual terms in the solutions to the twisted chiral ring equations, thereby allowing us to identify in an unambiguous way which quiver arises from a given contour prescription and vice-versa.  

This paper is organized as follows. In Section~\ref{sec:review} we review and 
extend our earlier work 
\cite{Ashok:2017odt, Ashok:2017bld,Ashok:2017lko}, and in particular we show how to map 
the oriented quiver to a 
particular contour by studying the solution of the chiral ring equations 
and the precise correspondence 
to the residues of the localization integrand. In Section~\ref{Seiberg Duality} we discuss 
the basics of 2d Seiberg duality and how it acts on the quiver theories we consider. 
In Section~\ref{Relating Quivers and Contours} we apply the duality moves to the oriented quiver of interest and show in detail (for the 4-node quiver), how it is possible to map each quiver to a particular integration contour on the localization side without explicitly solving the chiral ring equations. We also discuss how this integration contour can be specified in terms of a JK reference vector. 
In Section~\ref{allquiverslinear} we give a simple solution for the JK vector associated to a generic linear 2d/4d quiver with arbitrary number of nodes.
Finally, we summarize our main results in Section~\ref{summary} and collect the more technical material in the appendices. 

\section{Review of earlier work}
\label{sec:review}

To set the stage for the discussion in the next sections and also to introduce our notation, 
we briefly review the results obtained in our 
earlier work \cite{Ashok:2017lko} where we studied surface
operators both as monodromy defects in 4d and as coupled 2d/4d gauge theories. 

\subsection{Surface operators as monodromy defects}
\label{SOasMD}
As a monodromy defect, a surface operator in a 4d SU($N$) theory is specified by a partition 
of $N$, denoted by $\vec{n}=(n_1, n_2, \ldots n_M)$, which corresponds to the breaking of 
the gauge group to a Levi subgroup
\begin{equation}
{\mathbb L} = \mathrm{S}\left[\mathrm{U}(n_1)\times \mathrm{U}(n_2)\times \ldots 
\mathrm{U}(n_M) \right]
\end{equation}
at the location of the defect \cite{Gukov:2006jk, Gukov:2008sn}.
This also gives a natural partitioning of the classical Coulomb v.e.v.'s of the adjoint scalar $\Phi$ of the 
$\mathcal{N}=2$ SU($N$) theory as follows:
\begin{equation}
\label{asplit}
\langle \Phi \rangle =
\big\{a_1,\ldots,  a_{r_1}|\ldots \big|
a_{r_{I-1}+1}, \ldots a_{r_{I}}\big|
\ldots |a_{r_{M-1} +1}, \ldots, a_N \big\} ~.  
\end{equation}
Here we have defined the integers $r_I$ according to
\begin{equation}
\label{rI}
r_I = \sum_{J=1}^{I} n_J ~,
\end{equation}
so that the $I^{\text{th}}$ partition in \eqref{asplit} is of length $n_I$. 
Introducing the following set of numbers with cardinality $n_I$: 
\begin{equation}
\mathcal{N}_I \equiv \{ r_{I-1}+1,  r_{I-1}+2,\ldots, r_I \} ~,
\label{NI}
\end{equation}
we define the $n_I \times n_I$ block-diagonal matrices $\mathcal{A}_I$ according to
\begin{equation}
\label{Amatrix}
\mathcal{A}_I \equiv \diag\left(a_{s\in {\mathcal{N}}_I} \right) =  
\begin{pmatrix}
a_{r_{I-1}+1} &0 & 0 & \ldots \cr
0 & \ddots & 0 & \ldots\cr
\vdots &  \vdots & \ddots \cr
0 & 0 & \ldots & a_{r_{I}} 
\end{pmatrix}~.
\end{equation}
With these conventions, the splitting in \eqref{asplit} can be written as 
\begin{equation}
\langle\Phi \rangle = {\mathcal A}_1 \oplus {\mathcal A}_2 \oplus \ldots \oplus {\mathcal A}_M ~.
\end{equation}

The instanton partition function in the presence of such a surface operator, 
also known as the ramified instanton partition function, takes the following form 
\cite{Kanno:2011fw,Ashok:2017lko}:
\begin{equation}
Z_{\text{inst}}[\vec{n}] = \sum_{\{d_I\}}Z_{\{d_I\}}[\vec{n}]\quad
\mbox{with}~~~Z_{\{d_I\}}[\vec{n}]= \prod_{I=1}^M \Big[\frac{(-q_I)^{d_I}}{d_I!}
\int \prod_{\sigma=1}^{d_I} \frac{d\chi_{I,\sigma}}{2\pi\ii}\Big]~
z_{\{d_I\}}
\label{Zso4d5d}
\end{equation}
where
\begin{align}
z_{\{d_I\}} & = \,\prod_{I=1}^M \prod_{\sigma,\tau=1}^{d_I}\,
\frac{\left(\chi_{I,\sigma} - \chi_{I,\tau} + \delta_{\sigma,\tau}\right)
}{\left(\chi_{I,\sigma} - \chi_{I,\tau} + \epsilon_1\right)}
\times\prod_{I=1}^M \prod_{\sigma=1}^{d_I}\prod_{\rho=1}^{d_{I+1}}\,
\frac{\left(\chi_{I,\sigma} - \chi_{I+1,\rho} + \epsilon_1 + \hat\epsilon_2\right)}
{\left(\chi_{I,\sigma} - \chi_{I+1,\rho} + \hat\epsilon_2\right)}
\label{zexplicit4d}
\\
& ~~\times
\prod_{I=1}^M \prod_{\sigma=1}^{d_I} 
\frac{1}
{\prod_{s\in {\mathcal N}_I}\left(a_{s}-\chi_{I,\sigma} + \frac 12 (\epsilon_1 + \hat\epsilon_2)\right)}
\frac{1}
{\prod_{t\in {\mathcal N}_{I+1}}\left(\chi_{I,\sigma} - a_{t} + \frac 12 (\epsilon_1 + \hat\epsilon_2)\right)}~.
\nonumber
\end{align}
Here, the $M$ positive integers $d_I$ count the numbers of ramified
instantons in the various sectors, the variables $q_I$ are the ramified 
instanton weights, and the parameters $\epsilon_1$
and $\hat{\epsilon}_2=\epsilon_2/M$ specify the $\Omega$-background \cite{Nekrasov:2002qd,Nekrasov:2003rj} 
which is introduced to localize the integrals over the instanton moduli space\,\footnote{The
rescaling by a factor of $M$ in $\epsilon_2$ is due to a $\mathbb{Z}_M$-orbifold projection
that has to be performed in the ramified instanton case \cite{Kanno:2011fw}. Furthermore, in 
\eqref{zexplicit4d} the sub-index $I$ is always understood modulo $M$.}.

There is one more ingredient that is needed to calculate the partition function (\ref{Zso4d5d}), namely
the contour of integration for the $\chi_I$ variables. 
A convenient way to specify it and to select which poles of the 
integrand contribute and which do not, is to treat the Coulomb v.e.v.'s $a$ 
as real variables and assign an imaginary part to the $\Omega$-deformation parameters according to
\begin{equation}
0<\text{Im}(\hat\epsilon_2)\ll \text{Im}(\epsilon_1) \ll 1~.
\label{epsilon}
\end{equation}
Then, the contour is specified by integrating $\chi_{I,\sigma}$ in the upper or lower half-plane and
by choosing a definite order in the successive integrations.
Equivalently, as we will see in the following sections, the contour of integration can be selected
by specifying a Jeffrey-Kirwan reference vector \cite{JK1995}.

In the limit $\epsilon_1,\hat{\epsilon}_2\to 0$, the low-energy effective action of the gauge theory with 
the 2d defect is specified by two holomorphic functions: the prepotential $\mathcal{F}$ and the twisted 
chiral superpotential $\mathcal{W}$. Each of these functions can be written as a sum of the
classical, the one-loop, and the instanton contributions. The latter 
can be extracted from the ramified instanton partition function as follows \cite{Alday:2009fs, Alday:2010vg}:
\begin{equation}
\log \big(1+Z_{\text{inst}}\big) = -\frac{{\mathcal{F}}_{\text{inst}}}{\epsilon_1\hat\epsilon_2}  
+ \frac{{\mathcal{W}}_{\text{inst}}}{\epsilon_1} + \ldots
\label{FandW}
\end{equation}
where the ellipses refer to regular terms. In Appendix \ref{appA} we list 
the one-instanton contribution to ${\mathcal W}_{\text{inst}}$ calculated for various choices of contours
in the case $M=4$.
As we will show in the following, the different contour prescriptions can be given a precise 
meaning by associating them to specific 2d/4d quiver gauge theories. 

\subsection{Surface operators as coupled 2d/4d quivers}

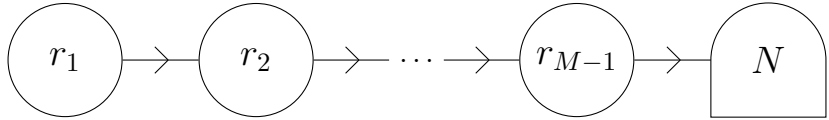
\begin{figure}[H]
\begin{center}
\begin{tikzpicture}[decoration={markings, mark=at position 0.6 with {\draw (-5pt,-5pt) -- (0pt,0pt);       
         \draw (-5pt,5pt) -- (0pt,0pt);}}]
  \matrix[row sep=10mm,column sep=5mm] {
      \node (g1)[gauge] {\Large $r_1$};  & & \node(g2)[gauge] {\Large $r_2$};  && \node(dots){ 
      $\ldots$};
      & & \node(glast)[gauge] {\Large $r_{M-1}$};& &\node(gfN)[gaugedflavor]{\Large $N$};\\
  };
\graph{(g1) --[postaction={decorate}](g2) --[postaction={decorate}](dots)--[postaction={decorate}](glast) 
--[postaction={decorate}](gfN);};
\end{tikzpicture}
\end{center}
\vspace{-0.5cm}
\caption{The quiver which describes the generic surface operator in pure SU$(N)$ gauge theory.}
\label{quiverpicgeneric1}
\end{figure}

The prepotential $\mathcal{F}$ governs the 4d gauge theory dynamics at a generic point on the Coulomb branch. The twisted chiral superpotential $\mathcal{W}$, instead, is best understood as the low-energy effective description of a 2d non-linear sigma model. 
For a surface operator with a Levi subgroup $\mathbb{L}$ in a 4d theory with a gauge
group $G$, the relevant sigma model is defined on the target space 
$G/\mathbb{L}$ \cite{Gukov:2006jk, Gukov:2008sn}. Such a space is, in general, a flag variety 
which can be realized as the low-energy limit of a GLSM 
\cite{Witten:1993yc, Hanany:1997vm}, whose gauge and matter content can be summarized in the 
quiver diagram of Fig.~\ref{quiverpicgeneric1}.
Each circular node represents a 2d gauge group $\mathrm{U}(r_I)$ where the ranks $r_I$ are 
as in \eqref{rI}, whereas the last node on the right hand side represents the 4d gauge 
group SU($N$) which acts as a flavour symmetry group for the $(M-1)^{\mathrm{th}}$ 
2d node. 
The arrows correspond to matter multiplets which are rendered massive by non-zero 
v.e.v's of the 
twisted scalars $\sigma^{(I)}$ of the $I^{\mathrm{th}}$ node and 
of the 4d adjoint scalar $\Phi$. 
The orientation of the arrows specifies whether the matter is in the 
fundamental (out-going) or in the anti-fundamental (in-going) representation.

The effective action for the twisted chiral multiplets is obtained 
by integrating out the massive 
matter multiplets and, thanks to supersymmetry, can be encoded in the effective twisted 
chiral superpotential. For the quiver of Fig.~\ref{quiverpicgeneric1}, this is given by:
\begin{equation}
\begin{aligned}
\mathcal{W} =& ~2\pi\ii \sum_{I=1}^{M-1}\sum_{s=1}^{r_I} \tau_I\,\sigma^{(I)}_s-\!
\sum_{I=1}^{M-2} \sum_{s=1}^{r_{I}} \sum_{t=1}^{r_{I+1}}
\varpi\big(\sigma^{(I)}_s - \sigma^{(I+1)}_t\big) 
-\!\sum_{s=1}^{r_{M-1}}\!\Big\langle 
\Tr \varpi\big(\sigma^{(M-1)}_s - \Phi\big) \Big\rangle
\end{aligned}
\label{tildeWoriginal}
\end{equation}
where
\begin{equation}
\label{defvarpi}
\varpi(x) = x\,\Big(\log \frac{x}{\mu} -1\Big)~,
\end{equation}
$\mu$ is the UV cut-off scale, and $\tau_I$ is the complexified 
FI parameter of the $I^{\text{th}}$ node at the scale $\mu$, namely
\begin{equation}
\tau_I=\frac{\theta_I}{2\pi}+\ii\,\zeta_I
\label{FI}
\end{equation}
with $\theta_I$ and $\zeta_I$ being, respectively, 
the $\theta$-parameter and the real FI parameter of the
$I^{\text{th}}$ gauge node. Finally, the angular brackets in the last term of \eqref{tildeWoriginal} 
correspond to a chiral correlator in the 4d SU($N$) theory. 
This correlator implies that the coupling between the 2d and 4d theory is via the resolvent of the SU($N$)
gauge theory \cite{Gaiotto:2013sma}, which in turn depends on the 4d dynamically generated
scale $\Lambda_{\text{4d}}$. 

The running of the FI parameters leads to introducing 2d low-energy 
scales $\Lambda_I$ at each 
node by the relation
\begin{equation}
\Lambda_I^{b_I}= \rme^{\,2\pi\ii \,\tau_I} \,\mu^{b_I}
\label{run}
\end{equation}
where $b_I$ is the corresponding $\beta$-function coefficient, which in this case is
\begin{equation}
b_I= n_I+ n_{I+1}~.
\label{betaI}
\end{equation}
Of course, we can rewrite (\ref{run}) as
\begin{equation}
\left|\frac{\Lambda_I}{\mu}\right| = \rme^{-2\pi\,\frac{\zeta_I}{b_I}}
\end{equation}
which implies that
\begin{equation}
\frac{\zeta_I}{b_I}>0~.
\end{equation}
Since for the quiver represented in Fig.~\ref{quiverpicgeneric1}, all $b_I$ are positive
(see (\ref{betaI})), we deduce that
\begin{equation}
\zeta_I>0~.
\label{zetapositive}
\end{equation}

Once the 4d Coulomb v.e.v.'s are given, the 2d Coulomb branch is completely lifted except for a finite 
number of discrete vacua. These are found by extremizing the twisted chiral superpotential 
$\mathcal{W}$, {\it{i.e.}} they are solutions of the twisted chiral ring equations 
\cite{Nekrasov:2009ui,Nekrasov:2009rc}
\begin{equation}
\label{TCR}
\exp\left(\frac{\partial\mathcal W}{\partial\sigma^{(I)}_s}\right)=1~.
\end{equation}
In order to make contact with the partition of the v.e.v.'s in \eqref{asplit},
we solve \eqref{TCR} about the following classical vacuum:
\begin{equation}
\label{linearclassicalvev}
\sigma^{(I)}_{\text{cl}} = {\mathcal A}_1\oplus {\mathcal A}_2\oplus \ldots \oplus {\mathcal A}_I ~.
\end{equation}
Once the solutions to the twisted chiral ring equations are obtained (order by order in the low-energy scales of the 2d/4d theories), we evaluate the effective twisted chiral superpotential $\mathcal{W}$ on
this particular solution, and verify that the non-perturbative contributions exactly coincide 
with the ${\mathcal W}_{\text{inst}}$ calculated using localization. 
In essence, this match provides a one-to-one map between 1/2-BPS defects in the 4d gauge theory and massive vacua in the coupled 2d/4d gauge theory. 

\subsection{A contour from the twisted chiral ring}
\label{TCRvsLocQ1JK1}

\begin{figure}[H]
\begin{centering}
\begin{tikzpicture}[decoration={
markings,
mark=at position 0.6 with {\draw (-5pt,-5pt) -- (0pt,0pt);
                \draw (-5pt,5pt) -- (0pt,0pt);}}]
  \matrix[row sep=10mm,column sep=5mm, ampersand replacement=\&] {
      \node(g1)[gauge] {$n_1$};  \& \& \node(g2)[gauge] {$n_1+n_2$}; \& \& \node(g3)[gauge,align=center] {$n_1 + n_2$\\$+\, n_3$}; 
      \& \&\node(gfN)[gaugedflavor]{$N$};\\
  };
\graph{(g1) --[postaction={decorate}](g2)--[postaction={decorate}](g3)--[postaction={decorate}](gfN);};
\end{tikzpicture}
\caption{The 4-node linear quiver that corresponds to the partition $[n_1, n_2, n_3, n_4]$.}
\label{Quiver1-4node}
\end{centering}
\end{figure}
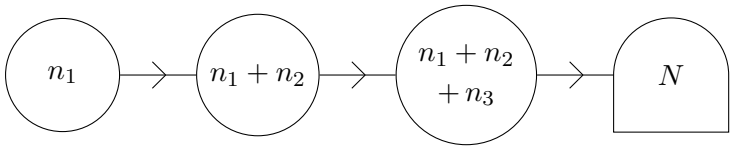

We now consider in detail the case $M=4$ corresponding to the quiver in Fig.~\ref{Quiver1-4node}.
This is the simplest example that is general enough to contain 
all relevant features of a generic linear quiver, and thus it serves as a prototypical case.
The twisted chiral ring equations \eqref{TCR} can be compactly written in terms of a characteristic 
gauge polynomial for each U$(r_I)$ node, given by
\begin{equation}
\mathcal{Q}_{I}(z) = \prod_{s=1}^{r_{I}} (z - \sigma^{(I)}_s)~,
\end{equation} 
and the characteristic polynomial of the 4d SU$(N)$ node, namely
\begin{equation}
\mathcal{P}_N(z) =z^N +\sum_{i=2}^N (-1)^{k}\, u_k \, z^{N-k} ~.
\label{PNdef}
\end{equation}
Here $u_k$ are the gauge invariant coordinates on the moduli space, 
which can be calculated at weak coupling using localization methods 
\cite{Bruzzo:2002xf,Losev:2003py,Flume:2004rp,Billo:2012st,Ashok:2016ewb}.
In terms of these polynomials, the twisted chiral equations \eqref{TCR} become
\cite{Ashok:2017lko}
\begin{equation}
\begin{aligned}
\cQ_{2}(\sigma^{(1)}_s) &= \Lambda_1^{n_1+n_2}~,\\
\cQ_{{3}}(\sigma^{(2)}_t) &= (-1)^{n_1} \,\Lambda_{2}^{n_2+n_3}\,\cQ_{{1}}(\sigma^{(2)}_t)~,\\
\cP_N(\sigma^{(3)}_u) &=(-1)^{n_1+n_2}\left(\Lambda_{3}^{n_3+n_4}\, \cQ_{{2}}(\sigma^{(3)}_u) 
+\frac{\Lambda_{\text{4d}}^{2N}}{\Lambda_{3}^{n_3+n_4}\, \cQ_{{2}}(\sigma^{(3)}_u)}\right)~,
\end{aligned}
\label{TCRQ1}
\end{equation}
for $s\in {\mathcal N}_1$, $t\in {\mathcal N}_1\cup {\mathcal N}_2$, 
and $u\in {\mathcal N}_1\cup {\mathcal N}_2 \cup {\mathcal N}_3$, respectively.
We look for solutions of these equations that are of the form
\begin{equation}
\sigma^{(I)}_{\star} = \sigma^{(I)}_{\text{cl}} + \delta\sigma^{(I)}~,
\end{equation}
where the classical part is as in \eqref{linearclassicalvev} for $I=1,2,3$. A detailed derivation 
of the solution at the one-instanton level is presented in Appendix.~\ref{appB}. 
Here we merely write the expressions for the non-vanishing first-order corrections, that are
\begin{equation}
\begin{aligned}
\delta\sigma^{(1)}_s &=\frac{\Lambda_1^{n_1+n_2}}
{\prod_{r\in\widehat{\mathcal{N}}_1\cup\mathcal{N}_2}(a_s-a_r)\phantom{\Big|}}
+\frac{(-1)^{n_2}\Lambda_{\text{4d}}^{2N}}{\Lambda_1^{n_1+n_2}\Lambda_2^{n_2+n_3}
\Lambda_3^{n_3+n_4}
\prod_{r\in\mathcal{N}_4\cup\widehat{\mathcal{N}}_1}(a_s-a_r)\phantom{\Big|}}~,\phantom{\bigg|}\\
\delta\sigma^{(2)}_s & =\,\delta\sigma^{(3)}_s= \frac{(-1)^{n_2}
\Lambda_{\text{4d}}^{2N}}{\Lambda_1^{n_1+n_2}\Lambda_2^{n_2+n_3}\Lambda_3^{n_3+n_4}
\prod_{r\in\mathcal{N}_4\cup\widehat{\mathcal{N}}_1}(a_s-a_r)\phantom{\Big|}}\phantom{\bigg|}
\end{aligned}
\label{deltasigma1}
\end{equation}
for $s\in\mathcal{N}_1$,
\begin{equation}
\delta\sigma^{(2)}_t=\frac{(-1)^{n_1}\Lambda_2^{n_2+n_3}}{
\prod_{r\in\widehat{\mathcal{N}}_2\cup\mathcal{N}_3}(a_t-a_r)\phantom{\Big|}}
\label{deltasigma2}
\end{equation}
for $t\in\mathcal{N}_2$, and
\begin{equation}
\delta\sigma^{(3)}_u=\frac{(-1)^{n_1+n_2}\Lambda_3^{n_3+n_4}}{
\prod_{r\in\widehat{\mathcal{N}}_3\cup\mathcal{N}_4}(a_u-a_r)\phantom{\Big|}}
\label{deltasigma3}
\end{equation}
for $u\in\mathcal{N}_3$. In these formulas, the symbol $\widehat{\mathcal{N}}_I$ means
that one has to omit from the set $\mathcal{N}_I$ the indices that would yield 
a vanishing denominator.

In \cite{Ashok:2017lko} it was shown that
\begin{equation}
\label{logLider}
\Tr \sigma^{(I)}_\star = 
\frac{1}{b_I}\Lambda_I \frac{\partial \mathcal W}{\partial {\Lambda_I}}\bigg|_{\sigma_\star}~.
\end{equation}  
Integrating in this relation, one can obtain the twisted superpotential in the chosen vacuum, 
which in the one-instanton approximation is
\begin{eqnarray}
\mathcal{W}\big|_{\sigma_\star}\!&=&\!\sum_{s\in\mathcal{N}_1}\frac{\Lambda_1^{n_1+n_2}}{
\prod_{r\in\widehat{\mathcal{N}}_1\cup\mathcal{N}_2}(a_s-a_r)\phantom{\Big|}}+\!
\sum_{t\in\mathcal{N}_2}\frac{(-1)^{n_1}\Lambda_2^{n_2+n_3}}{
\prod_{r\in\widehat{\mathcal{N}}_2\cup\mathcal{N}_3}(a_t-a_r)\phantom{\Big|}}+\!
\sum_{u\in\mathcal{N}_3}\frac{(-1)^{n_1+n_2}\Lambda_3^{n_3+n_4}}{
\prod_{r\in\widehat{\mathcal{N}}_3\cup\mathcal{N}_4}(a_u-a_r)\phantom{\Big|}}\nonumber\\
&&\hspace{2cm}+\sum_{s\in\mathcal{N}_1}\frac{(-1)^{n_2+1}\Lambda_{\text{4d}}^{2N}}
{\Lambda_1^{n_1+n_2}\Lambda_2^{n_2+n_3}\Lambda_3^{n_3+n_4}
\prod_{r\in\mathcal{N}_4\cup\widehat{\mathcal{N}}_1}(a_s-a_r)\phantom{\Big|}}
~. \label{Wquiver1}
\end{eqnarray}
We now compare this expression with the result of the localization analysis
at the one ramified instanton level. {From} \eqref{Zso4d5d} 
and \eqref{zexplicit4d}, specified to
the partition $[n_1,\ldots,n_4]$, we find 
\begin{equation}
\label{Zoneinst}
Z_{1-\text{inst}} = -\sum_{I=1}^4 q_I\!
 \int \!\frac{d\chi_I}{2\pi \ii}\, \frac{1}{\epsilon_1}\prod_{s\in\mathcal{N}_I}
\frac{1}{\left(a_{s}-\chi_I + \frac 12 (\epsilon_1 + \hat\epsilon_2)\right)}
\prod_{t\in\mathcal{N}_{I+1}}\frac{1}{\left(\chi_I 
- a_{t} + \frac 12 (\epsilon_1 + \hat\epsilon_2)\right)}~.
\end{equation}
In view of the prescription \eqref{epsilon}, it is clear that the number 
of poles that contribute to a given $\chi_I$-integral depends 
upon whether we close the contour in the 
upper or lower half-planes. Closing the contour in the upper half-plane 
leads to $n_I$ poles 
that contribute, while closing the contour in the lower half-plane leads to $n_{I+1}$ poles that contribute.
Furthermore, the mass dimensions of each $q_I$ is fixed to be $n_I+n_{I+1}$, since the partition function itself is dimensionless. These two facts immediately help us in relating the localization results with the chiral ring analysis%
\footnote{In a purely 2d context, a relation between the solution of chiral ring equations for certain quiver theories and contour integrals has been noticed in \cite{Orlando:2010uu}.}. Indeed, the dimensional argument allows us to express the ramified
instanton counting parameters in terms of the 2d effective scales as follows \cite{Ashok:2017lko}\,\footnote{The signs have been chosen to match the two superpotentials exactly.}:
\begin{equation}
\begin{aligned}
q_1&=(-1)^{n_1}\Lambda_1^{n_1+n_2}~,\quad q_2=(-1)^{n_1+n_2}\Lambda_2^{n_2+n_3}~,\\
q_3&=(-1)^{n_1+n_2+n_3}\Lambda_3^{n_3+n_4}~,\quad q_4=\frac{(-1)^{n_2+n_4}\Lambda_{\text{4d}}^{2N}}{\Lambda_1^{n_1+n_2}\Lambda_2^{n_2+n_3}\Lambda_3^{n_3+n_4}}~.
\end{aligned}
\label{qlambda}
\end{equation}
Using \eqref{run}, the first three $q_I$ can also be written in terms of the bare complexified FI 
parameters $\tau_I$ of the three 2d nodes as
\begin{equation}
\begin{aligned}
q_1&=\rme^{2 \pi \ii\, \tau_1}\,(-1)^{n_1} \,\mu^{n_1+n_2}~,\\
q_2&=\rme^{2 \pi \ii\, \tau_2}\,(-1)^{n_1+n_2}\,\mu^{n_2+n_3}~,\\
q_3&=\rme^{2 \pi \ii\, \tau_3}\,(-1)^{n_1+n_2+n_3}\,\mu^{n_3+n_4}~.
\label{tauqI}
\end{aligned}
\end{equation}
Once the identification \eqref{qlambda} is made, we can match the number and the structure 
of the terms that appear in \eqref{Wquiver1} by closing the contours for $\chi_1$, $\chi_2$ and $\chi_3$ in the upper half-plane, and the contour of $\chi_4$ in the lower half-plane. 
We denote this choice of contours as $\big(\chi_1|_+,\chi_2|_+,\chi_3|_+,\chi_4|_-\big)$. 
Indeed, computing the corresponding residues and extracting the one-instanton
twisted superpotential from \eqref{FandW} and \eqref{Zoneinst}, we find
\begin{equation}
\begin{aligned}
\label{Wloc+++-}
\mathcal{W}_{1-\text{inst}}&=\sum_{s\in\mathcal{N}_1}\frac{(-1)^{n_1}q_1}{
\prod_{r\in\widehat{\mathcal{N}}_1\cup\mathcal{N}_2}(a_s-a_r)\phantom{\Big|}}
+\sum_{t\in\mathcal{N}_2}\frac{(-1)^{n_2}q_2}{
\prod_{r\in\widehat{\mathcal{N}}_2\cup\mathcal{N}_3}(a_t-a_r)\phantom{\Big|}}\\
&\quad+\sum_{u\in\mathcal{N}_3}\frac{(-1)^{n_3}q_3}{
\prod_{r\in\widehat{\mathcal{N}}_3\cup\mathcal{N}_4}(a_u-a_r)\phantom{\Big|}}
+\sum_{s\in\mathcal{N}_1}
\frac{(-1)^{n_4+1}q_4}{\prod_{r\in\mathcal{N}_4\cup\widehat{\mathcal{N}}_1}(a_s-a_r)\phantom{\Big|}}
\end{aligned}
\end{equation}
which, term by term, exactly matches the superpotential \eqref{Wquiver1} obtained by solving the
twisted chiral ring equations.

\section{2d Seiberg duality}
\label{Seiberg Duality}

The notion of Seiberg duality in 4d gauge theories \cite{Seiberg:1994pq} can be generalized to
two dimensions (see for example \cite{Benini:2014mia}). Thus, by applying 2d Seiberg duality it is
possible to obtain distinct quiver theories in the UV that have the same IR behaviour.

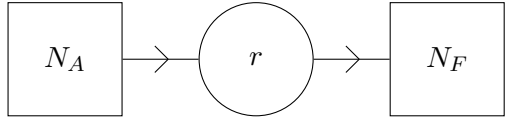
\begin{figure}[H]
\begin{centering}
\begin{tikzpicture}[decoration={
markings,
mark=at position 0.6 with {\draw (-5pt,-5pt) -- (0pt,0pt);
                \draw (-5pt,5pt) -- (0pt,0pt);}}]
  \matrix[row sep=10mm,column sep=5mm, ampersand replacement=\&] {
      \node(g1)[flavor] {$N_A$};  \& \& \node(g2)[gauge] {$r$}; 
      \& \& \node(g3)[flavor]{$N_F$};\\
  };
\graph{(g1) --[postaction={decorate}](g2)--[postaction={decorate}](g3);};
\end{tikzpicture}
\caption{A single 2d gauge node of rank $r$ with $N_F$ fundamental and $N_A$ anti-fundamental flavours attached to it.}
\label{PreBasicDuality}
\end{centering}
\end{figure}

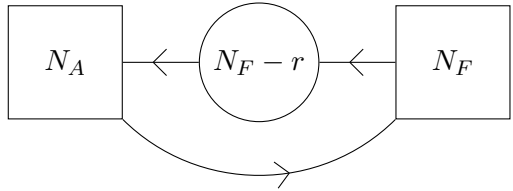
\begin{figure}[H]
\begin{centering}
\begin{tikzpicture}[decoration={
markings,
mark=at position 0.6 with {\draw (-5pt,-5pt) -- (0pt,0pt);
                \draw (-5pt,5pt) -- (0pt,0pt);}}]
  \matrix[row sep=10mm,column sep=5mm,ampersand replacement=\&] {
      \node(g1)[flavor] {$N_A$};  \& \& \node(g2)[gauge] {$N_F-r$}; 
      \& \&\node(g3)[flavor]{$N_F$};\\
  };
\graph{(g2) --[postaction={decorate}](g1) (g3)--[postaction={decorate}](g2);};
\graph{(g1) --[out=-45,in=-135,postaction={decorate}](g3)};
\end{tikzpicture}
\caption{The theory obtained after a 2d Seiberg duality on the gauge node in Fig.~\ref{PreBasicDuality}.}
\label{PostBasicDuality}
\end{centering}
\end{figure}

Let us first consider the simplest case, shown  in Fig.~\ref{PreBasicDuality}.
\noindent
This is a 2d U($r$) gauge theory with $N_F$ fundamental flavours and $N_A$ 
anti-fundamental flavours. For definiteness we take $N_F > N_A$, and call this system ``theory A''.
Its classical twisted superpotential is simply
\begin{align}
\mathcal{W}^{\text{A}}_{\text{cl}}=2\pi\ii\tau \,\Tr \sigma~.
\end{align}
We now perform a Seiberg duality, and obtain ``theory B'', which is described by the quiver in 
Fig.~\ref{PostBasicDuality}.
\noindent
Under the duality, the rank of the gauge group changes as
\begin{align}
r\longrightarrow r'&=\text{max}(N_F,N_A)-r
=N_F-r ~,
\end{align}
and the roles of the fundamental and anti-fundamental flavours are exchanged as denoted by the reversal 
of the arrows.
The classical  twisted superpotential for ``theory B'' is\,\footnote{In addition, an ordinary superpotential term is generated, but 
it plays no role in our discussion.}
\begin{align}
\label{dualW}
\mathcal{W}^{\text{B}}_{\text{cl}}=-2\pi\ii\,\tau\,\Tr \sigma'+2\pi \ii\,\tau\sum_{f=1}^{N_F}m_f
\end{align}
where $\sigma'$ denotes the twisted chiral superfield in the vector multiplet of the dualized node and 
$m_f$ are the twisted masses that completely break the flavour symmetry to its Cartan subgroup. 

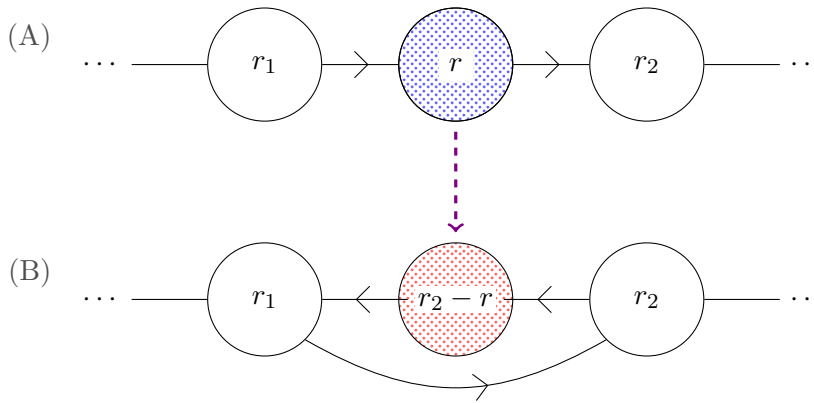
\begin{figure}[H]
\begin{center}
\begin{tikzpicture}[decoration={
markings,
mark=at position 0.6 with {\draw (-5pt,-5pt) -- (0pt,0pt);
                \draw (-5pt,5pt) -- (0pt,0pt);}}]
  \matrix(quiver)[row sep=10mm,column sep=5mm] {
      \node(AdotsL){$\cdots$} ;&&\node(Ag1)[gauge] {$r_1$};  & & \node(Ag2)[gauge] {$r$}; 
      & &\node(Ag3)[gauge]{$r_{2}$};&&\node(AdotsR){$\cdots$} ;\\[.6cm]
      \node(BdotsL){$\cdots$} ;&& \node(Bg1)[gauge] {$r_1$};  & & \node(Bg2)[] {$r_2-r$}; 
      & &\node(Bg3)[gauge]{$r_{2}$};&&\node(BdotsR){$\cdots$} ;\\
  };
\graph[edges={postaction={decorate}}]{
(Ag1) --(Ag2)--(Ag3);
(Bg3) --(Bg2)--(Bg1);
(Bg1)--[left anchor=south east, right anchor=south west, bend right,looseness=1.1](Bg3);
};
\graph[]{
(AdotsL) --(Ag1);
(Ag3) --(AdotsR);
(BdotsL) --(Bg1);
(Bg3) --(BdotsR);
};

\begin{scope}[on background layer]
\node(quiverA) [fill=white,fit=(AdotsL) (AdotsR)] {};
\node[black!70,anchor=east] at (quiverA.north west) {(A)};
\node(quiverB) [fill=white,fit=(BdotsL) (BdotsR)] {};
\node[black!70,anchor=east] at (quiverB.north west) {(B)};
\node(dualin1)[gauge,outer sep=4pt,draw, pattern=crosshatch dots,pattern color=blue!60] at (Ag2){$\phantom{r}$};
\node[fill=white] at (Ag2){$\phantom{r}$};
\node(dualout1)[gauge,outer sep=4pt,draw, pattern=crosshatch dots,pattern color=red!60] at (Bg2){$\phantom{r_2-r}$};
\node[fill=white,align=center,inner sep=1pt] at (Bg2){$\phantom{r_2-r}$};
\graph[edges={violet,very thick,dashed}]{(dualin1)->(dualout1);};
\end{scope}
\end{tikzpicture}
\caption{2d Seiberg duality on a node with both fundamental and anti-fundamental matter with 
$r_2>r_1$. The rank of the dualized node is max$(r_1, r_2)-r = r_2-r$. 
The blue and red colours indicate the node before and after the duality.}
\label{SD2a}
\end{center}
\end{figure}

We now apply this basic duality rule to the quiver theories that describe surface operators. Since
for a given 2d node the flavour symmetry is realized by the adjacent nodes, we can encounter three 
kinds of configurations. The first one is when 
we dualize a gauge node with both fundamental and anti-fundamental fields
in an oriented sequence, as shown in Fig.~\ref{SD2a}.
Before the duality, the classical superpotential for the three relevant nodes is
\begin{align}
\mathcal{W}^{\text{A}}_{\text{cl}}=
\ldots
+2\pi\ii \,\tau_{1}\,\Tr\sigma^{(1)}
+2\pi\ii \,\tau\,\Tr\sigma
+2\pi\ii\,\tau_{2}\,\Tr\sigma^{(2)}
+\ldots~,
\label{Wcl}
\end{align}
while, after duality, it becomes 
\begin{align}
\label{Wdual2}
\mathcal{W}^{\text{B}}_{\text{cl}}=
\ldots
+2\pi\ii\,\tau_{1}\,\Tr\sigma^{(1)}
-2\pi\ii\,\tau\,\Tr{\sigma'}
+2\pi\ii\,(\tau_{2}+\tau)\,\Tr\sigma^{(2)}
+\ldots~.
\end{align}
Here we have taken into account the fact that the role of the twisted masses for the dualized node
is played by the $\sigma$-variables of the $r_2$ node. This explains why the FI parameter $\tau_2$ is
shifted by $\tau$.

\begin{figure}[H]
\begin{center}
\begin{tikzpicture}[decoration={
markings,
mark=at position 0.6 with {\draw (-5pt,-5pt) -- (0pt,0pt);
                \draw (-5pt,5pt) -- (0pt,0pt);}}]
  \matrix(quiver)[row sep=10mm,column sep=5mm] {
      \node(AdotsL){$\cdots$} ;&&\node(Ag1)[gauge] {$r_1$};  & & \node(Ag2)[gauge] {$r$}; 
      & &\node(Ag3)[gauge]{$r_{2}$};&&\node(AdotsR){$\cdots$} ;\\[.6cm]
      \node(BdotsL){$\cdots$} ;&& \node(Bg1)[gauge] {$r_1$};  & & \node(Bg2)[gauge,align=center] {$r_1+r_2$\\$-\, r$}; 
      & &\node(Bg3)[gauge]{$r_{2}$};&&\node(BdotsR){$\cdots$} ;\\
  };
\graph[edges={postaction={decorate}}]{
(Ag2) --(Ag1);(Ag2)--(Ag3);
(Bg1) --(Bg2); (Bg3)--(Bg2);
};
\graph[]{
(AdotsL) --(Ag1);
(Ag3) --(AdotsR);
(BdotsL) --(Bg1);
(Bg3) --(BdotsR);
};
\begin{scope}[on background layer]
\node(quiverA) [fill=white,fit=(AdotsL) (AdotsR)] {};
\node[black!70,anchor=east] at (quiverA.north west) {(A)};
\node(quiverB) [fill=white,fit=(BdotsL) (BdotsR)] {};
\node[black!70,anchor=east] at (quiverB.north west) {(B)};
\node(dualin1)[gauge,outer sep=4pt,draw, pattern=crosshatch dots,pattern color=blue!60] at (Ag2){$\phantom{r}$};
\node[fill=white] at (Ag2){$\phantom{r}$};
\node(dualout1)[gauge,outer sep=4pt,draw, pattern=crosshatch dots,pattern color=red!60,align=center] at (Bg2){$\phantom{r_1+r_2}$\\$\phantom{-\, r}$};
\node[fill=white,align=center,inner sep=1pt] at (Bg2){$\phantom{r_1+r_2}$\\$\phantom{-\, r}$};
\graph[edges={violet,very thick,dashed}]{(dualin1)->(dualout1);};
\end{scope}
\end{tikzpicture}
\caption{2d Seiberg duality on a node with only chiral fundamental matter realized by adjacent 2d gauge nodes. In this case there are no mesonic fields introduced in this case.}
\label{SD1}
\end{center}
\end{figure}
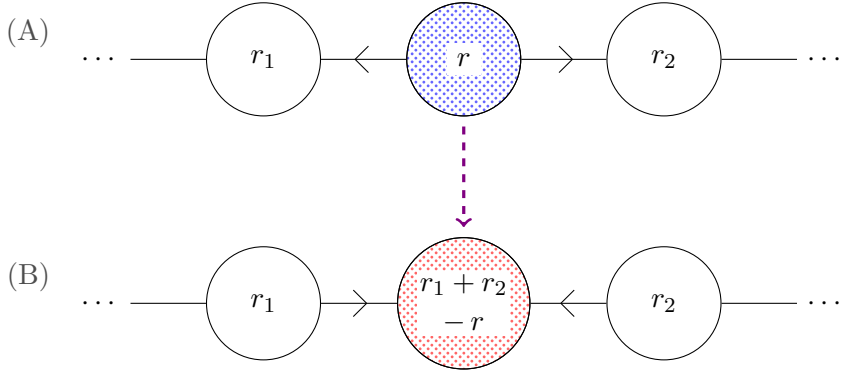

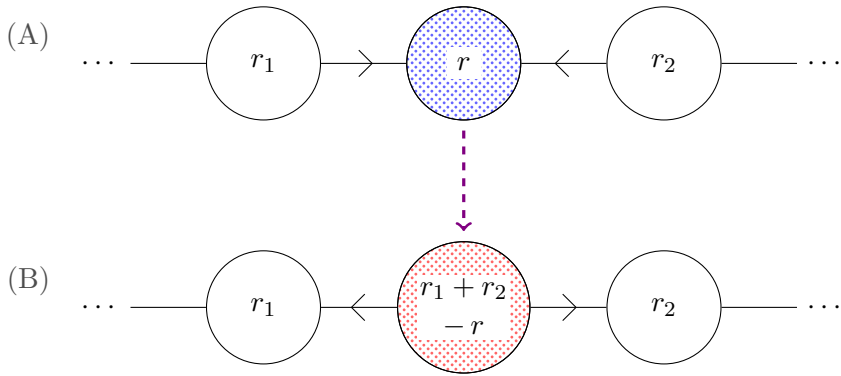
\begin{figure}[H]
\begin{center}
\begin{tikzpicture}[decoration={
markings,
mark=at position 0.6 with {\draw (-5pt,-5pt) -- (0pt,0pt);
                \draw (-5pt,5pt) -- (0pt,0pt);}}]
  \matrix(quiver)[row sep=10mm,column sep=5mm] {
      \node(AdotsL){$\cdots$} ;&&\node(Ag1)[gauge] {$r_1$};  & & \node(Ag2)[gauge] {$r$}; 
      & &\node(Ag3)[gauge]{$r_{2}$};&&\node(AdotsR){$\cdots$} ;\\[.6cm]
      \node(BdotsL){$\cdots$} ;&& \node(Bg1)[gauge] {$r_1$};  & & \node(Bg2)[gauge,align=center] {$r_1+r_2$\\$-\, r$}; 
      & &\node(Bg3)[gauge]{$r_{2}$};&&\node(BdotsR){$\cdots$} ;\\
  };
\graph[edges={postaction={decorate}}]{
(Ag1) --(Ag2);(Ag3)--(Ag2);
(Bg2) --(Bg1); (Bg2)--(Bg3);
};
\graph[]{
(AdotsL) --(Ag1);
(Ag3) --(AdotsR);
(BdotsL) --(Bg1);
(Bg3) --(BdotsR);
};
\begin{scope}[on background layer]
\node(quiverA) [fill=white,fit=(AdotsL) (AdotsR)] {};
\node[black!70,anchor=east] at (quiverA.north west) {(A)};
\node(quiverB) [fill=white,fit=(BdotsL) (BdotsR)] {};
\node[black!70,anchor=east] at (quiverB.north west) {(B)};
\node(dualin1)[gauge,outer sep=4pt,draw, pattern=crosshatch dots,pattern color=blue!60] at (Ag2){$\phantom{r}$};
\node[fill=white] at (Ag2){$\phantom{r}$};
\node(dualout1)[gauge,outer sep=4pt,draw, pattern=crosshatch dots,pattern color=red!60,align=center] at (Bg2){$\phantom{r_1+r_2}$\\$\phantom{-\, r}$};
\node[fill=white,align=center,inner sep=1pt] at (Bg2){$\phantom{r_1+r_2}$\\$\phantom{-\, r}$};
\graph[edges={violet,very thick,dashed}]{(dualin1)->(dualout1);};
\end{scope}
\end{tikzpicture}
\caption{2d Seiberg duality on a node with only anti-chiral fundamental matter realized by adjacent 2d gauge nodes. There are no mesonic fields introduced in this case.}
\label{SDonlyAF}
\end{center}
\end{figure}

The second possibility is when we dualize a node with only fundamental matter, as shown in 
Fig.~\ref{SD1}.
In this case the classical superpotential before the duality is still given by \eqref{Wcl}, but after the duality
it becomes
\begin{align}
\label{Wdual1}
\mathcal{W}^{\text{B}}_{\text{cl}}=
\ldots
+2\pi\ii\,
(\tau_{1}+\tau)\,\Tr\sigma^{(1)}
-2\pi\ii\,\tau\,\Tr{\sigma'}
+2\pi\ii\,(\tau_{2}+\tau)\,\Tr\sigma^{(2)}
+\ldots
\end{align}
because both adjacent nodes provide fundamental matter for the dualized node, and hence 
both FI parameters $\tau_1$ and $\tau_2$ get shifted by $\tau$.
In the third possibility, we dualize a node that has only anti-fundamental matter as shown in Fig.~\ref{SDonlyAF}.
In this case the classical superpotential before the duality is given again by \eqref{Wcl}, but 
after the duality it becomes
\begin{align}
\label{WdualAF}
\mathcal{W}^{\text{B}}_{\text{cl}}=
\ldots
+2\pi\ii\,\tau_{1}\,\Tr\sigma^{(1)}
-2\pi\ii\,\tau\,\Tr{\sigma'}
+2\pi\ii\,\tau_{2}\,\Tr\sigma^{(2)}
+\ldots
\end{align}
with no shifts in $\tau_1$ and $\tau_2$ since the dualized node has no fundamental matter.

\section{Relating quivers and contours}
\label{Relating Quivers and Contours}

In this section we discuss different 2d/4d theories related by Seiberg duality
to the oriented quiver represented in Fig.~\ref{Quiver1-4node}. 
To any of these theories we can
associate a system of twisted chiral ring equations that are distinct from the ones we have 
discussed in Section~\ref{TCRvsLocQ1JK1}. However, being related by 
Seiberg duality, there is 
a simple one-to-one map among them and their solutions. Then, a natural question arises: how is this 
duality map reflected on the localization side?

To answer this question, consider again the oriented quiver 
of Fig.~\ref{Quiver1-4node}, which we now denote by $Q_0$.
{From} it we can generate equivalent quivers by dualizing any of the 2d nodes. We first
carry out a very specific sequence of dualities that are shown in Fig.~\ref{[n1,n2,n3,n4]chain}:
at each step of the duality chain, the node being dualized 
has only fundamental matter. Therefore, Seiberg duality always acts as 
in \eqref{Wdual1}\,\footnote{The same sequence of dualities has also been mentioned
in \cite{Gorsky:2017hro}.}. 

For each quiver in the chain, we can proceed as we did in Section~\ref{TCRvsLocQ1JK1} for $Q_0$.
We integrate out the matter multiplets to obtain the effective twisted chiral superpotential, 
derive from it the twisted chiral ring equations, solve them about a particular massive vacuum 
order by order in the strong coupling scales, evaluate the superpotential 
on the corresponding vacuum and 
finally compare the result with the ramified instanton calculation with a specific integration 
contour for the $\chi_I$ variables. In this program, the choice of the classical 
vacuum is the first important piece of information which we have to provide.

\subsection*{Classical vacuum}

The classical twisted superpotential for the quiver $Q_0$ is 
\begin{align}
\label{W1}
\mathcal{W}_{\text{cl}}^{\,Q_0}=2\pi\ii\,\tau_1\,
\text{Tr}\,\sigma^{(1)}+2\pi\ii\,\tau_2\,\text{Tr}\,\sigma^{(2)}+2\pi\ii\,\tau_3\,\text{Tr}\,\sigma^{(3)}~.
\end{align}
Applying to it the duality rule \eqref{Wdual1}, we obtain the classical superpotential for the quiver 
$Q_1$. With a further duality we obtain the classical superpotential for the quiver $Q_2$ and so
on along the duality chain of Fig.~\ref{[n1,n2,n3,n4]chain}. Explicitly these superpotentials are\,\footnote{For ease of notation we use the same 
symbol $\sigma^{(I)}$ to denote the chiral superfield before and after the duality.}:
\begin{equation}
\label{Wlist}
\begin{aligned}
\mathcal{W}_{\text{cl}}^{\,Q_1}&=-2\pi\ii\,\tau_1\,\text{Tr}\,\sigma^{(1)}+2\pi\ii\,(\tau_1+\tau_2)\,
\text{Tr}\,\sigma^{(2)}+2\pi\ii\,\tau_3\,\text{Tr}\,\sigma^{(3)}~,\phantom{\Big|}\\
\mathcal{W}_{\text{cl}}^{\,Q_2}&=2\pi\ii\,\tau_2\,\text{Tr}\,\sigma^{(1)}
-2\pi\ii\,(\tau_1+\tau_2)\,\text{Tr}\,\sigma^{(2)}+
2\pi\ii\,(\tau_1+\tau_2+\tau_3)\,\text{Tr}\,\sigma^{(3)}~,\phantom{\Big|}\\
\mathcal{W}_{\text{cl}}^{\,Q_4}&=2\pi\ii\,\tau_2\,\text{Tr}\,\sigma^{(1)}+2\pi\ii\,\tau_3\,
\text{Tr}\,\sigma^{(2)}-2\pi\ii\,(\tau_1+\tau_2+\tau_3)\,\text{Tr}\,\sigma^{(3)}~,\phantom{\Big|}\\
\mathcal{W}_{\text{cl}}^{\,Q_5}&=-2\pi\ii\,\tau_2\,\text{Tr}\,\sigma^{(1)}
+2\pi\ii\,(\tau_2+\tau_3)\,\text{Tr}\,\sigma^{(2)}-
2\pi\ii\,(\tau_1+\tau_2+\tau_3)\,\text{Tr}\,\sigma^{(3)}~,\phantom{\Big|}\\
\mathcal{W}_{\text{cl}}^{\,Q_6}&=2\pi\ii\,\tau_3\,\text{Tr}\,\sigma^{(1)}-
2\pi\ii\,(\tau_2+\tau_3)\,\text{Tr}\,\sigma^{(2)}
-2\pi\ii\,\tau_1\text{Tr}\,\sigma^{(3)}~,\phantom{\Big|}\\
\mathcal{W}_{\text{cl}}^{\,Q_7}&=-2\pi\ii\,\tau_3\,\text{Tr}\,\sigma^{(1)}-2\pi\ii\,\tau_2\,
\text{Tr}\,\sigma^{(2)}-2\pi\ii\,\tau_1\,\text{Tr}\,\sigma^{(3)}~.\phantom{\Big|}
\end{aligned}
\end{equation}

\begin{figure}[H]
\begin{centering}
\begin{tikzpicture}[decoration={
markings,
mark=at position 0.6 with {\draw (-5pt,-5pt) -- (0pt,0pt);
                \draw (-5pt,5pt) -- (0pt,0pt);}}]
  \matrix(quiver)[row sep=10mm,column sep=5mm] {
      \node(Ag1)[gauge] {$n_1$};  & & \node(Ag2)[gauge] {$n_1+n_2$}; & & \node(Ag3)[gauge,align=center] {$n_1 + n_2$\\$+\, n_3$}; 
      & &\node(Agf)[gaugedflavor]{$N$};\\
      \node(Bg1)[gauge] {$n_2$};  & & \node(Bg2)[gauge] {$n_1+n_2$}; & & \node(Bg3)[gauge,align=center] {$n_1 + n_2$\\$+\, n_3$}; 
      & &\node(Bgf)[gaugedflavor]{$N$};\\
      \node(Cg1)[gauge] {$n_2$};  & & \node(Cg2)[gauge] {$n_2+n_3$}; & & \node(Cg3)[gauge,align=center] {$n_1 + n_2$\\$+\, n_3$}; 
      & &\node(Cgf)[gaugedflavor]{$N$};\\
      \node(Dg1)[gauge] {$n_2$};  & & \node(Dg2)[gauge] {$n_2+n_3$}; & & \node(Dg3)[gauge,align=center] {$n_2 + n_3$\\$+\, n_4$}; 
      & &\node(Dgf)[gaugedflavor]{$N$};\\
      \node(Eg1)[gauge] {$n_3$};  & & \node(Eg2)[gauge] {$n_2+n_3$}; & & \node(Eg3)[gauge,align=center] {$n_2 + n_3$\\$+\, n_4$}; 
      & &\node(Egf)[gaugedflavor]{$N$};\\
      \node(Fg1)[gauge] {$n_3$};  & & \node(Fg2)[gauge] {$n_3+n_4$}; & & \node(Fg3)[gauge,align=center] {$n_2 + n_3$\\$+\, n_4$}; 
      & &\node(Fgf)[gaugedflavor]{$N$};\\
      \node(Gg1)[gauge] {$n_4$};  & & \node(Gg2)[gauge] {$n_3+n_4$}; & & \node(Gg3)[gauge,align=center] {$n_2 + n_3$\\$+\, n_4$}; 
      & &\node(Ggf)[gaugedflavor]{$N$};\\
  };
\graph[edges={postaction={decorate}}]{
(Ag1) --(Ag2)--(Ag3)--(Agf);
(Bg2) --(Bg1); (Bg2)--(Bg3)--(Bgf);
(Cg1) --(Cg2); (Cg3)--(Cg2); (Cg3)--(Cgf);
(Dg1) --(Dg2)--(Dg3);  (Dgf)--(Dg3);
(Eg2) --(Eg1); (Eg2)--(Eg3); (Egf)--(Eg3);
(Fg1) --(Fg2); (Fg3)--(Fg2); (Fgf)--(Fg3);
(Gg2) --(Gg1); (Gg3)--(Gg2); (Ggf)--(Gg3);
};

\begin{scope}[on background layer]
\node(quiverA) [fill=white,fit=(Ag1) (Agf)] {};
\node[black!70,anchor=east] at (quiverA.north west) {$Q_0$};
\node(quiverB) [fill=white,fit=(Bg1) (Bgf)] {};
\node[black!70,anchor=east] at (quiverB.north west) {$Q_1$};
\node(quiverC) [fill=white,fit=(Cg1) (Cgf)] {};
\node[black!70,anchor=east] at (quiverC.north west) {$Q_2$};
\node(quiverD) [fill=white,fit=(Dg1) (Dgf)] {};
\node[black!70,anchor=east] at (quiverD.north west) {$Q_4$};
\node(quiverE) [fill=white,fit=(Eg1) (Egf)] {};
\node[black!70,anchor=east] at (quiverE.north west) {$Q_5$};
\node(quiverF) [fill=white,fit=(Fg1) (Fgf)] {};
\node[black!70,anchor=east] at (quiverF.north west) {$Q_6$};
\node(quiverG) [fill=white,fit=(Gg1) (Ggf)] {};
\node[black!70,anchor=east] at (quiverG.north west) {$Q_7$};
\node(dualin1)[gauge,outer sep=4pt,draw, pattern=crosshatch dots,pattern color=blue!60] at (Ag1){$\phantom{n_1}$};
\node[fill=white] at (Ag1){$\phantom{n_1}$};
\node(dualout1)[gauge,outer sep=4pt,draw, pattern=crosshatch dots,pattern color=red!60] at (Bg1){$\phantom{n_2}$};
\node[fill=white] at (Bg1){$\phantom{n_2}$};
\node(dualin2)[gauge,outer sep=4pt,pattern=crosshatch dots,pattern color=blue!60] at (Bg2){$\phantom{n_1+n_2}$};
\node[fill=white,inner sep=2pt] at (Bg2){$\phantom{n_1+n_2}$};
\node(dualout2)[gauge,outer sep=4pt,draw, pattern=crosshatch dots,pattern color=red!60] at (Cg2){$\phantom{n_2+n_3}$};
\node[fill=white,inner sep=2pt] at (Cg2){$\phantom{n_2+n_3}$};
\node(dualin3)[gauge,outer sep=4pt,align=center,draw, pattern=crosshatch dots,pattern color=blue!60] at (Cg3){$\phantom{n_1 + n_2}$\\$\phantom{+\, n_3}$};
\node[fill=white,align=center,inner sep=1pt] at (Cg3){$\phantom{n_1 + n_2}$\\$\phantom{+\, n_3}$};
\node(dualout3)[gauge,outer sep=4pt,align=center,draw, pattern=crosshatch dots,pattern color=red!60] at (Dg3){$\phantom{n_2 + n_3}$\\$\phantom{+\, n_4}$};
\node[fill=white,align=center,inner sep=1pt] at (Dg3){$\phantom{n_2 + n_3}$\\$\phantom{+\, n_4}$};
\node(dualin4)[gauge,outer sep=4pt,draw, pattern=crosshatch dots,pattern color=blue!60] at (Dg1){$\phantom{n_2}$};
\node[fill=white] at (Dg1){$\phantom{n_2}$};
\node(dualout4)[gauge,outer sep=4pt,draw, pattern=crosshatch dots,pattern color=red!60] at (Eg1){$\phantom{n_3}$};
\node[fill=white] at (Eg1){$\phantom{n_3}$};
\node(dualin5)[gauge,outer sep=4pt,pattern=crosshatch dots,pattern color=blue!60] at (Eg2){$\phantom{n_2+n_3}$};
\node[fill=white,inner sep=2pt] at (Eg2){$\phantom{n_2+n_3}$};
\node(dualout5)[gauge,outer sep=4pt,draw, pattern=crosshatch dots,pattern color=red!60] at (Fg2){$\phantom{n_3+n_4}$};
\node[fill=white,inner sep=2pt] at (Fg2){$\phantom{n_3+n_4}$};
\node(dualin6)[gauge,outer sep=4pt,align=center,draw, pattern=crosshatch dots,pattern color=blue!60] at (Fg1){$\phantom{n_3}$};
\node[fill=white,align=center] at (Fg1){$\phantom{n_3}$};
\node(dualout6)[gauge,outer sep=4pt,align=center,draw, pattern=crosshatch dots,pattern color=red!60] at (Gg1){$\phantom{n_4}$};
\node[fill=white,align=center] at (Gg1){$\phantom{n_4}$};
\graph[edges={violet,very thick,dashed}]{(dualin1)->(dualout1) (dualin2)->(dualout2) (dualin3)->(dualout3) (dualin4)->(dualout4) (dualin5)->(dualout5) (dualin6)->(dualout6);};
\end{scope}
\end{tikzpicture}
\vspace{0.5cm}
\caption{A sequence of Seiberg dualities obtained by dualizing the node that has only 
fundamental flavours attached to it at each step. The node that is dualized is indicated 
by the blue arrow. The reason why in the list of names for the quivers we skipped $Q_3$ 
will become clear later on.}
\label{[n1,n2,n3,n4]chain}
\end{centering}
\end{figure}
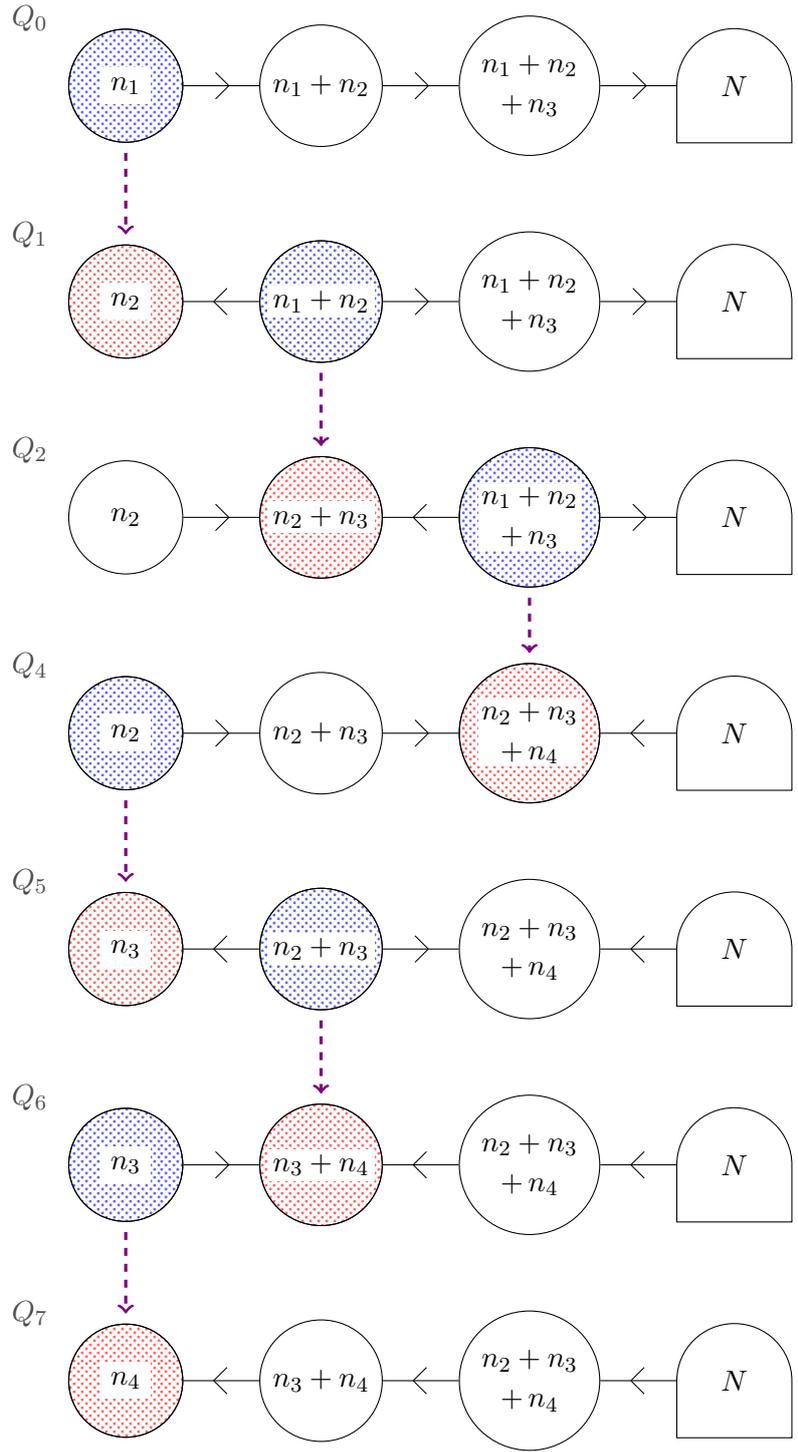
{From} these expressions we can read the map between the FI parameters of 
any quiver and those of the initial quiver $Q_0$. For example, for $Q_1$ we have
\begin{equation}
\label{tauefftauQ2}
\tau_1^{Q_1} = -\tau_1~, \qquad \tau_2^{Q_1} = \tau_1 + \tau_2~ , 
\qquad \tau_3^{Q_1} = \tau_3~,
\end{equation}
while for the quiver $Q_2$ we have
\begin{equation}
\label{tauefftauQ3}
\tau_1^{Q_2} = \tau_2~, \qquad \tau_2^{Q_2} = -\tau_1 - \tau_2~ , 
\qquad \tau_3^{Q_2} = \tau_1+\tau_2+\tau_3~.
\end{equation}

The next step is to identify the classical vacuum for each quiver. We already know that for 
$Q_0$ the vacuum that respects the partition $[n_1, \ldots, n_4]$ associated to the
surface operator, is (see \eqref{linearclassicalvev})
\begin{equation}
\sigma^{(1)}_{\text{cl}} = {\mathcal A}_1~,\qquad 
\sigma^{(2)}_{\text{cl}} = {\mathcal A}_1\oplus{\mathcal A}_2~,\qquad 
\sigma^{(3)}_{\text{cl}} = {\mathcal A}_1\oplus{\mathcal A}_2\oplus{\mathcal A}_3~.
\label{vacuumQ1}
\end{equation}
Since Seiberg duality is an exact infrared equivalence, the classical superpotentials 
of two dual quivers, evaluated in the respective vacua, should be identical.
This requirement immediately fixes the structure of the classical vacuum 
for all quivers.
For instance, for $Q_1$ one can check that
\begin{equation}
\sigma^{(1)}_{\text{cl}} = {\mathcal A}_2~,\qquad 
\sigma^{(2)}_{\text{cl}} = {\mathcal A}_1\oplus{\mathcal A}_2~,\qquad
\sigma^{(3)}_{\text{cl}} = {\mathcal A}_1\oplus{\mathcal A}_2\oplus{\mathcal A}_3~,
\label{vacuumQ2}
\end{equation}
leads to the desired match; indeed
\begin{eqnarray}
{\mathcal W}_{\text{cl}}^{\,Q_1} &=& -2\pi\ii\,\tau_1 \,\Tr {\mathcal A}_2 
+ 2\pi\ii\,(\tau_1+\tau_2)\,
\big(\Tr {\mathcal A}_1+\Tr {\mathcal A}_2\big)+ 2\pi\ii\,\tau_3\,
\big(\Tr{\mathcal A}_1+\Tr{\mathcal A}_2+\Tr {\mathcal A}_3\big)\nonumber\phantom{\Big|}\\
&=&2\pi\ii\,\tau_1 \,\Tr{\mathcal A}_1 + 2\pi\ii\,\tau_2\,
\big(\Tr {\mathcal A}_1+\Tr {\mathcal A}_2\big) 
+ 2\pi\ii\,\tau_3\,\big(\Tr{\mathcal A}_1+\Tr{\mathcal A}_2+
\Tr {\mathcal A}_3\big)\nonumber\phantom{\Big|}\\
&=& {\mathcal W}_{\text{cl}}^{\,Q_0} ~.\phantom{\Big|}
\end{eqnarray}
This calculation can be easily generalized to all other quivers in the duality chain and the results are
summarized in Tab.~\ref{vevs1}.
\begin{table}[H]
\begin{centering}
\small{
\begin{tabular}{|c|c|c|c|}
\hline
\,\,Quiver \phantom{\bigg|}& $\sigma^{(1)}_{\text{cl}}$ & $\sigma^{(2)}_{\text{cl}}$ 
& $\sigma^{(3)}_{\text{cl}}$ \\
\hline
\hline
$\,\,Q_0 \phantom{\Big|}$& ${\mathcal A}_1$ & ${\mathcal A}_1\oplus {\mathcal A}_2$ 
& ${\mathcal A}_1\oplus {\mathcal A}_2 \oplus {\mathcal A}_3$  \\
\hline
$\,\,Q_1 \phantom{\Big|}$&  ${\mathcal A}_2$ & ${\mathcal A}_1\oplus {\mathcal A}_2$ 
& ${\mathcal A}_1\oplus {\mathcal A}_2 \oplus {\mathcal A}_3$  \\
\hline
$\,\,Q_2 \phantom{\Big|}$&  ${\mathcal A}_2$ & ${\mathcal A}_2\oplus {\mathcal A}_3$ 
& ${\mathcal A}_1\oplus {\mathcal A}_2 \oplus {\mathcal A}_3$  \\
\hline
$\,\,Q_4 \phantom{\Big|}$&  ${\mathcal A}_2$ & ${\mathcal A}_2\oplus {\mathcal A}_3$ 
& ${\mathcal A}_2\oplus {\mathcal A}_3 \oplus {\mathcal A}_4$  \\
\hline
$\,\,Q_5 \phantom{\Big|}$&  ${\mathcal A}_3$ & ${\mathcal A}_2\oplus {\mathcal A}_3$ 
& ${\mathcal A}_2\oplus {\mathcal A}_3 \oplus {\mathcal A}_4$  \\
\hline
$\,\,Q_6 \phantom{\Big|}$&  ${\mathcal A}_3$ & ${\mathcal A}_3\oplus {\mathcal A}_4$ 
& ${\mathcal A}_2\oplus {\mathcal A}_3 \oplus {\mathcal A}_4$  \\
\hline
$\,\,Q_7 \phantom{\Big|}$ &  ${\mathcal A}_4$ & ${\mathcal A}_3\oplus {\mathcal A}_4$ 
& ${\mathcal A}_2\oplus {\mathcal A}_3 \oplus {\mathcal A}_4$  \\
\hline
\end{tabular}
}
\caption{For each of the quivers in Fig.~\ref{[n1,n2,n3,n4]chain}, we list the classical 
expectation values of the twisted chiral fields in each of the three 2d nodes. 
Using them in the classical twisted chiral superpotentials given in \eqref{Wlist}, one finds 
identical expressions.}
\label{vevs1}
\end{centering}
\end{table}

\subsection*{The $q$ vs $\Lambda$ map}
The next necessary ingredient is the relation between the ramified instanton parameters $q_I$ and the 
strong coupling scales $\Lambda_I^{Q_i}$ of a given quiver. 

For the first quiver $Q_0$, the $q$ vs $\Lambda$ map was already
derived and written in \eqref{qlambda}. If we now consider the second quiver $Q_1$, from the
running of the FI parameters we find
\begin{equation}
\begin{aligned}
\big(\Lambda^{Q_1}_1\big)^{-n_1-n_2}
\,&=\, \rme^{2\pi\ii\, \tau_1^{Q_1}}\,\mu^{-n_1-n_2}~,\\
\big(\Lambda^{Q_1}_2\big)^{n_1+2n_2+n_3}
\,&=\, \rme^{2\pi\ii\, \tau_2^{Q_1}}\,\mu^{n_1+2n_2+n_3}~,\\
\big(\Lambda^{Q_1}_3\big)^{n_3+n_4}\,&=\, \rme^{2\pi\ii\, \tau_3^{Q_1}}\,\mu^{n_3+n_4}~.
\end{aligned}
\label{LQmu1}
\end{equation}
Using the relations \eqref{tauefftauQ2} and the definitions \eqref{tauqI}, it is easy to 
obtain (up to inessential signs) the $q$ vs $\Lambda$ map in this case, namely
\begin{equation}
\label{Q2qmap}
q_1 \sim \big(\Lambda^{Q_1}_1\big)^{n_1+n_2}~,
\quad q_2 \sim \frac{\big(\Lambda^{Q_1}_2\big)^{n_1+2n_2+n_3}}{\big(\Lambda^{Q_1}_1\big)^{n_1+n_2}}
~,\quad q_3 \sim \big(\Lambda^{Q_1}_3\big)^{n_3+n_4}~.
\end{equation}
Applying the same procedure to $Q_2$ and using \eqref{tauefftauQ3}, we find
\begin{equation}
q_1 \sim \frac{\big(\Lambda^{Q_2}_2\big)^{n_1+2n_2+n_3}}{\big(\Lambda^{Q_2}_1\big)^{n_2+n_3}}~,\quad 
q_2 \sim \big(\Lambda^{Q_2}_1\big)^{n_2+n_3}~,\quad 
q_3 \sim \frac{\big(\Lambda^{Q_2}_3\big)^{N+n_2+n_3}}{\big(\Lambda^{Q_2}_2\big)^{n_1+2n_2+n_3}}~.
\end{equation}
Repeating this analysis for all quivers of Fig.~\ref{[n1,n2,n3,n4]chain}, we obtain the results 
collected in Tab.~\ref{mapTable}. 
\begin{table}[H]
\begin{centering}
\small{
\begin{tabular}{|c|c|c|c|}
\hline
\,\,Quiver\phantom{\Big|} & $q_1$ & $q_2$ & $q_3$ 
\\
\hline
\hline
$\,Q_0\phantom{\bigg|}$& $\Lambda_1^{n_1+n_2}$ 
& $\Lambda_2^{n_2+n_3}$ & $\Lambda_3^{n_3+n_4}$ 
\\
\hline
$\,Q_1\phantom{\Bigg|}$& $\big(\Lambda^{Q_1}_1\big)^{n_1+n_2}$ 
& $\frac{\big(\Lambda^{Q_1}_2\big)^{n_1+2n_2+n_3\phantom{\big|}}}{\big(\Lambda^{Q_1}_1\big)^{n_1+n_2}}$ & $\big(\Lambda^{Q_1}_3\big)^{n_3+n_4}$ 
\\
\hline
$\,Q_2\phantom{\Bigg|}$& $\frac{\big(\Lambda^{Q_2}_2\big)^{n_1+2n_2+n_3\phantom{\big|}}}
{\big(\Lambda^{Q_2}_1\big)^{n_2+n_3}}$ 
& $\big(\Lambda^{Q_2}_1\big)^{n_2+n_3}$ 
& $\frac{\big(\Lambda^{Q_2}_3\big)^{N+n_2+n_3\phantom{\big|}}}{\big(\Lambda^{Q_2}_2\big)^{n1+2n_2+n_3}}$ 
\\
\hline
$\,Q_4\phantom{\Bigg|}$ & 
$\frac{\big(\Lambda^{Q_4}_3\big)^{N+n_2+n_3\phantom{\big|}}}{\big(\Lambda^{Q_4}_1\big)^{n_2+n_3} \big(\Lambda^{Q_4}_2\big)^{n_3+n_4}}$ 
& $\big(\Lambda^{Q_4}_1\big)^{n_2+n_3}$ & $\big(\Lambda^{Q_4}_2\big)^{n_3+n_4}$ 
\\
\hline
$\,Q_5\phantom{\Bigg|}$ & 
$\frac{\big(\Lambda^{Q_5}_3\big)^{N+n_2+n_3\phantom{\big|}}}{\big(\Lambda^{Q_5}_2\big)^{n_2+2n_3+n_4}}$ & $\big(\Lambda^{Q_5}_1\big)^{n_2+n_3}$ 
& $\frac{\big(\Lambda^{Q_5}_2\big)^{n_2+2n_3+n_4\phantom{\big|}}}{\big(\Lambda^{Q_5}_1\big)^{n_2+n_3}}$ 
\\
\hline
$\,Q_6\phantom{\Bigg|}$ & $\big(\Lambda^{Q_6}_3\big)^{n_1+n_2}$ & 
$\frac{\big(\Lambda^{Q_6}_2\big)^{n_2+2n_3+n_4\phantom{\big|}}}{\big(\Lambda^{Q_6}_1\big)^{n_3+n_4}}$ & $\big(\Lambda^{Q_6}_1\big)^{n_3+n_4}$ 
\\
\hline
$\,Q_7\phantom{\bigg|}$ &$\big(\Lambda^{Q_7}_3\big)^{n_1+n_2}$ & 
$\big(\Lambda^{Q_7}_2\big)^{n_2+n_3}$ & $\big(\Lambda^{Q_7}_1\big)^{n_3+n_4}$ 
\\
\hline
\end{tabular}
}
\caption{For each quiver of Fig.~\ref{[n1,n2,n3,n4]chain}, we list the $q$ vs $\Lambda$ 
map (up to sign factors, which can be found in Appendix~\ref{appB}). The exponent of each strong coupling scale is determined by the 
number of effective flavours at that node in the quiver and is related to the $\beta$-function coefficient
of the corresponding FI parameter. 
}
\label{mapTable}
\end{centering}
\end{table}
We finally recall that the following relation 
\begin{equation}
q_4 \sim \frac{\Lambda_{\text{4d}}^{2N}}{q_1\,q_2\,q_3}
\label{q4}
\end{equation}
holds for all quivers.

{From} Tab.~\ref{mapTable}, we observe that except for the oriented quivers $Q_0$ and $Q_7$, in
all other cases the contributions of a single ramified instanton can be proportional 
to a ratio of strong coupling scales. It would be interesting 
to understand the origin of this fact from the perspective of vortex solutions 
in 2d quivers with bi-fundamental matter. However, for our present purposes 
it is important to keep in mind
that the ramified instanton partition function is a power series in $q_I$. This 
means that, except for the quivers $Q_0$ and $Q_7$, we are forced to have some 
hierarchy among the scales 
$\Lambda_I^{Q_i}$ in order for the $q$ vs $\Lambda$ map to be consistent with 
the power series expansion of the ramified instanton partition function. 
For instance for the quiver
$Q_1$, we see from Table~\ref{mapTable} that if we want that both $q_1$ and $q_2$ 
be ``small'', it is necessary to have
\begin{equation}
1 ~>~ \left| \frac{\Lambda^{Q_1}_1}{\mu} \right|^{n_1+n_2}~ \gg ~\left|
\frac{\Lambda^{Q_1}_2}{\mu} \right|^{n_1+2n_2+n_3} ~.
\label{scaleQ1}
\end{equation}
Using (\ref{LQmu1}) and the fact that the $\beta$-function coefficient of the first node is negative, we can easily see that (\ref{scaleQ1}) is equivalent to 
\begin{equation}
0<-\zeta_1^{Q_1} \ll \zeta_2^{Q_1}~.
\label{newzetaQ1}
\end{equation}
Notice that this inequality follows from the
duality relations (\ref{tauefftauQ2}): indeed, $\zeta_1^{Q_1}=-\zeta_1$ and
$\zeta_2^{Q_1}=\zeta_1+\zeta_2$, with $\zeta_I>0$ 
as indicated in (\ref{zetapositive}).

In a similar way, for quiver $Q_2$ we see from Tab.~\ref{mapTable} that in order 
for the instanton weights $q_I$ to be ``small'', we must have
\begin{equation}
1 ~>~ \left| \frac{\Lambda^{Q_2}_1}{\mu} \right|^{n_2+n_3}~ \gg ~\left|
\frac{\Lambda^{Q_2}_2}{\mu} \right|^{n_1+2n_2+n_3} 
~ \gg ~
\left|\frac{\Lambda^{Q_2}_3}{\mu} \right|^{N+n_2+n_3} ~,
\label{scaleQ2}
\end{equation}
which, taking into account the signs of the $\beta$-function coefficients, in this case implies that
\begin{equation}
0<\zeta_1^{Q_2} \ll -\zeta_2^{Q_2} \ll \zeta_3^{Q_2}~.
\label{zetaQ2}
\end{equation}
Again we can check that this hierarchy just follows from the duality relations
(\ref{tauefftauQ3}), since $\zeta_1^{Q_2}=\zeta_2$, $\zeta_2^{Q_2}=-\zeta_1-\zeta_2$ and $\zeta_3^{Q_2}=\zeta_1+\zeta_2+\zeta_3$, with $\zeta_I>0$.

We can repeat this analysis for all linear quivers of the sequence, and always find the same pattern: when a hierarchy of scales is needed in order to have a meaningful ramified instanton expansion, this is automatically guaranteed
by the duality relations among the real FI parameters of the various quivers.
Moreover, the 4d low-energy scale $\Lambda_{\text{4d}}$ is always the smallest scale in view of (\ref{q4}).

\subsection{Contour prescriptions for dual quivers}
We now address the question of how the non-perturbative superpotential associated to each quiver 
can be obtained from the ramified instanton partition function \eqref{Zso4d5d} using a 
suitable contour prescription for the $\chi_I$-integrals. In Section~\ref{TCRvsLocQ1JK1} we answered
this question for the oriented quiver $Q_0$ by comparing each term of the solution of the chiral
ring equations with the localization results. Here we provide a general argument that allows 
one to derive the appropriate contour prescription for any quiver of the duality chain, without 
explicitly solving the twisted chiral ring equations and integrating them in.
We perform a detailed analysis at the one-instanton level, but our conclusions are valid also at 
higher instantons.

Let us first consider only the three 2d nodes and neglect for the moment the contribution of the
4d node by setting $\Lambda_{\text{4d}}\to 0$ and hence, according to \eqref{q4}, 
$q_4\to 0$. Using the partition function \eqref{Zoneinst}, the
one-instanton superpotential in this case can be written as
\begin{equation}
{\mathcal W}_{\text{1-inst}}= \sum_{I=1}^3 q_I\, w_I
\end{equation}
where
\begin{equation}
w_I = - \lim_{\epsilon_1,\hat\epsilon_2\to 0}\int \!\frac{d\chi_I}{2\pi \ii}\, \prod_{s\in\mathcal{N}_I}
\frac{1}{\left(a_{s}-\chi_I + \frac 12 (\epsilon_1 + \hat\epsilon_2)\right)}
\prod_{t\in\mathcal{N}_{I+1}}\frac{1}{\left(\chi_I 
- a_{t} + \frac 12 (\epsilon_1 + \hat\epsilon_2)\right)}~.
\label{wI}
\end{equation}
{From} this we immediately see that $w_I$ can have either $n_I$ or $n_{I+1}$ terms depending 
on whether the $\chi_I$-contour is closed in the upper or lower half plane, respectively.
On the other hand, exploiting the relation \cite{Ashok:2017lko}
\begin{equation}
\label{logLider2}
\Tr \sigma^{(I)}_{\star}= 
\Tr \big(\sigma^{(I)}_{\text{cl}}+\delta\sigma^{(I)}\big) = 
\frac{1}{b_I}\Lambda_I \frac{\partial \mathcal W}{\partial {\Lambda_I}}\bigg|_{\sigma_\star}~,
\end{equation}  
and the maps in Tab.~\ref{mapTable}, we can understand which type of ramified instantons 
contributes to each term proportional to $\Tr \sigma^{(I)}$. 
For example, for $Q_0$ using the
map \eqref{qlambda}, we find
\begin{equation}
\Tr\delta\sigma^{(1)}= q_1\, w_1~,\quad
\Tr\delta\sigma^{(2)}= q_2\, w_2~,\quad
\Tr\delta\sigma^{(3)}= q_3\, w_3~.
\end{equation}
These relations establish a natural correspondence between the nodes of the quiver 
and the instanton counting parameters $q_I$ and the corresponding $\chi_I$ fields for $I=1,2,3$: indeed, the first node is associated to $\chi_1$, the second 
node to $\chi_2$ and the third node to $\chi_3$. Furthermore, exploiting the fact that
$\delta\sigma^{(I)}$ must have the same structure of the classical part $\sigma^{(I)}_{\text{cl}}$
and hence that their entries can only arise in any of the blocks that make up the rank of 
the corresponding 2d gauge node, we conclude that we have to close 
the integration contour in the upper-half plane for all $\chi_I$, so that
$\Tr\delta\sigma^{(1)}$ has $n_1$ contributions,
$\Tr\delta\sigma^{(2)}$ has $n_2$ contributions and $\Tr\delta\sigma^{(3)}$ has $n_3$ 
contributions. We indicate this choice of integration contour with the notation 
$\big(\chi_1|_+,\chi_2|_+,\chi_3|_+\big)$. 
In this way we have retrieved the same contour prescription of Section~\ref{TCRvsLocQ1JK1}, 
without explicitly solving the twisted chiral ring equations.

The same strategy can be used for the other quivers of the duality chain. Let us consider for
example $Q_1$. {From} \eqref{logLider2} and the map \eqref{Q2qmap},
we find
\begin{equation}
\Tr\delta\sigma^{(1)} =  q_1\, w_1-q_2\, w_2 ~,\quad
\Tr\delta\sigma^{(2)} =  q_2\, w_2 ~,\quad
\Tr\delta\sigma^{(3)} =  q_3\, w_3 ~.
\end{equation}
In this case, the correspondence between the second node and $\chi_2$ and between 
the third node and $\chi_3$ is again obvious, but since now there are two $w_I$ contributing 
to the first trace, we need to use the hierarchy of scales (\ref{scaleQ1}) to disentangle the linear 
combination. In particular we see that the contribution proportional to $q_2$ is 
sub-dominant and thus can be neglected
at leading order. This allows us to conclude that the first node must be 
unambiguously associated to $\chi_1$.
However, the number of terms contributing to $\Tr\delta\sigma^{(1)}$ must be $n_2$, since
for $Q_1$ we have $\sigma^{(1)}_{\text{cl}}=\mathcal{A}_2$ (see \eqref{vacuumQ2}). Thus, the
$\chi_1$-integral should be closed in the lower half-plane to provide this number of terms, while the
integrations over $\chi_2$ and $\chi_3$ must be carried out in the upper half-plane as before. 
In conclusion, to $Q_1$ we assign the contour prescription
$\big(\chi_1|_-,\chi_2|_+,\chi_3|_+\big)$. It can be checked that with this choice the
localization results perfectly agree, term by term, with the solution of the appropriate chiral ring equations
(see Appendix~\ref{appB} for details).

Comparing the classical superpotentials $\mathcal{W}_{\text{cl}}^{\,Q_0}$ and 
$\mathcal{W}_{\text{cl}}^{\,Q_1}$ given in \eqref{W1} and \eqref{Wlist}, we notice that 
an indication for the flipping of the $\chi_1$ integration contour between $Q_0$ 
and $Q_1$ 
can be traced to the change in sign of the term containing $\Tr \sigma^{(1)}$, or equivalently to the 
change in sign of the $\beta$-function coefficient and of the FI parameter
of the first node under the duality map 
from $Q_0$ to $Q_1$. 
We propose that this is in fact the rule, and that it is the sign of the $\beta$-function coefficient
for a given node (or of its FI parameter) that determines whether the contour of integration 
for the corresponding $\chi$ variable has to be closed in the upper or in the 
lower half-plane.

As a simple and non-trivial check of this proposal we consider the quiver $Q_2$. Here, using the 
$q$ vs $\Lambda$ map of Tab.~\ref{mapTable} into \eqref{logLider2}, 
we find
\begin{equation}
\label{tracesQ3}
\Tr\delta\sigma^{(1)}=  q_2\, w_2-q_1\, w_1~,\quad
\Tr\delta\sigma^{(2)}=  q_1\, w_1~,\quad
\Tr\delta\sigma^{(3)}=  q_3\, w_3~.
\end{equation}
{From} the second and third relations respectively, we see that $\chi_1$ is associated to the second 
node and $\chi_3$ to the third node. To decide which $\chi$-variable is associated to the first 
node, we again exploit the hierarchy of scales (\ref{scaleQ2}), which for the case at hand implies 
that $q_1$ is sub-dominant with respect to $q_2$.
Thus, the $q_1$-term in the first relation of \eqref{tauefftauQ3} can be neglected at leading order, 
implying that $\chi_2$ must be associated to the first node.
Notice that it is the second node of $Q_2$ that has a negative $\beta$-function, and hence a negative FI parameter, and so it is again $\chi_1$ that has to be 
integrated in the lower half-plane. We then conclude
that to the quiver $Q_2$ we must assign the contour prescription
$\big(\chi_2|_+,\chi_1|_-,\chi_3|_+\big)$.
A similar analysis can be done for all other quivers of the sequence in Fig.~\ref{[n1,n2,n3,n4]chain}.

Let us now turn to the contour for the last integration variable $\chi_4$. To specify it, we have 
to switch on the dynamics on the 4d node of the quiver, since the corresponding
parameter $q_4$ is non-zero only when
$\Lambda_{\text{4d}}$ is non-zero (see \eqref{q4}). Thus, $q_4$ and hence $\chi_4$
cannot be associated to any of the 2d nodes and must be related to the 4d node. 
By observing the duality chain, we see that the third node, which is the only 2d node connected 
to the 4d node, is dualized precisely once. Until this point the 4d node provides fundamental matter to 
the third 2d node, while from this point on it provides anti-fundamental matter. 
Given that we know that for the initial quiver $Q_0$ the variable $\chi_4$ has to be
integrated in the lower half-plane, we are naturally led to propose that
the contour for $\chi_4$ remains in the lower plane $(-)$ until the 
third node is dualized, {\it{i.e.}} for $Q_0$, $Q_1$ and $Q_2$, and then it flips to the upper 
half-plane $(+)$, remaining unchanged for the rest of the duality chain, 
{\it{i.e.}} for $Q_4$, $Q_5$, $Q_6$ and $Q_7$. We have verified the validity of 
this proposal by explicitly solving the twisted chiral ring equations for all seven quivers  
to obtain the corresponding twisted superpotentials, and checking that 
these agree term by term with what the ramified instanton partition function yields with
the proposed integration prescriptions (see Appendix~\ref{appB} for details).
Our results on the contour assignments for the various quivers
are summarized in Tab.~\ref{contoursTable}.%
\footnote{We remark that the results for the last quiver $Q_7$ coincide with
those derived in Ref.~\cite{Ashok:2017lko}, once the nodes are numbered in the opposite order}

\begin{table}[H]
\begin{centering}
\small{
\begin{tabular}{|c|c|c|c|c|}
\hline
\,\,Quiver\phantom{\bigg|} 
& $\text{sgn}(b_1^{Q_i})$ 
& $\text{sgn}(b_2^{Q_i})$
& $\text{sgn}(b_3^{Q_i})$ 
& contour prescription\\
\hline
\hline
$\,\,Q_0\phantom{\Big|}$ 
&$+$ &  $+$  & $+$ & $\big(\chi_1|_+,\chi_2|_+,\chi_3|_+,\chi_4|_-\big)$  \\
\hline
$\,\,Q_1\phantom{\Big|}$ 
&$-$ &  $+$  & $+$ & $\big(\chi_1|_-,\chi_2|_+,\chi_3|_+,\chi_4|_-\big)$  \\
\hline
$\,\,Q_2\phantom{\Big|}$ 
&$+$ &  $-$  & $+$ & $\big(\chi_2|_+,\chi_1|_-,\chi_3|_+,\chi_4|_-\big)$  \\
\hline
$\,\,Q_4\phantom{\Big|}$ 
&$+$ &  $+$  & $-$ & $\big(\chi_2|_+,\chi_3|_+,\chi_1|_-,\chi_4|_+\big)$  \\
\hline
$\,\,Q_5\phantom{\Big|}$ 
&$-$ &  $+$  & $-$ & $\big(\chi_2|_-,\chi_3|_+,\chi_1|_-,\chi_4|_+\big)$  \\
\hline
$\,\,Q_6\phantom{\Big|}$ 
&$+$ &  $-$  & $-$ & $\big(\chi_3|_+,\chi_2|_-,\chi_1|_-,\chi_4|_+\big)$  \\
\hline
$\,\,Q_7\phantom{\Big|}$ 
&$-$ &  $-$  & $-$ & $\big(\chi_3|_-,\chi_2|_-,\chi_1|_-,\chi_4|_+\big)$  \\
\hline
\end{tabular}
}
\caption{For each quiver $Q_i$ in Fig.~\ref{[n1,n2,n3,n4]chain}, we list the 
signs of the 
$\beta$-function coefficients $b_I^{Q_i}$ for the three 2d nodes, which are
also the signs of the corresponding FI parameters $\zeta_I^{Q_i}$. These
signs determine 
whether the integration contour for the 
corresponding $\chi$-variable has to be closed in the upper ($+$) or lower ($-$) half-plane.
The last column displays the contour prescription from which we can also read which $\chi$-variable
is associated to which node of the quiver. 
The variable $\chi_4$ is always the last one to be integrated.}
\label{contoursTable}
\end{centering}
\end{table}

\subsection{The Jeffrey-Kirwan prescription for dual quivers }
\label{JKdiscussion}

At one-instanton it is sufficient to specify whether the contours of integration for 
$\chi_{I}$ are closed in the upper or lower half-planes to completely specify the prescription. 
However, at higher instantons this may be no longer sufficient since also the 
order in which the integrations are performed may become relevant to have a 
one-to-one correspondence between the terms appearing in the superpotential derived
from the twisted chiral ring equations and the residues contributing in the localization
integrals.

An elegant way to fully specify the contour of integration for all variables 
(including the order in which they are integrated) is using the 
Jeffrey-Kirwan (JK) residue prescription \cite{JK1995} (see also, for example, 
\cite{Benini:2013nda, Hori:2014tda, Gorsky:2017hro} for recent applications to gauge theories). 
The essential point of this prescription is that the set of poles chosen by a contour 
is completely specified by the so-called JK reference vector $\eta$. 

As we have seen before, for the oriented quiver $Q_0$ the variable $\chi_4$ associated to the
4d gauge node has to be integrated as the last one in the lower-half plane, while 
the variables $\chi_1$, $\chi_2$ and $\chi_3$, associated to the first, 
second and third node respectively, have to be integrated in the upper-half plane 
but no particular order of integration is required in this case.
This means that the JK vector for the quiver $Q_0$ can be written as
\begin{equation}
\label{JK1}
\eta_{Q_0}=-\zeta_1\,\chi_1-\zeta_2\,\chi_2-\zeta_3\,\chi_3+\zeta_4\,\chi_4
\end{equation}
where $\zeta_I$, with $I=1,2,3$, are the FI parameters of the three 2d nodes of the quiver and
$\zeta_4$ is a positive real number such that
\begin{equation}
\zeta_4~\gg~\zeta_I
\label{zeta4zeta}
\end{equation}
for $I=1,2,3$. As remarked in (\ref{zetapositive}), the FI parameters are positive, so 
that, given our sign conventions, the vector (\ref{JK1}) indeed selects a contour 
in the upper-half plane for $\chi_I$ with $I=1,2,3$.
The JK prescription corresponding to (\ref{JK1}) requires that these integrals are successively performed according to the magnitudes of $\zeta_I$. However, the order of integration does not 
affect the final result, and thus this prescription always gives the correct answer no matter how
the FI parameters are ordered.
The inequality (\ref{zeta4zeta}) implies, instead, that the
integral over $\chi_4$ is the last one to be performed and, because of the $+$ sign in the
last term of $\eta_{Q_0}$, this integral must be
computed along a contour in the lower-half plane.

Let us now consider the quiver $Q_1$. In this case, the JK vector that selects the appropriate
contour of integration can be written as
\begin{equation}
\label{JK2}
\begin{aligned}
\eta_{Q_1}&=+\big|\zeta_1^{Q_1}\big|\,\chi_1-\big|\zeta_2^{Q_1}\big|\,\chi_2
-\big|\zeta_3^{Q_1}\big|\,\chi_3+\zeta_4\,\chi_4
\\
&=-\zeta_1^{Q_1}\,\chi_1-\zeta_2^{Q_1}\,\chi_2-\zeta_3^{Q_1}\,\chi_3+\zeta_4\,\chi_4
\end{aligned}
\end{equation}
where $\zeta_I^{Q_1}$, with $I=1,2,3$, are the FI parameters of the the three 2d nodes of $Q_1$ 
and $\zeta_4$ is a positive number such that
\begin{equation}
\zeta_4~\gg~\big|\zeta_I^{Q_1}\big|~.
\label{zeta4zetaQ1}
\end{equation}
We recall that in the quiver $Q_1$ the FI parameters satisfy 
the inequality (\ref{newzetaQ1}). Consequently,
the JK vector (\ref{JK2}) implies that the integral over $\chi_1$ must be computed in the lower-half plane before the integral over $\chi_2$, which instead must be computed along a contour in 
the upper-half plane. The last integral is the one over $\chi_4$ which must be computed
along a contour in the lower half-plane. 
The order of integration over $\chi_1$ and $\chi_2$ is crucial at 
higher instantons to achieve a one-to-one correspondence between the superpotential
obtained from the chiral ring equations and the one computed using the ramified instantons.
Some details on this fact at the two-instanton level are provided in Appendix~\ref{app2instantons}.

For the quiver $Q_2$ one can see that the appropriate integration contour corresponds to the
following JK vector
\begin{equation}
\label{JK3}
\begin{aligned}
\eta_{Q_2} &=  -\big|\zeta_1^{Q_2}\big|\, \chi_2+\big|\zeta_2^{Q_2}\big|\, \chi_1
-\big|\zeta_3^{Q_2}\big|\, \chi_3+\zeta_4\, \chi_4\\
&=  -\zeta_1^{Q_2}\, \chi_2-\zeta_2^{Q_2}\, \chi_1
-\zeta_3^{Q_2}\, \chi_3+\zeta_4\, \chi_4
\end{aligned}
\end{equation}
where the FI parameters satisfy the inequality (\ref{zetaQ2}) and the last parameter $\zeta_4$ is such that
\begin{equation}
\zeta_4~\gg~\big|\zeta_I^{Q_2}\big|~.
\label{zeta4zetaQ2}
\end{equation}
Using this, we can see a precise correlation with the prescription 
$\big(\chi_2|_+,\chi_1|_-,\chi_3|_+,\chi_4|_-\big)$ which we discussed above for $Q_2$. Notice that
in this case the integrals are performed in a specific order, starting form $\chi_2$
and finishing with $\chi_4$. This order is essential at higher instantons to obtain a perfect match, term by term, between
the results from the chiral ring equations and those from localization (see Appendix~\ref{app2instantons}
for some details at the two-instanton level).

This procedure can be systematically applied to all quivers in the duality chain of Fig.~\ref{[n1,n2,n3,n4]chain}, and the corresponding
JK reference vectors are listed in Tab.~\ref{JKlist}. 
\begin{table}[H]
\begin{centering}
\small{
\begin{tabular}{|c|c|}
\hline
\,\,Quiver\phantom{\Big|} 
& JK vector  \\
\hline
\hline
$\,\,Q_0\phantom{\bigg|}$ & $~-\zeta_1\,\chi_1-\zeta_2\,\chi_2-\zeta_3\,\chi_3 
+ \zeta_4\,\chi_4~$    \\
\hline
$\,\,Q_1\phantom{\bigg|}$&  $~-\zeta_1^{Q_1}\,\chi_1-\zeta_2^{Q_1}\,\chi_2
-\zeta_3^{Q_1}\,\chi_3 
+ \zeta_4\,\chi_4~$    \\
\hline
$\,\,Q_2\phantom{\bigg|}$& $~-\zeta_1^{Q_2}\,\chi_2-\zeta_2^{Q_2}\,\chi_1
-\zeta_3^{Q_2}\,\chi_3 
+ \zeta_4\,\chi_4~$   \\
\hline
$\,\,Q_4\phantom{\bigg|}$& $~-\zeta_1^{Q_4}\,\chi_2-\zeta_2^{Q_4}\,\chi_3 
-\zeta_3^{Q_4}\,\chi_1
- \zeta_4\,\chi_4 ~$   \\
\hline
$\,\,Q_5\phantom{\bigg|}$& $~-\zeta_1^{Q_5}\,\chi_2-\zeta_2^{Q_5}\,\chi_3 -
\zeta_3^{Q_5}\,\chi_1-\zeta_4\,\chi_4~$    \\
\hline
$\,\,Q_6\phantom{\bigg|}$ & $~-\zeta_1^{Q_6}\,\chi_3-\zeta_2^{Q_6}\,\chi_2
-\zeta_3^{Q_6}\,\chi_1 - \zeta_4\,\chi_4~$    \\
\hline
$\,\,Q_7\phantom{\bigg|}$ &  $~-\zeta_1^{Q_7}\,\chi_3-\zeta_2^{Q_7}\,\chi_2-
\zeta_3^{Q_7}\,\chi_1 -\zeta_4\,\chi_4~$     \\
\hline
\end{tabular}
}
\caption{For each quiver we list the JK reference vector that picks the appropriate 
contour on the localization side. The parameter $\zeta_4$ is always positive and bigger in magnitude than any of the FI parameters. If $\zeta_I^{Q_i}>0$ the associated $\chi$-variable is integrated along a contour in the upper-half plane, while 
if  $\zeta_I^{Q_i}<0$ it is integrated in the lower-half plane, in agreement with the
prescription in the last column of Tab.~\ref{contoursTable}.}
\label{JKlist}
\end{centering}
\end{table}

\subsection{New quivers and the corresponding contours}
\label{New Quivers and their Contours}

\begin{figure}
\begin{center}
\begin{tikzpicture}[decoration={
markings,
mark=at position 0.6 with {\draw (-5pt,-5pt) -- (0pt,0pt);
                \draw (-5pt,5pt) -- (0pt,0pt);}}]
  \matrix(quiver)[row sep=10mm,column sep=5mm] {
      \node(Ag1)[gauge] {$n_1$};  & & \node(Ag2)[gauge] {$n_1+n_2$}; & & \node(Ag3)[gauge,align=center] {$n_1 + n_2$\\$+\, n_3$}; 
      & &\node(Agf)[gaugedflavor]{$N$};\\
      \node(Bg1)[gauge] {$n_1$};  & & \node(Bg2)[gauge] {$n_3$}; & & \node(Bg3)[gauge,align=center] {$n_1 + n_2$\\$+\, n_3$}; 
      & &\node(Bgf)[gaugedflavor]{$N$};\\
      \node(Cg1)[gauge] {$n_3$};  & & \node(Cg2)[gauge] {$n_2+n_3$}; & & \node(Cg3)[gauge,align=center] {$n_1 + n_2$\\$+\, n_3$}; 
      & &\node(Cgf)[gaugedflavor]{$N$};\\
      \node(Dg1)[gauge] {$n_2$};  & & \node(Dg2)[gauge] {$n_2+n_3$}; & & \node(Dg3)[gauge,align=center] {$n_1 + n_2$\\$+\, n_3$}; 
      & &\node(Dgf)[gaugedflavor]{$N$};\\
      \node(Eg1)[gauge] {$n_2$};  & & \node(Eg2)[gauge] {$n_2+n_3$}; & & \node(Eg3)[gauge,align=center] {$n_2 + n_3$\\$+\, n_4$}; 
      & &\node(Egf)[gaugedflavor]{$N$};\\
      \node(Fg1)[gauge] {$n_4$};  & & \node(Fg2)[gauge] {$n_2$}; & & \node(Fg3)[gauge,align=center] {$n_2 + n_3$\\$+\, n_4$}; 
      & &\node(Fgf)[gaugedflavor]{$N$};\\
      \node(Gg1)[gauge] {$n_4$};  & & \node(Gg2)[gauge] {$n_3+n_4$}; & & \node(Gg3)[gauge,align=center] {$n_2 + n_3$\\$+\, n_4$}; 
      & &\node(Ggf)[gaugedflavor]{$N$};\\
  };
\graph[edges={postaction={decorate}}]{
(Ag1) --(Ag2)--(Ag3)--(Agf);
(Bg3)--(Bg2) --(Bg1); (Bg3)--(Bgf);
(Bg1)--[bend right,looseness=1.2](Bg3);
(Cg3)--(Cg2)--(Cg1); (Cg3)--(Cgf);
(Dg1) --(Dg2);(Dg3)--(Dg2); (Dg3)--(Dgf);
(Eg1)--(Eg2)--(Eg3); (Egf)--(Eg3);
(Fg1) --(Fg2)--(Fg3); (Fgf)--(Fg3);
(Fg3)--[bend left,looseness=1.2](Fg1);
(Ggf)--(Gg3)--(Gg2) --(Gg1);
};

\begin{scope}[on background layer]
\node(quiverA) [fill=white,fit=(Ag1) (Agf)] {};
\node[black!70,anchor=east] at (quiverA.north west) {$Q_0$};
\node(quiverB) [fill=white,fit=(Bg1) (Bgf)] {};
\node[black!70,anchor=east] at (quiverB.north west) {$\widehat{Q}_{1}$};
\node(quiverC) [fill=white,fit=(Cg1) (Cgf)] {};
\node[black!70,anchor=east] at (quiverC.north west) {$Q_{3}$};
\node(quiverD) [fill=white,fit=(Dg1) (Dgf)] {};
\node[black!70,anchor=east] at (quiverD.north west) {$Q_2$};
\node(quiverE) [fill=white,fit=(Eg1) (Egf)] {};
\node[black!70,anchor=east] at (quiverE.north west) {$Q_4$};
\node(quiverF) [fill=white,fit=(Fg1) (Fgf)] {};
\node[black!70,anchor=east] at (quiverF.north west) {$\widehat{Q}_{5}$};
\node(quiverG) [fill=white,fit=(Gg1) (Ggf)] {};
\node[black!70,anchor=east] at (quiverG.north west) {$Q_7$};
\node(dualin1)[gauge,outer sep=4pt,draw, pattern=crosshatch dots,pattern color=blue!60] at (Ag2){$\phantom{n_1+n_2}$};
\node[fill=white,inner sep=2pt] at (Ag2){$\phantom{n_1+n_2}$};
\node(dualout1)[gauge,outer sep=4pt,draw, pattern=crosshatch dots,pattern color=red!60] at (Bg2){$\phantom{n_2}$};
\node[fill=white] at (Bg2){$\phantom{n_3}$};
\node(dualin2)[gauge,outer sep=4pt,pattern=crosshatch dots,pattern color=blue!60] at (Bg1){$\phantom{n_1}$};
\node[fill=white] at (Bg1){$\phantom{n_1}$};
\node(dualout2)[gauge,outer sep=4pt,draw, pattern=crosshatch dots,pattern color=red!60] at (Cg2){$\phantom{n_2+n_3}$};
\node[fill=white,inner sep=2pt] at (Cg2){$\phantom{n_2+n_3}$};
\node(dualin3)[gauge,outer sep=4pt,align=center,draw, pattern=crosshatch dots,pattern color=blue!60] at (Cg1){$\phantom{n_3}$};
\node[fill=white,align=center] at (Cg1){$\phantom{n_3}$};
\node(dualout3)[gauge,outer sep=4pt,align=center,draw, pattern=crosshatch dots,pattern color=red!60] at (Dg1){$\phantom{n_2}$};
\node[fill=white,align=center] at (Dg1){$\phantom{n_2}$};
\node(dualin4)[gauge,outer sep=4pt,draw, pattern=crosshatch dots,pattern color=blue!60,align=center] at (Dg3){$\phantom{n_1 + n_2}$\\$\phantom{+\, n_3}$};
\node[fill=white,align=center,inner sep=1pt] at (Dg3){$\phantom{n_1 + n_2}$\\$\phantom{+\, n_3}$};
\node(dualout4)[gauge,outer sep=4pt,draw, pattern=crosshatch dots,pattern color=red!60,align=center] at (Eg3){$\phantom{n_2 + n_3}$\\$\phantom{+\, n_4}$};
\node[fill=white,align=center,inner sep=1pt] at (Eg3){$\phantom{n_2 + n_3}$\\$\phantom{+\, n_4}$};
\node(dualin5)[gauge,outer sep=4pt,pattern=crosshatch dots,pattern color=blue!60] at (Eg2){$\phantom{n_2+n_3}$};
\node[fill=white,inner sep=2pt] at (Eg2){$\phantom{n_2+n_3}$};
\node(dualout5)[gauge,outer sep=4pt,draw, pattern=crosshatch dots,pattern color=red!60] at (Fg1){$\phantom{n_4}$};
\node[fill=white] at (Fg1){$\phantom{n_4}$};
\node(dualin6)[gauge,outer sep=4pt,align=center,draw, pattern=crosshatch dots,pattern color=blue!60] at (Fg2){$\phantom{n_2}$};
\node[fill=white,align=center] at (Fg2){$\phantom{n_2}$};
\node(dualout6)[gauge,outer sep=4pt,align=center,draw, pattern=crosshatch dots,pattern color=red!60] at (Gg2){$\phantom{n_3+n_4}$};
\node[fill=white,align=center,inner sep=2pt] at (Gg2){$\phantom{n_3+n_4}$};
\graph[edges={violet,very thick,dashed}]{(dualin1)->(dualout1) (dualin2)->(dualout2) (dualin3)->(dualout3) (dualin4)->(dualout4) (dualin5)->(dualout5) (dualin6)->(dualout6);};
\end{scope}
\end{tikzpicture}
\caption{Another chain of dualities to proceed from $Q_0$ to $Q_7$.}
\label{AnotherSequence}
\end{center}
\end{figure}
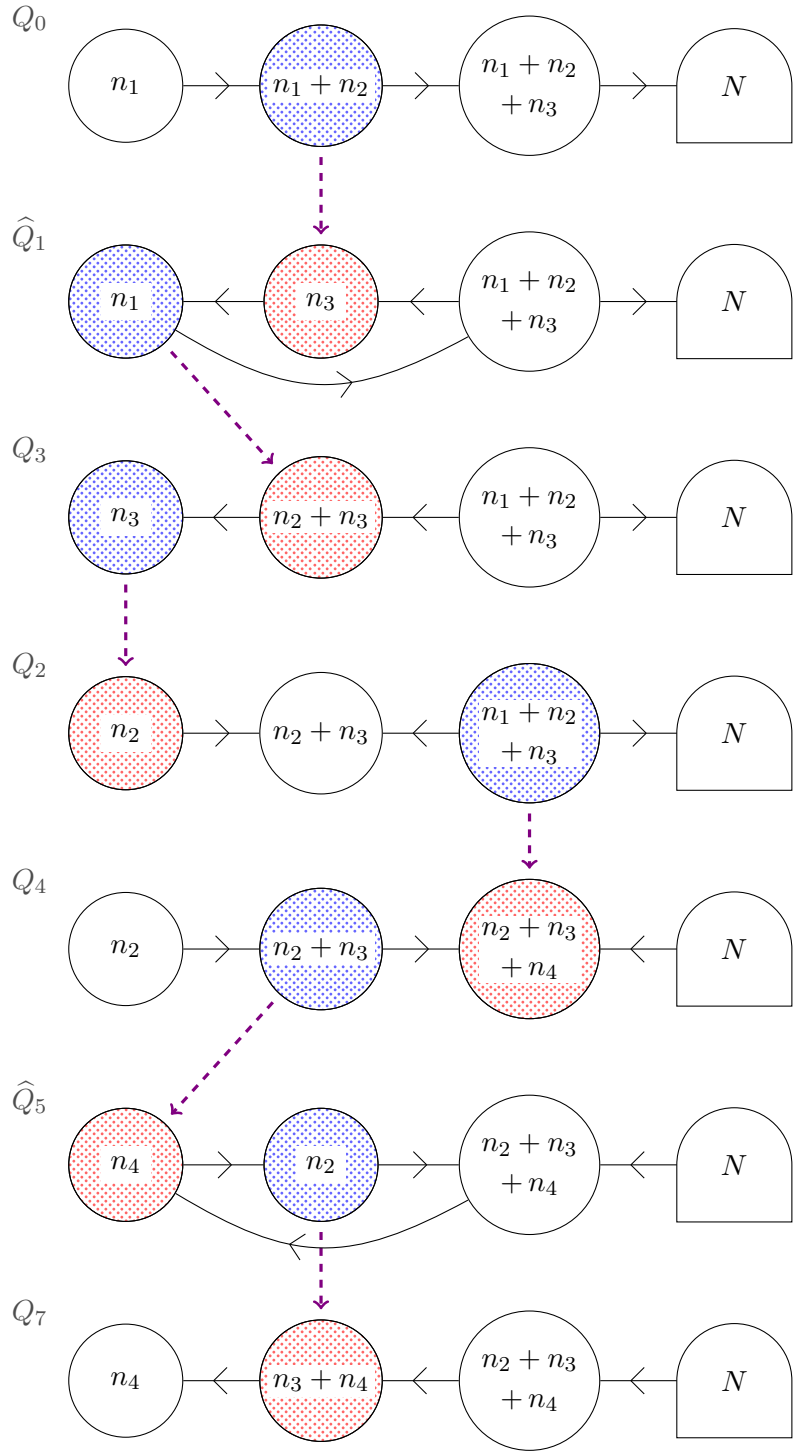

The chain of Seiberg dualities shown in Fig.~\ref{[n1,n2,n3,n4]chain} is of a very special kind, since
the 2d gauge node being dualized at each step always has only fundamental flavours attached 
to it. 
This ensures that the resulting quivers are always linear. We now relax this condition and consider 
an alternative duality chain with the same initial and final points, but in which 
we start by dualizing the second node of the quiver $Q_0$ that has both fundamental 
and anti-fundamental flavours attached to it. 
This duality leads to the quiver $\widehat{Q}_{1}$ which contains a loop, as  shown 
in Fig.~\ref{AnotherSequence}. 
Proceeding all the way down as indicated in this figure,
we encounter the quivers $Q_2$ and $Q_4$, which were also part of the earlier sequence, 
but we also find two new quivers, which we call $Q_{3}$ and $\widehat{Q}_{5}$. 
The latter, like $\widehat{Q}_{1}$, contains a loop.

We can repeat the same analysis as before and derive the contour
prescription for all quivers in this sequence, including the non-linear ones.
The first step is obtaining the classical part of the superpotential. Starting from 
$\mathcal{W}_{\text{cl}}^{Q_0}$
given in \eqref{W1} and applying the duality rule \eqref{Wdual2} to the second node, we find
that the classical part of the superpotential for $\widehat{Q}_{1}$ is
\begin{equation}
\mathcal{W}_{\text{cl}}^{\widehat{Q}_{1}}=2\pi\ii\,\tau_1\,
\text{Tr}\,\sigma^{(1)}-2\pi\ii\,\tau_2\,\text{Tr}\,\sigma^{(2)}+2\pi\ii\,(\tau_2+\tau_3)\,\text{Tr}\,\sigma^{(3)}~.
\label{WQ2A}
\end{equation}
If we now dualize the first node of $\widehat{Q}_{1}$ we obtain a new linear quiver $Q_{3}$. Here it
is natural to relabel the nodes in such a way that the dualized node corresponds to $I=2$,
thus respecting the order shown in Fig.~\ref{AnotherSequence}. 
Taking this into account and applying the duality map to \eqref{WQ2A}, we then obtain
\begin{equation}
\mathcal{W}_{\text{cl}}^{Q_{3}}=-2\pi\ii\,\tau_2\,
\text{Tr}\,\sigma^{(1)}-2\pi\ii\,\tau_1\,\text{Tr}\,\sigma^{(2)}
+2\pi\ii\,(\tau_1+\tau_2+\tau_3)\,\text{Tr}\,\sigma^{(3)}~.
\label{WQ2B}
\end{equation}
In the next two duality steps we find the quivers
$Q_2$ and $Q_4$ whose classical superpotentials
are given in \eqref{Wlist}. Dualizing the second node of $Q_4$, we obtain the non-linear quiver
$\widehat{Q}_{5}$, whose classical superpotential is
\begin{equation}
\mathcal{W}_{\text{cl}}^{\widehat{Q}_{5}}=-2\pi\ii\,\tau_3\,
\text{Tr}\,\sigma^{(1)}+2\pi\ii\,\tau_2\,\text{Tr}\,\sigma^{(2)}
-2\pi\ii\,(\tau_1+\tau_2)\,\text{Tr}\,\sigma^{(3)}~.
\label{WQ5A}
\end{equation}
Here we have again renamed indices in such a way that the labelling of the $\sigma$-variables follows
the same order in which the gauge nodes are drawn in Fig.~\ref{AnotherSequence}. 

Next, we determine the classical vacuum for the quivers in this duality chain by equating 
the classical twisted chiral superpotentials for each dual pairs. In Tab.~\ref{tablevevs21}
we report the results for the three new quivers $\widehat{Q}_{1}$, $Q_{3}$ and $\widehat{Q}_{5}$ of 
this sequence.

\begin{table}[H]
\begin{centering}
\small{
\begin{tabular}{|c|c|c|c|}
\hline
\,\,Quiver \phantom{\Big|} & $\sigma^{(1)}_{\text{cl}}$ & $\sigma^{(2)}_{\text{cl}}$ & $\sigma^{(3)}_{\text{cl}}$ \\
\hline
\hline
$\,\,\widehat{Q}_{1} \phantom{\Big|}$&  ${\mathcal A}_1$ & ${\mathcal A}_3$ & ${\mathcal A}_1\oplus {\mathcal A}_2 \oplus {\mathcal A}_3$  \\
\hline
$\,\,Q_{3} \phantom{\Big|}$ &  ${\mathcal A}_3$ & ${\mathcal A}_2\oplus {\mathcal A}_3$ & ${\mathcal A}_1\oplus {\mathcal A}_2 \oplus {\mathcal A}_3$  \\
\hline
$\,\,\widehat{Q}_{5} \phantom{\Big|}$&  ${\mathcal A}_4$ & ${\mathcal A}_2$ & ${\mathcal A}_2\oplus {\mathcal A}_3 \oplus {\mathcal A}_4$  \\
\hline
\end{tabular}
}
\caption{For the quivers $\widehat{Q}_{1}$, $Q_{3}$ and $\widehat{Q}_{5}$ drawn in Fig.~\ref{AnotherSequence},
we list the classical expectation values of the twisted chiral fields in each of the 2d nodes, about which one finds the solution to the twisted chiral ring. Using this vacuum, along with the FI couplings in the classical twisted chiral superpotentials for each quiver, one finds identical expressions 
at leading order. The vacuum for the other quivers of the duality chain, namely $Q_0$, $Q_2$, $Q_4$ and $Q_7$, can be read from Tab.~\ref{vevs1}.}
\label{tablevevs21}
\end{centering}
\end{table}

Using this information and following the same procedure described above, 
we can find the $q$ vs $\Lambda$ map and the 
contour prescription that has to be used in the localization formula
in order to match term-by-term the superpotential with the one 
obtained from solving the twisted chiral ring equations.
Of course, we do not repeat the derivation of these results since the calculations are a straightforward
generalization of what we did for the other duality chain, and we simply collect our findings
for the three new quivers $\widehat{Q}_{1}$, $Q_{3}$ and $\widehat{Q}_{5}$ 
in Tab.~\ref{JK2ndchain}.
We have checked the validity of our proposal up to two instantons, while 
some details on the results at the one-instanton level
can be found in Appendix~\ref{appB}.

\begin{table}
\begin{centering}
\small{
\begin{tabular}{|c|c|c|c|c|c|}
\hline
{\,Quiver}
& $q_1$ & $q_2$ & $q_3$ 
& JK vector \phantom{\Big|} \!\! \!\! \\
\hline
\hline
$\widehat{Q}_{1}$& $\big(\Lambda_1^{\widehat{Q}_{1}}\big)^{n_1+n_2}$ 
& $\big(\Lambda_2^{\widehat{Q}_{1}}\big)^{n_2+n_3}$ & 
$\frac{\big(\Lambda_3^{\widehat{Q}_{1}}\big)^{n_2+2n_3+n_4\phantom{\big|}}}{\big(\Lambda_2^{\widehat{Q}_{1}}
\big)^{n_2+n_3}}$ 
&  $-\zeta_1^{\widehat{Q}_{1}}\chi_1-\zeta_2^{\widehat{Q}_{1}}\chi_2
-\zeta_3^{\widehat{Q}_{1}}\chi_3 +\zeta_4\,\chi_4$    \\
\hline
$Q_{3}$& $\big(\Lambda_2^{Q_3}\big)^{n_1+n_2}$ & $\big(\Lambda_1^{Q_3}\big)^{n_2+n_3}$ 
& $\frac{\big(\Lambda_3^{Q_3}\big)^{N+n_2+n_3\phantom{\big|}}}{\big(\Lambda_1^{Q_3}\big)^{n_2+n_3}\big(\Lambda_2^{Q_3}\big)^{n_1+n_2}}$ 
& $-\zeta_1^{Q_3}\chi_2-\zeta_2^{Q_3}\chi_1-\zeta_3^{Q_3}\chi_3 + \zeta_4\chi_4$    \\
\hline
$\widehat{Q}_{5}$ & $\frac{\big(\Lambda_3^{\widehat{Q}_5}\big)^{n_1+2n_2+n_3\phantom{\big|}}}{\big(\Lambda_2^{\widehat{Q}_5}\big)^{n_2+n_3}}$ & $\big(\Lambda_2^{\widehat{Q}_5}\big)^{n_2+n_3}$ & $\big(\Lambda_1^{\widehat{Q}_5}\big)^{n_3+n_4}$ 
& $-\zeta_1^{\widehat{Q}_5}\chi_3-\zeta_2^{\widehat{Q}_5}\chi_2 
- \zeta_3^{\widehat{Q}_5}\chi_1-\zeta_4\,\chi_4$    \\
\hline
\end{tabular}
}
\caption{For the quivers $\widehat{Q}_{1}$, $Q_{3}$ and $\widehat{Q}_{5}$ drawn in Fig.~\ref{AnotherSequence}, we list the relations (up to signs) between the ramified instanton counting parameters $q_I$ and the strong 
coupling scales $\Lambda_I$, and also the JK reference vector that selects the contour prescription
needed to compute the ramified instanton partition function using the localization formula.}
\label{JK2ndchain}
\end{centering}
\end{table}

\section{Proposal for generic linear quivers}
\label{allquiverslinear}
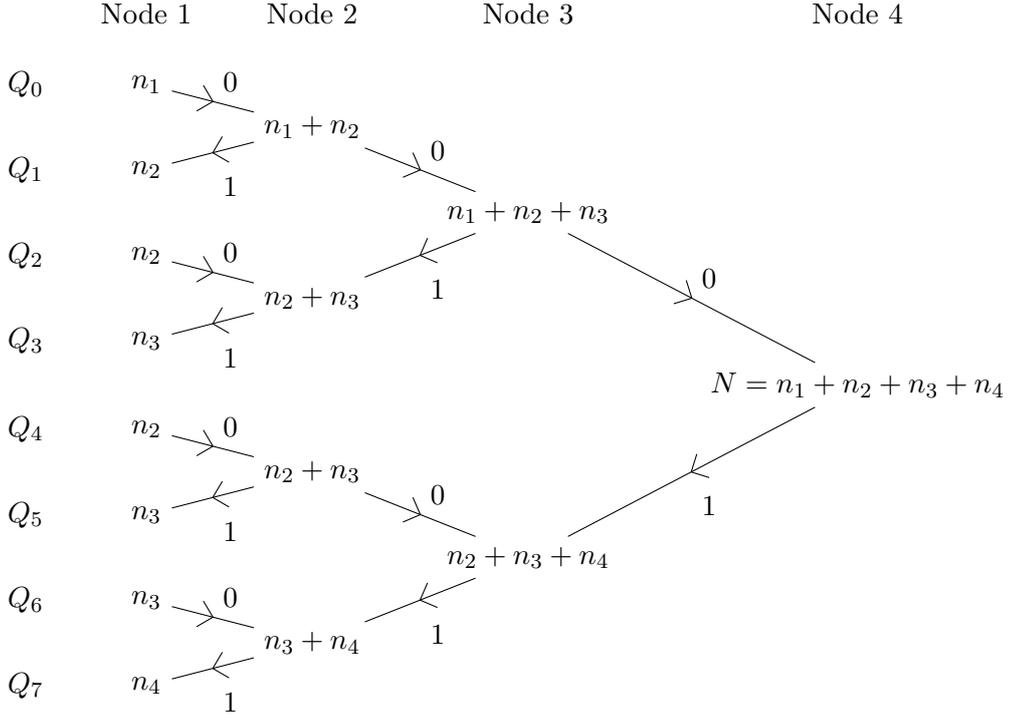
\begin{figure}[H]
\begin{centering}
\begin{tikzpicture}[decoration={
markings,
mark=at position 0.5 with {\draw (-5pt,-5pt) -- (0pt,0pt);
                \draw (-5pt,5pt) -- (0pt,0pt);}}]
  \matrix(table)[column sep=5mm] {
      && \node{Node 1};&[+.5em]   \node{Node 2}; &[+1em] \node{Node 3};&[+1.5em] \node{Node 4};\\ [+1em]     
      &\node{$Q_0$};& \node(q1n1){$n_1$};&  & & \\
      &&  &\node(q12n2){$n_1+n_2$};  & & \\
      &\node{$Q_1$};& \node(q2n1){$n_2$};&  & & \\
      && &  &  \node(qtopn3){$n_1+n_2+n_3$};& \\
      &\node{$Q_2$};& \node(q3n1){$n_2$};&  & & \\
      && & \node(q32bn2){$n_2+n_3$};  & & \\
      &\node{$Q_3$};& \node(q2bn1){$n_3$};&  & & \\
      && &  & & \node(qmidn4){$N=n_1+n_2+n_3+n_4$}; \\
      &\node{$Q_4$};& \node(q4n1){$n_2$};&  & & \\
      && &  \node(q45n2){$n_2+n_3$}; & & \\
      &\node{$Q_5$};& \node(q5n1){$n_3$};&  & & \\
      && &  & \node(qbotn3){$n_2+n_3+n_4$}; & \\
      &\node{$Q_6$};& \node(q6n1){$n_3$};&  & & \\
      && &  \node(q67n2){$n_3+n_4$}; & & \\
      &\node{$Q_7$};& \node(q7n1){$n_4$};&  & & \\
             };
      
  \graph[edge label=$0$,edges={postaction={decorate}}]{
  (q1n1)--(q12n2); (q3n1)--(q32bn2); (q4n1)--(q45n2); (q6n1)--(q67n2); 
  (q12n2)--(qtopn3);(q45n2)--(qbotn3);
  (qtopn3)--(qmidn4);
  };

 \graph[edge label=$1\phantom{\Big)}$,edges={postaction={decorate}}]{
  (q12n2)--(q2n1); (q32bn2)--(q2bn1); (q45n2)--(q5n1); (q67n2)--(q7n1); 
  (qtopn3)--(q32bn2);(qbotn3)--(q67n2);
  (qmidn4)--(qbotn3);
  };
\end{tikzpicture}
\caption{The linear quivers that are Seiberg-dual to the oriented quiver $Q_0$. 
To each link we associate $0$ or $1$ depending whether it is rightward or leftward.}
\label{4nodelinearlist}
\end{centering}
\end{figure}

The detailed analysis of the previous section shows that in the 4-node case there are eight linear quivers related to each other by duality: the seven ones found in the sequence of Fig.~\ref{[n1,n2,n3,n4]chain},
and the quiver $Q_3$ in the sequence of Fig.~\ref{AnotherSequence}. If we consider these eight linear
quivers all together, a nice structure emerges as illustrated in Fig.~\ref{4nodelinearlist} where we
exhibit the ranks of the nodes of the various quivers and their connections. 
We recall that the ranks of the nodes of the initial oriented quiver $Q_0$ can be obtained from the vector
$\vec{n}=(n_1,n_2,n_3,n_4)$ as discussed in Section~\ref{sec:review} (see (\ref{rI})). 
Then, given the action of Seiberg duality, it is easy to realize that the ranks of the nodes of the other
quivers can be obtained from vectors that are a permutation of the entries of $\vec{n}$. For example,
for the quiver $Q_2$ the ranks can be obtained from $(n_2,n_3,n_1,n_4)$, while for quiver $Q_6$ they
are obtained from $(n_3,n_4,n_2,n_1)$. It is not difficult to realize that all these permuted vectors
can be written as
\begin{equation}
P_2^{s_3}\,P_3^{s_2}\, P_4^{s_1} \, \vec{n}~,
\end{equation}
where $s_i=0,1$ and $P_k$ is the cyclic permutation on the first $k$ elements out of 4. 
In matrix form, we have
\begin{equation}
		P_{2} = 
		\begin{pmatrix}
			~0 & ~1 & ~0 & ~0 ~\cr
			~1 & ~0 & ~0 & ~0 ~\cr
			~0 & ~0 & ~1 & ~0~ \cr
			~0 & ~0 & ~0 & ~1~
		\end{pmatrix}~,~~~
		P_{3} = 
		\begin{pmatrix}
			~0 & ~1 & ~0 & ~0~ \cr
			~0 & ~0 & ~1 & ~0~ \cr
			~1 & ~0 & ~0 & ~0~ \cr
			~0 & ~0 & ~0 & ~1~
		\end{pmatrix}~,~~~
		P_{4} = 
		\begin{pmatrix}
			~0 & ~1 & ~0 & ~0~ \cr
			~0 & ~0 & ~1 & ~0~ \cr
			~0 & ~0 & ~0 & ~1~ \cr
			~1 & ~0 & ~0 & ~0~
		\end{pmatrix}~.		
	\end{equation}
We therefore see that each linear quiver $Q_i$ can be labelled by the set $\vec{s}=(s_1,s_2,s_3)$ identifying
the permutation 
\begin{equation}
P[\vec{s}\,]=P_2^{s_3}\, P_3^{s_2}\, P_4^{s_1} 
\end{equation}
which determines the ranks of the various nodes. For example, the quiver $Q_3$ corresponds to $(0,1,1)$ and the quiver $Q_5$ to $(1,0,1)$.For any quiver, its corresponding $\vec{s}$ can be 
easily read from Fig.~\ref{4nodelinearlist} by looking at the labels $0$ and $1$ on the links connecting
the nodes, starting from the rightmost one and moving leftwards.
Notice that, with the conventions we have chosen, 
the quiver $Q_i$ turns out to be labelled by the vector $\vec{s}$ that 
represents the number $i$ written in binary notation. 

The permutation $P[\vec{s}\,]$ can be represented in an irreducible way in terms of $3\times 3$ matrices
as follows
\begin{equation}
\widehat{P}\,[\vec{s}\,]=\widehat{P}_2^{s_3}\, \widehat{P}_3^{s_2}\, \widehat{P}_4^{s_1} 
\end{equation}
where
\begin{equation}
		\widehat{P}_{2} = 
		\begin{pmatrix}
			-1& ~0 & ~0 ~\cr
			~1 & ~1 & ~0~  \cr
			~0 & ~0 & ~1~ & 
		\end{pmatrix}~,~~~
		\widehat{P}_{3} = 
		\begin{pmatrix}
			~0 & ~1 & ~0 ~ \cr
			-1 & -1 & ~0 ~\cr
			~1 & ~1& ~1~ 
		\end{pmatrix}~,~~~
		\widehat{P}_{4} = 
		\begin{pmatrix}
			~0 & ~1 & ~0~  \cr
			~0 & ~0 & ~1~ \cr
			-1 & -1 & -1	~	
			\end{pmatrix}~.		
	\end{equation}
This defines the action on the FI couplings. Indeed, if we introduce the vector $\vec{\zeta}=(\zeta_1,\zeta_2,\zeta_3)$ with the FI parameters of the first quiver $Q_0$, then it is easy to check
that
\begin{equation}
\big(\zeta_1^{Q_i}\,,\,\zeta_2^{Q_i}\,,\,\zeta_3^{Q_i}\big) =\widehat{P}\,[\vec{s}\,]\,\vec{\zeta}~.
\label{zetas}
\end{equation}
For example, for $Q_4$ we have 
$\widehat{P}\,[(1,0,0)]\,\vec{\zeta} =
\widehat{P}_4\,\vec{\zeta} =\big(\,\zeta_2\,,\,\zeta_3\,,\,-\zeta_1-\zeta_2-\zeta_3\,\big)~,$
which indeed are the FI parameters of $Q_4$, 
as one can see from the superpotential $\mathcal{W}_{\text{cl}}^{\,Q_4}$ in (\ref{Wlist}).

This formalism can be nicely used also to describe how the variables $\chi_I$ appearing in the localization
integrals are associated to the various nodes of the quiver. From the detailed analysis of Section~\ref{Relating Quivers and Contours}, we see that $\chi_4$ is always associated to the last 4d node of the
quiver, while the other three variables $\chi_1$, $\chi_2$ and $\chi_3$ are associated to the first 
three 2d nodes in a permutation determined by the $q$ vs $\Lambda$ map. Moreover, we see that two 
quivers whose vectors $\vec{s}$ only differ by the value of $s_3$ have the same permutation
and that this permutation involves only cyclic rearrangements of the first two or the first three variables described by $P_2$ and $P_3$.  In particular, introducing 
the vector $\vec{\chi}=(\chi_1,\chi_2,\chi_3,\chi_4)$, we can check that
\begin{equation}
\vec{\chi}\,[\vec{s}\,]= P_2^{s_2}\, P_3^{s_1}\,\vec{\chi}
\label{chis}
\end{equation}
correctly describes the correspondence between the nodes of the quiver and the $\chi$-variables.
For example, for $Q_6$ we find that $\vec{\chi}\,[(1,1,0)]=P_2\, P_3\,\vec{\chi}=(\chi_3,\chi_2,\chi_1,\chi_4)$, which is indeed the correct sequence of $\chi$-variables for $Q_6$ as one can see from Tab.~\ref{contoursTable}.

We are now in the position of using this formalism to write the JK reference vector for any linear quiver in
a compact form. To this aim, we first extend the three-component vector (\ref{zetas}) by adding to it a
fourth component according to
\begin{equation}
\vec{\zeta}\,[\vec{s}\,] = \big(\widehat{P}\,[\vec{s}\,]\,\vec{\zeta}\,,\,\pm\,\zeta_4\big)\,=\,
\big(\zeta_1^{Q_i}\,,\,\zeta_2^{Q_i},\zeta_3^{Q_i}\,,\,\pm\,\zeta_4\big)
\label{zetas1}
\end{equation}
Here $\zeta_4$ is a positive parameter that is always bigger than $\big|\zeta_I^{Q_i}\big|$ for $I=1,2,3$. The sign in (\ref{zetas1}) depends whether the 4d node of the quiver provides fundamental ($+$) or anti-fundamental ($-$) matter to the last 2d node. By considering the detailed structure of the various quivers, we see that in the
first four quivers from $Q_0$ to $Q_3$ the 4d node provide anti-fundamental matter, while in the last four ones from $Q_4$ to $Q_7$ it provides fundamental flavors. This means that the sign in (\ref{zetas1}) can also be written as
$(-1)^{s_1+1}$. With these positions, it is easy to realize that the JK vectors described in the previous
section can all be compactly written as follows:
\begin{equation}
\eta_{Q_i}=
-\vec{\zeta}\,[\vec{s}\,]\cdot\vec{\chi}\,[\vec{s}\,]~.
\end{equation}

This analysis can be extended to linear quivers with $M$ nodes
in a straightforward manner. In this case we have $(M-1)$ binary choices corresponding 
to $2^{M-1}$ linear quivers that are related to each other by 
Seiberg duality. Therefore, they can be labelled by a vector $\vec{s}=(s_1,s_2,\cdots,s_{M-1})$
with $s_i=0,1$.
For each choice, the ranks of the $M$ nodes are determined by the permutation
\begin{equation}
\begin{aligned}
P[\vec{s}\,]=P_2^{s_{M-1}}\, P_3^{s_{M-2}}\ldots P_{M-1}^{s_2}\, P_M^{s_{1}}
\end{aligned}
\end{equation}
where $P_k$ permutes the first $k$ numbers, while the FI parameters of the $(M-1)$ nodes are
obtained using $\widehat{P}\,[\vec{s}\,]$, which represents the permutation $P[\vec{s}]$ in an 
irreducible way in an $(M-1)$-dimensional space. Generalizing (\ref{zetas1}) to the $M$-node case
in an obvious way, and defining
\begin{equation}
\vec{\chi}\,[\vec{s}\,]= P_2^{s_{M-2}}\, P_3^{s_{M-3}}\ldots P_{M-2}^{s_2}\, 
P_{M-1}^{s_1}\, \vec{\chi}
\label{chisM}
\end{equation}
where $\vec{\chi}=(\chi_1,\ldots,\chi_M)$, it is natural to propose that the JK reference vector for a generic quiver $Q_i$ is
\begin{equation}
\eta_{Q_i}=
-\vec{\zeta}\,[\vec{s}\,]\cdot\vec{\chi}\,[\vec{s}\,] = - \sum_{I=1}^{M-1}
\zeta_I^{Q_i}\,\chi_{\alpha(I)}\pm \zeta_4\,\chi_4
\label{etaM}
\end{equation}
where $\alpha(I)$ is determined by the permutation in (\ref{chisM}). We have verified in several examples the validity of this proposal.
 
\section{Summary of results}
\label{summary}

In this paper we have discussed in detail the relation between two distinct realizations of surface operators: as monodromy defects and as coupled 2d/4d quiver gauge theories. The main features of these
two points of view and their relations are summarized in Tab.~\ref{dictionary}. 
\begin{table}[H]
\begin{centering}
\begin{tabular}{|c||c|}
\hline
\,\,Monodromy defect\phantom{\Big|} & 2d/4d quiver models  \\
\hline
\hline
Partition of $N$: $(n_1, n_2, \ldots, n_M)$ \phantom{\Big|} & Ranks of 2d gauge nodes  \\
\hline
4d Coulomb v.e.v.'s \phantom{\Big|}& 2d twisted masses \\
\hline
Partition of Coulomb v.e.v.'s  \phantom{\Big|}& ~~Classical (massive) vacuum~~\\
\hline
Ramified instanton counting parameters & 2d/4d strong coupling scales \\
$q_I$, $q_M$ & $\Lambda_I$, $\Lambda_{\text{4d}}$ \\
\hline 
${\mathcal W}_{\text{inst}}(a, q)$ \phantom{\Big|}& $\left. {\mathcal W}(\sigma, a, 
\Lambda_I, \Lambda_{\text{4d}}) \right|_{\sigma_{\star}}$\\
\hline
Contour prescription \phantom{\Big|}& 2d Seiberg duality frame\\
\hline
\end{tabular}
\caption{The dictionary between the various features of surface operators in the
two descriptions, as monodromy defects and as coupled 2d/4d quivers.}
\label{dictionary} 
\end{centering}
\end{table}
Establishing a precise correspondence between different integration 
contour prescriptions in the
ramified instanton partition function for a monodromy defect and different quiver theories related 
to each other by a Seiberg duality has been the main focus of our 
present work. Dual quivers have different ultraviolet realizations
but share the same infrared physics and thus the (massive) vacua of their low-energy
theories can be mapped onto each other. These massive vacua 
are obtained by extremizing the effective twisted chiral superpotential of the 2d/4d quiver. The evaluation of the effective 
superpotential in a particular vacuum is in turn mapped to the twisted superpotential which
is extracted from the ramified instanton partition function with a specific contour of integration. 

For surface operators in pure ${\mathcal N}=2$ gauge theories, like the ones we have considered
in this paper, residue theorem ensures that one always obtains the same superpotential irrespective of the 
contour of integration chosen. Nevertheless, by a careful study of the individual residues 
that contribute to the superpotential, we have been able to map distinct contours on the 
localization side to distinct Seiberg-dual 2d quivers coupled to the same 
4d SU$(N)$ flavour group. The duality frame one chooses affects the details of the other entries 
in the table above, such as the choice of the classical vacuum and the map between the ramified instanton 
counting parameters $q_I$ and the strong coupling scales $\Lambda_I$. We initially restricted ourselves 
to systems with four nodes to exhibit our explicit results, but in the end we have generalized
our analysis to linear quivers with an arbitrary number of nodes providing the map
between the data of the quiver and the corresponding JK prescription, which takes a universal form.

There is one caveat to our analysis. All quivers we have studied so far, have only a single 2d node that is 
connected to the flavour node that is gauged in 4d. It is only for such cases that the coupling of the 2d 
degrees of freedom to the 4d theory via its resolvent gives results that are consistent with those obtained 
using localization methods in the monodromy defect approach. It would be very interesting to understand 
whether quivers with more
2d nodes connected to the 4d node also have an interpretation as surface operators in a 4d gauge theory. 
Furthermore, there are many worthwhile but yet unexplored directions to pursue, such as the extension of 
our analysis to (conformal) SQCD models for which the integrands of ramified instanton partition function 
may have non-vanishing residues at infinity, or the lift of our techniques to five dimensions to study 
surface operators from the point of view of 3d/5d coupledsystems, with possible Chern-Simons 
interactions. We leave these extensions and generalizations to future work.
\vskip 1.5cm
\noindent {\large {\bf Acknowledgments}}
\vskip 0.2cm
We would like to thank Stefano Cremonesi and Amihay Hanany for many useful 
discussions. S.K.A. would especially like to thank the Physics Department of the University of 
Torino and the Torino Section of INFN for hospitality during the final stages of this work.
M.B, M.F., R.R.J. and A.L. would like to thank the ``Galileo Galilei Institute for Theoretical 
Physics'' in Florence for hospitality.

The work of M.B., M.F., R.R.J. and A.L. is partially supported by the MIUR PRIN Contract 
2015MP2CX4 ``Non-perturbative Aspects Of Gauge Theories And Strings''. The research of A.L. is original and receives financial support
from the Universit\`a del Piemonte Orientale.

\vskip 1cm
\begin{appendix}
\section{Localization results at one-instanton level}
\label{appA}
In this Appendix we collect the localization results at the one-instanton level for the different 
contours of integrations, corresponding to the different quivers discussed 
in Sections~\ref{Relating Quivers and Contours} and 
\ref{New Quivers and their Contours}.
The twisted superpotential extracted from the partition function \eqref{Zso4d5d} is expressed
as a sum of residues, and at the one-instanton level it can be easily derived from \eqref{Zoneinst}.

\subsection*{Integration contour $\big(\chi_1|_+,\chi_2|_+,\chi_3|_+,\chi_4|_-\big)$:} 
\begin{equation}
\begin{aligned}
\label{C1}
\mathcal{W}_{1-\text{inst}}&=\sum_{s\in\mathcal{N}_1}
\frac{(-1)^{n_1}q_1}{\prod_{r\in\widehat{\mathcal{N}}_1\cup\mathcal{N}_2}(a_s-a_r)
\phantom{\Big|}}
+\sum_{t\in\mathcal{N}_2}
\frac{(-1)^{n_2}q_2}{\prod_{r\in\widehat{\mathcal{N}}_2\cup\mathcal{N}_3}(a_t-a_r)
\phantom{\Big|}}\\
&\quad+\sum_{u\in\mathcal{N}_3}
\frac{(-1)^{n_3}q_3}{\prod_{r\in\widehat{\mathcal{N}}_3\cup\mathcal{N}_4}(a_u-a_r)
\phantom{\Big|}}
+\sum_{s\in\mathcal{N}_1}\frac{(-1)^{n_4+1}q_4}{\prod_{r\in\mathcal{N}_4\cup
\widehat{\mathcal{N}}_1}(a_s-a_r)\phantom{\Big|}}~.
\end{aligned}
\end{equation}

\subsection*{Integration contour $\big(\chi_1|_-,\chi_2|_+,\chi_3|_+,\chi_4|_-\big)$:} 
\begin{equation}
\begin{aligned}
\label{C23}
\mathcal{W}_{1-\text{inst}}&=\sum_{t\in\mathcal{N}_2}
\frac{(-1)^{n_1+1}q_1}{\prod_{r\in\mathcal{N}_1
\cup\widehat{\mathcal{N}}_2}(a_t-a_r)\phantom{\Big|}}
+\sum_{t\in\mathcal{N}_2}
\frac{(-1)^{n_2}q_2}{\prod_{r\in\widehat{\mathcal{N}}_2\cup\mathcal{N}_3}(a_t-a_r)
\phantom{\Big|}}\\
&\quad+\sum_{u\in\mathcal{N}_3}
\frac{(-1)^{n_3}q_3}{\prod_{r\in\widehat{\mathcal{N}}_3\cup\mathcal{N}_4}(a_u-a_r)
\phantom{\Big|}}
+\sum_{s\in\mathcal{N}_1}\frac{(-1)^{n_4+1}q_4}{\prod_{r\in\mathcal{N}_4\cup
\widehat{\mathcal{N}}_1}(a_s-a_r)\phantom{\Big|}}~.
\end{aligned}
\end{equation}

\subsection*{Integration contour $\big(\chi_2|_+,\chi_1|_-,\chi_3|_+,\chi_4|_-\big)$:} 
The one-instanton superpotential for this integration contour is the same as 
in \eqref{C23} since at this order there is no difference between the two cases.

\subsection*{Integration contour $\big(\chi_2|_+,\chi_3|_+,\chi_1|_-,\chi_4|_+\big)$:} 
\begin{equation}
\begin{aligned}
\label{C4}
\mathcal{W}_{1-\text{inst}}&=\sum_{t\in\mathcal{N}_2}
\frac{(-1)^{n_1+1}q_1}{\prod_{r\in\mathcal{N}_1
\cup\widehat{\mathcal{N}}_2}(a_t-a_r)\phantom{\Big|}}
+\sum_{t\in\mathcal{N}_2}
\frac{(-1)^{n_2}q_2}{\prod_{r\in\widehat{\mathcal{N}}_2\cup\mathcal{N}_3}(a_t-a_r)
\phantom{\Big|}}\\
&\quad+\sum_{u\in\mathcal{N}_3}
\frac{(-1)^{n_3}q_3}{\prod_{r\in\widehat{\mathcal{N}}_3\cup\mathcal{N}_4}(a_u-a_r)
\phantom{\Big|}}
+\sum_{v\in\mathcal{N}_4}\frac{(-1)^{n_4}q_4}{\prod_{r\in\widehat{\mathcal{N}}_4\cup
\mathcal{N}_1}(a_v-a_r)\phantom{\Big|}}~.
\end{aligned}
\end{equation}

\subsection*{Integration contour $\big(\chi_2|_-,\chi_3|_+,\chi_1|_-,\chi_4|_+\big)$:} 
\begin{equation}
\begin{aligned}
\label{C56}
\mathcal{W}_{1-\text{inst}}&=\sum_{t\in\mathcal{N}_2}
\frac{(-1)^{n_1+1}q_1}{\prod_{r\in\mathcal{N}_1
\cup\widehat{\mathcal{N}}_2}(a_t-a_r)\phantom{\Big|}}
+\sum_{u\in\mathcal{N}_3}
\frac{(-1)^{n_2+1}q_2}{\prod_{r\in\mathcal{N}_2\cup\widehat{\mathcal{N}}_3}(a_t-a_r)
\phantom{\Big|}}\\
&\quad+\sum_{u\in\mathcal{N}_3}
\frac{(-1)^{n_3}q_3}{\prod_{r\in\widehat{\mathcal{N}}_3\cup\mathcal{N}_4}(a_u-a_r)
\phantom{\Big|}}
+\sum_{v\in\mathcal{N}_4}\frac{(-1)^{n_4}q_4}{\prod_{r\in\widehat{\mathcal{N}}_4\cup
\mathcal{N}_1}(a_v-a_r)\phantom{\Big|}}~.
\end{aligned}
\end{equation}

\subsection*{Integration contour $\big(\chi_3|_+,\chi_2|_-,\chi_1|_-,\chi_4|_+\big)$:} 

The one-instanton superpotential for this integration contour is the same as in \eqref{C56} since at this 
order there is no difference between the two cases.

\subsection*{Integration contour $\big(\chi_3|_-,\chi_2|_-,\chi_1|_-,\chi_4|_+\big)$:} 
\begin{equation}
\begin{aligned}
\label{C7}
\mathcal{W}_{1-\text{inst}}&=\sum_{t\in\mathcal{N}_2}
\frac{(-1)^{n_1+1}q_1}{\prod_{r\in\mathcal{N}_1
\cup\widehat{\mathcal{N}}_2}(a_t-a_r)\phantom{\Big|}}
+\sum_{u\in\mathcal{N}_3}
\frac{(-1)^{n_2+1}q_2}{\prod_{r\in\mathcal{N}_2\cup\widehat{\mathcal{N}}_3}(a_u-a_r)
\phantom{\Big|}}\\
&\quad+\sum_{v\in\mathcal{N}_4}
\frac{(-1)^{n_3+1}q_3}{\prod_{r\in\mathcal{N}_3\cup\widehat{\mathcal{N}}_4}(a_v-a_r)
\phantom{\Big|}}
+\sum_{v\in\mathcal{N}_4}\frac{(-1)^{n_4}q_4}{\prod_{r\in\widehat{\mathcal{N}}_4\cup
\mathcal{N}_1}(a_v-a_r)\phantom{\Big|}}~.
\end{aligned}
\end{equation}

\subsection*{Integration contour $\big(\chi_1|_+,\chi_2|_-,\chi_3|_+,\chi_4|_-\big)$:} 
\begin{equation}
\begin{aligned}
\label{C2A}
\mathcal{W}_{1-\text{inst}}&=\sum_{s\in\mathcal{N}_1}
\frac{(-1)^{n_1}q_1}{\prod_{r\in\widehat{\mathcal{N}}_1
\cup\mathcal{N}_2}(a_s-a_r)\phantom{\Big|}}
+\sum_{u\in\mathcal{N}_3}
\frac{(-1)^{n_2+1}q_2}{\prod_{r\in\mathcal{N}_2\cup\widehat{\mathcal{N}}_3}(a_u-a_r)
\phantom{\Big|}}\\
&\quad+\sum_{u\in\mathcal{N}_3}
\frac{(-1)^{n_3}q_3}{\prod_{r\in\widehat{\mathcal{N}}_3\cup\mathcal{N}_4}(a_u-a_r)
\phantom{\Big|}}
+\sum_{s\in\mathcal{N}_1}\frac{(-1)^{n_4+1}q_4}{\prod_{r\in\mathcal{N}_4\cup
\widehat{\mathcal{N}}_1}(a_s-a_r)\phantom{\Big|}}~.
\end{aligned}
\end{equation}

\subsection*{Integration contour $\big(\chi_2|_-,\chi_1|_-,\chi_3|_+,\chi_4|_-\big)$:} 
\begin{equation}
\begin{aligned}
\label{C2B}
\mathcal{W}_{1-\text{inst}}&=\sum_{t\in\mathcal{N}_2}
\frac{(-1)^{n_1+1}q_1}{\prod_{r\in\mathcal{N}_1
\cup\widehat{\mathcal{N}}_2}(a_t-a_r)\phantom{\Big|}}
+\sum_{u\in\mathcal{N}_3}
\frac{(-1)^{n_2+1}q_2}{\prod_{r\in\mathcal{N}_2\cup\widehat{\mathcal{N}}_3}(a_u-a_r)
\phantom{\Big|}}\\
&\quad+\sum_{u\in\mathcal{N}_3}
\frac{(-1)^{n_3}q_3}{\prod_{r\in\widehat{\mathcal{N}}_3\cup\mathcal{N}_4}(a_u-a_r)
\phantom{\Big|}}
+\sum_{s\in\mathcal{N}_1}\frac{(-1)^{n_4+1}q_4}{\prod_{r\in\mathcal{N}_4\cup
\widehat{\mathcal{N}}_1}(a_s-a_r)\phantom{\Big|}}~.
\end{aligned}
\end{equation}

\subsection*{Integration contour $\big(\chi_3|_-,\chi_2|_+,\chi_1|_-,\chi_4|_+\big)$:} 
\begin{equation}
\begin{aligned}
\label{C5A}
\mathcal{W}_{1-\text{inst}}&=\sum_{t\in\mathcal{N}_2}
\frac{(-1)^{n_1+1}q_1}{\prod_{r\in\mathcal{N}_1
\cup\widehat{\mathcal{N}}_2}(a_t-a_r)\phantom{\Big|}}
+\sum_{t\in\mathcal{N}_2}
\frac{(-1)^{n_2}q_2}{\prod_{r\in\widehat{\mathcal{N}}_2\cup\mathcal{N}_3}(a_t-a_r)
\phantom{\Big|}}\\
&\quad+\sum_{v\in\mathcal{N}_4}
\frac{(-1)^{n_3+1}q_3}{\prod_{r\in\mathcal{N}_3\cup\widehat{\mathcal{N}}_4}(a_v-a_r)
\phantom{\Big|}}
+\sum_{u\in\mathcal{N}_4}\frac{(-1)^{n_4}q_4}{\prod_{r\in\widehat{\mathcal{N}}_4\cup
\mathcal{N}_1}(a_u-a_r)\phantom{\Big|}}~.
\end{aligned}
\end{equation}

\section{Chiral ring equations and superpotentials at the one-instanton level}
\label{appB}

\subsection*{Quiver $Q_0$}
We begin by considering the first quiver $Q_0$ of the two duality chains 
of Fig.~\ref{[n1,n2,n3,n4]chain} and Fig.~\ref{AnotherSequence}, namely
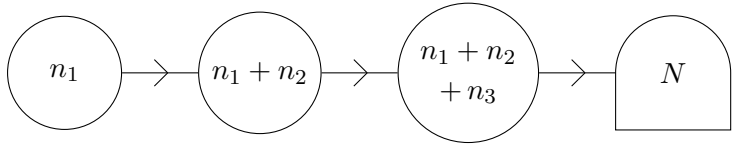
\begin{figure}[H]
\begin{center}
\begin{tikzpicture}[decoration={
markings,
mark=at position 0.6 with {\draw (-5pt,-5pt) -- (0pt,0pt);
                \draw (-5pt,5pt) -- (0pt,0pt);}}]
  \matrix[row sep=10mm,column sep=5mm, ampersand replacement=\&] {
      \node(g1)[gauge] {$n_1$};  \& \& \node(g2)[gauge] {$n_1+n_2$}; \& \&  \node(g3)[gauge,align=center] {$n_1+n_2$\\$+\, n_3$}; 
      \& \&\node(gfN)[gaugedflavor]{$N$};\\
  };
\graph{(g1) --[postaction={decorate}](g2)--[postaction={decorate}](g3)--[postaction={decorate}](gfN);};
\end{tikzpicture}
\label{Quiver1-4node_app}
\end{center}
\end{figure}
\vspace{-0.5cm}
\noindent
The corresponding chiral ring equations have already been written 
in Section~\ref{TCRvsLocQ1JK1}, but we rewrite them here for convenience
\begin{subequations}
\begin{align}
\cQ_{2}(\sigma^{(1)}_s) &= \Lambda_1^{n_1+n_2}~,
\label{TCRQ1_a_app}\\
\cQ_{{3}}(\sigma^{(2)}_t) &= (-1)^{n_1} \,\Lambda_{2}^{n_2+n_3}\,\cQ_{{1}}(\sigma^{(2)}_t)~,
\label{TCRQ1_b_app}\\
\cP_N(\sigma^{(3)}_u) &=(-1)^{n_1+n_2}\left(\Lambda_{3}^{n_3+n_4}\, \cQ_{{2}}(\sigma^{(3)}_u) 
+\frac{\Lambda_{\text{4d}}^{2N}}{\Lambda_{3}^{n_3+n_4}\, \cQ_{{2}}(\sigma^{(3)}_u)}\right)~,
\label{TCRQ1_c_app}
\end{align}
\label{TCRQ1_app}
\end{subequations}
for $s\in {\mathcal N}_1$, $t\in {\mathcal N}_1\cup {\mathcal N}_2$, 
and $u\in {\mathcal N}_1\cup {\mathcal N}_2 \cup {\mathcal N}_3$, respectively.

We look for solutions of these equations that are 
of the form $\sigma^{(I)}_{\star} = \sigma^{(I)}_{\text{cl}} 
+ \delta\sigma^{(I)}$
with the classical vacuum given in the first row of Tab.~\ref{vevs1}.
We work at the lowest order in the quantum fluctuations, proportional to%
\footnote{According to Table \ref{mapTable} and eq. (\ref{q4}) this corresponds to the lowest order
in the $q_I$ parameters.}
$\Lambda_1^{n_1+n_2}$, $\Lambda_{2}^{n_2+n_3}$, $\Lambda_{3}^{n_3+n_4}$ 
and $\Lambda_{\text{4d}}^{2N}/(\Lambda_1^{n_1+n_2}\Lambda_{2}^{n_2+n_3}\Lambda_{3}^{n_3+n_4})$.
With this Ansatz, equation \eqref{TCRQ1_a_app} gives
\begin{equation}
\label{TCRQ1eqn1}
\delta\sigma^{(1)}_s-\delta\sigma^{(2)}_s=\frac{\Lambda_1^{n_1+n_2}}{\prod_{r\in
\widehat{\mathcal{N}}_1\cup\mathcal{N}_2}(a_s-a_r)\phantom{\Big|}} 
\end{equation}
for $s\in {\mathcal N}_1$, while equation \eqref{TCRQ1_b_app} yields
\begin{equation}
\delta\sigma^{(2)}_t-\delta\sigma^{(3)}_t =\frac{(-1)^{n_1+1}\Lambda_1^{n_1+n_2}
\Lambda_2^{n_2+n_3}}{\prod_{u\in\widehat{\mathcal{N}}_1\cup\mathcal{N}_2
\cup\mathcal{N}_3}(a_t-a_u)\phantom{\Big|}
\prod_{r\in\mathcal{N}_2}(a_t-a_r)\phantom{\Big|}}
\label{TCRQ1eqn1a}
\end{equation}
when $t\in {\mathcal N}_1$, and 
\begin{equation}
\delta\sigma^{(2)}_t-\delta\sigma^{(3)}_t =\frac{(-1)^{n_1}\Lambda_2^{n_2+n_3}}
{\prod_{r\in\widehat{\mathcal{N}}_2\cup\mathcal{N}_3}(a_t-a_r)\phantom{\Big|}}
\label{TCRQ1eqn1b}
\end{equation}
when $t\in {\mathcal N}_2$.
The right hand side of \eqref{TCRQ1eqn1a} appears as a higher-order term 
and hence one could naively think that it may be discarded. However, 
one should not do that, since it contributes to the lowest-order term in
\eqref{TCRQ1_c_app}.

Let us now consider \eqref{TCRQ1_c_app}. First of all, we observe that, since we are 
at the lowest order in the ramified instanton expansion, the quantum polynomial 
$\cP_N(z)$ can be replaced with its classical counterpart
\begin{equation}
\cP_N(z)=\prod_{i=1}^N(z-a_i)~.
\end{equation}
Then, we proceed to solve \eqref{TCRQ1_c_app} block by block. In the first
block when $u\in \mathcal{N}_1$ and $\cQ_2(\sigma^{(3)}_u)$ has a zero, 
it is the term proportional to $\Lambda_{\text{4d}}^{2N}$ in the right hand side 
of \eqref{TCRQ1_c_app} that contributes to lowest order, and one finds
\begin{equation}
\prod_{r\in\widehat{\mathcal{N}}_1\cup\mathcal{N}_2
\cup\mathcal{N}_3\cup\mathcal{N}_4}\!\!(a_u-a_r)\,\delta\sigma^{(3)}_u
=\frac{(-1)^{n_1+n_2}\Lambda_{\text{4d}}^{2N}}{\Lambda_3^{n_3+n_4}
\prod_{r\in\widehat{\mathcal{N}}_1\cup
\mathcal{N}_2}(a_u-a_r)\phantom{\Big|}\big(\delta\sigma^{(3)}_u-
\delta\sigma^{(2)}_u\big)\phantom{\Big|}}~.
\end{equation}
Inserting \eqref{TCRQ1eqn1a}, we find
\begin{equation}
\label{TCR3N1}
\delta\sigma^{(3)}_u=\frac{(-1)^{n_2}\Lambda_{\text{4d}}^{2N}}{\Lambda_1^{n_1+n_2}\Lambda_2^{n_2+n_3}\Lambda_3^{n_3+n_4}
\prod_{r\in\widehat{\mathcal{N}}_1\cup\mathcal{N}_4}(a_u-a_r)\phantom{\Big|}}~.
\end{equation}

In the second block when $u\in\mathcal{N}_2$ and $\cQ_2(\sigma^{(3)}_u)$ has a zero, 
it is again the term proportional to $\Lambda_{\text{4d}}^{2N}$ in the right hand side 
of \eqref{TCRQ1_c_app} that can contribute 
to the solution at the lowest order. Indeed, we have
\begin{equation}
\prod_{r\in\mathcal{N}_1\cup\widehat{\mathcal{N}}_2
\cup\mathcal{N}_3\cup\mathcal{N}_4}\!\!(a_u-a_r)\,\delta\sigma^{(3)}_u=\frac{(-1)^{n_1+n_2}\Lambda_{\text{4d}}^{2N}}{\Lambda_3^{n_3+n_4}
\prod_{r\in\mathcal{N}_1\cup\widehat{\mathcal{N}}_2}
(a_u-a_r)\phantom{\Big|}\big(\delta\sigma^{(3)}_u-\delta\sigma^{(2)}_u\big)
\phantom{\Big|}}~.
\end{equation}
Substituting \eqref{TCRQ1eqn1b}, we get
\begin{equation}
\label{TCR3N2}
\delta\sigma^{(3)}_u=\frac{(-1)^{n_2+1}\Lambda_{\text{4d}}^{2N}}{\Lambda_2^{n_2+n_3}\Lambda_3^{n_3+n_4}\prod_{r\in\mathcal{N}_1\cup\widehat{\mathcal{N}}_2
\cup\mathcal{N}_4}(a_u-a_r)\phantom{\Big|}
\prod_{s\in\mathcal{N}_1}(a_u-a_s)}~.
\end{equation}
This term, however, is of higher order and thus can be neglected at the one-instanton
level.

Finally, in the third block when $u\in\mathcal{N}_3$ and $\cQ_2(\sigma^{(3)}_u)$ 
has no zeroes, it is the term proportional to $\Lambda_3^{n_3+n_4}$ 
in the right hand side of \eqref{TCRQ1_c_app} that contributes. Indeed,
we find
\begin{equation}
\label{TCR3N3}
\delta\sigma^{(3)}_u=\frac{(-1)^{n_1+n_2}\Lambda_3^{n_3+n_4}}
{\prod_{r\in\widehat{\mathcal{N}}_3\cup\mathcal{N}_4}(a_u-a_q)\phantom{\Big|}}~.
\end{equation}

Having obtained the explicit first-order expression for $\delta\sigma^{(3)}$, we can use it 
in \eqref{TCRQ1eqn1a} and \eqref{TCRQ1eqn1b} to derive the first-order expression
for $\delta\sigma^{(2)}$. Explicitly we have
\begin{equation}
\delta\sigma^{(2)}_t=\frac{(-1)^{n_2}\Lambda_{\text{4d}}^{2N}}{\Lambda_1^{n_1+n_2}
\Lambda_2^{n_2+n_3}\Lambda_3^{n_3+n_4}\prod_{r\in\widehat{\mathcal{N}}_1
\cup\mathcal{N}_4}(a_t-a_r)\phantom{\Big|}}
\end{equation}
for $t\in\mathcal{N}_1$, and
\begin{equation}
\delta\sigma^{(2)}_t=\frac{(-1)^{n_1}\Lambda_2^{n_2+n_3}}
{\prod_{r\in\widehat{\mathcal{N}}_2\cup\mathcal{N}_3}(a_t-a_r)\phantom{\Big|}}
\end{equation}
for $t\in\mathcal{N}_2$. 
Further substituting these results in \eqref{TCRQ1eqn1}, we get the first-order expression
for $\delta\sigma^{(1)}$, namely
\begin{equation}
\delta\sigma^{(1)}_s=\frac{\Lambda_1^{n_1+n_2}}{\prod_{r\in
\widehat{\mathcal{N}}_1\cup\mathcal{N}_2}(a_s-a_r)\phantom{\Big|}}
+\frac{(-1)^{n_2}\Lambda_{\text{4d}}^{2N}}{\Lambda_1^{n_1+n_2}
\Lambda_2^{n_2+n_3}\Lambda_3^{n_3+n_4}
\prod_{r\in\widehat{\mathcal{N}}_1\cup\mathcal{N}_4}
(a_s-a_r)\phantom{\Big|}}
\end{equation}
for $s\in\mathcal{N}_1$.

Using this explicit solution in \eqref{logLider}  and integrating in, we obtain that
the twisted superpotential in the vacuum is given by
\begin{eqnarray}
\mathcal{W}\big|_{\sigma_\star}\!&=&\!\sum_{s\in\mathcal{N}_1}
\frac{\Lambda_1^{n_1+n_2}}{
\prod_{r\in\widehat{\mathcal{N}}_1\cup\mathcal{N}_2}(a_s-a_r)\phantom{\Big|}}+\!
\sum_{t\in\mathcal{N}_2}\frac{(-1)^{n_1}\Lambda_2^{n_2+n_3}}{
\prod_{r\in\widehat{\mathcal{N}}_2\cup\mathcal{N}_3}(a_t-a_r)\phantom{\Big|}}+\!
\sum_{u\in\mathcal{N}_3}\frac{(-1)^{n_1+n_2}\Lambda_3^{n_3+n_4}}{
\prod_{r\in\widehat{\mathcal{N}}_3\cup\mathcal{N}_4}(a_u-a_r)\phantom{\Big|}}
\nonumber\\
&&\hspace{2cm}+\sum_{s\in\mathcal{N}_1}\frac{(-1)^{n_2+1}\Lambda_{\text{4d}}^{2N}}
{\Lambda_1^{n_1+n_2}\Lambda_2^{n_2+n_3}\Lambda_3^{n_3+n_4}
\prod_{r\in\mathcal{N}_4\cup\widehat{\mathcal{N}}_1}(a_s-a_r)\phantom{\Big|}}
~. \label{Wquiver1app}
\end{eqnarray}
which, term by term, matches the localization result \eqref{C1} if the $q$ vs $\Lambda$
map is
\begin{equation}
q_1=(-1)^{n_1}\Lambda_1^{n_1+n_2}~,\quad 
q_2=(-1)^{n_1+n_2}\Lambda_2^{n_2+n_3}~,\quad
q_3=(-1)^{n_1+n_2+n_3}\Lambda_3^{n_3+n_4}
\label{qlambda1app}
\end{equation} 
with $q_1\,q_2\,q_3\,q_4=(-1)^N\Lambda_{\text{4d}}^{2N}$.

\subsection*{Quiver $Q_1$}
Let us now consider the quiver $Q_1$:
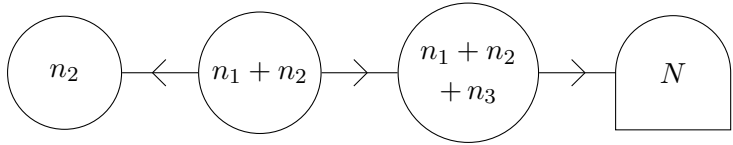
\begin{figure}[H]
\begin{center}
\begin{tikzpicture}[decoration={
markings,
mark=at position 0.6 with {\draw (-5pt,-5pt) -- (0pt,0pt);
                \draw (-5pt,5pt) -- (0pt,0pt);}}]
  \matrix[row sep=10mm,column sep=5mm] {
      \node(g1)[gauge] {$n_2$};  & & \node(g2)[gauge] {$n_1+n_2$}; & & \node(g3)[gauge,align=center] {$n_1+n_2$\\$+\, n_3$}; 
      & &\node(gfN)[gaugedflavor]{$N$};\\
  };
\graph{(g2) --[postaction={decorate}](g1) (g2)--[postaction={decorate}](g3)--[postaction={decorate}](gfN);};
\end{tikzpicture}
\label{Q2diagram}
\end{center}
\end{figure}
\vspace{-0.5cm}
\noindent
The corresponding twisted chiral ring equations are
\begin{equation}
\begin{aligned}
\cQ_2( \sigma_s^{(1)}) &= (-1)^{n_1+n_2} \Lambda_1^{n_1+n_2}~,\\
\cQ_1( \sigma_t^{(2)}) \, \cQ_3( \sigma_t^{(2)}) &= \Lambda_2^{n_1+2n_2 + n_3}~,\\
\cP_N(\sigma^{(3)}_u) &= (-1)^{n_1+n_2}\left(\Lambda_3^{n_3+n_4}\cQ_2(\sigma^{(3)}_u) 
+ \frac{\Lambda_{\text{4d}}^{2N}}{\Lambda_3^{n_3+n_4}\cQ_2(\sigma^{(3)}_u)}\right)~,
\end{aligned}
\label{4nodeQ2TCR}
\end{equation}
for $s\in {\mathcal N}_2$, $t\in {\mathcal N}_1\cup {\mathcal N}_2$, 
and $u\in {\mathcal N}_1\cup {\mathcal N}_2 \cup {\mathcal N}_3$, respectively.
Here, to avoid clutter, we have denoted the low-energy scales $\Lambda_I^{Q_1}$ simply as
$\Lambda_I$.
The solution of these equations about the classical vacuum indicated in the second
row of Tab.~\ref{vevs1} is a generalization of what we have discussed in the 
previous subsection for the quiver $Q_0$, and thus we do not repeat it here.
Instead, we write the result of substituting this solution into \eqref{logLider}
and integrating in, which yields the twisted superpotential in the vacuum, namely
\begin{eqnarray}
\mathcal{W}\big|_{\sigma_\star}\!&=&\!\sum_{t\in\mathcal{N}_2}
\frac{(-1)^{n_1+n_2+1}\Lambda_1^{n_1+n_2}}{
\prod_{r\in\mathcal{N}_1\cup\widehat{\mathcal{N}}_2}(a_t-a_r)\phantom{\Big|}}+
\sum_{t\in\mathcal{N}_2}\frac{(-1)^{n_1+n_2+1}\Lambda_2^{n_1+2n_2+n_3}}{
\Lambda_1^{n_1+n_2}
\prod_{r\in\widehat{\mathcal{N}}_2\cup\mathcal{N}_3}(a_t-a_r)\phantom{\Big|}}
\label{Wquiver2app} \\
&&+\!\!
\sum_{u\in\mathcal{N}_3}\frac{(-1)^{n_1+n_2}\Lambda_3^{n_3+n_4}}{
\prod_{r\in\widehat{\mathcal{N}}_3\cup\mathcal{N}_4}(a_u-a_r)\phantom{\Big|}}
+\!\!\sum_{s\in\mathcal{N}_1}\frac{(-1)^{n_1+n_2}\Lambda_{\text{4d}}^{2N}}
{\Lambda_2^{n_1+2n_2+n_3}\Lambda_3^{n_3+n_4}
\prod_{r\in\mathcal{N}_4\cup\widehat{\mathcal{N}}_1}(a_s-a_r)\phantom{\Big|}}
~. \nonumber
\end{eqnarray}
It is easy to see that this exactly matches, term by term, the superpotential \eqref{C23}
obtained from localization, if the following $q$ vs $\Lambda$ map is used
\begin{equation}
q_1=(-1)^{n_2}\Lambda_1^{n_1+n_2},~~ q_2=
\frac{(-1)^{n_1+1}\Lambda_2^{n_1 + 2 n_2 + n_3}}{\Lambda_1^{n_1+n_2}}\,~~ q_3=(-1)^{n_1+n_2+n_3}\Lambda_3^{n_3+n_4}
\end{equation}
with $q_1\,q_2\,q_3\,q_4=(-1)^N\Lambda_{\text{4d}}^{2N}$.

\subsection*{Quiver $Q_2$}
We now consider the quiver $Q_2$, namely
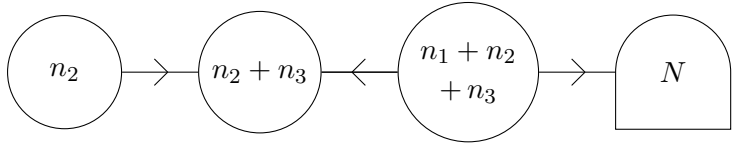
\begin{figure}[H]
\begin{center}
\begin{tikzpicture}[decoration={
markings,
mark=at position 0.6 with {\draw (-5pt,-5pt) -- (0pt,0pt);
                \draw (-5pt,5pt) -- (0pt,0pt);}}]
  \matrix[row sep=10mm,column sep=5mm] {
      \node(g1)[gauge] {$n_2$};  & & \node(g2)[gauge] {$n_2+n_3$}; & & \node(g3)[gauge,align=center] {$n_1+n_2$\\$+\, n_3$}; 
      & &\node(gfN)[gaugedflavor]{$N$};\\
  };
\graph{(g1) --[postaction={decorate}](g2) (g3)--[postaction={decorate}](g2)--(g3) -- [postaction={decorate}](gfN);};
\end{tikzpicture}
\label{Q3-4 nodes}
\end{center}
\end{figure}
\vspace{-0.5cm}
\noindent
The corresponding twisted chiral ring equations are
\begin{equation}
\label{4nodeQ3TCR}
\begin{aligned}
\cQ_2(\sigma^{(1)}_s)&=\Lambda_1^{n_2+n_3}~,\\
\cQ_1(\sigma^{(2)}_t) \,\cQ_3(\sigma^{(2)}_t) &= 
(-1)^{n_1+n_3}\Lambda_2^{n_1+2n_2+n_3}~,\\
\cP_N(\sigma^{(3)}_u) &=
\frac{\Lambda_3^{N+n_2+n_3}}{\cQ_2(\sigma^{(3)}_u)} 
+ \frac{\Lambda_{\text{4d}}^{2N}\,\cQ_2(\sigma^{(3)}_u)}{\Lambda_3^{N+n_2+n_3}}.
\end{aligned}
\end{equation}
for $s\in\mathcal{N}_2$, $t\in\mathcal{N}_2\cup\mathcal{N}_3$ and
$u\in\mathcal{N}_1\cup\mathcal{N}_2\cup\mathcal{N}_3$, respectively.
Again, to avoid clutter, we have denoted $\Lambda_I^{Q_2}$ simply as $\Lambda_I$.
Solving around the vacuum indicated in Tab.~\ref{vevs1}, plugging the solution into
\eqref{logLider} and integrating in, we find
\begin{eqnarray}
\mathcal{W}\big|_{\sigma_\star}\!&=&\!\sum_{t\in\mathcal{N}_2}
\frac{(-1)^{n_1+n_3}\Lambda_2^{n_1+2n_2+n_3}}{
\Lambda_1^{n_2+n_3}
\prod_{r\in\mathcal{N}_1\cup\widehat{\mathcal{N}}_2}(a_t-a_r)\phantom{\Big|}}+
\sum_{t\in\mathcal{N}_2}\frac{\Lambda_1^{n_2+n_3}}{
\prod_{r\in\widehat{\mathcal{N}}_2\cup\mathcal{N}_3}(a_t-a_r)\phantom{\Big|}}
\label{Wquiver3app} \\
&&+\!\!
\sum_{u\in\mathcal{N}_3}\frac{(-1)^{n_1+n_3+1}\Lambda_3^{N+n_2+n_3}}{
\Lambda_2^{n_1+2n_2+n_3}\prod_{r\in\widehat{\mathcal{N}}_3\cup\mathcal{N}_4}
(a_u-a_r)\phantom{\Big|}}
-\!\!\sum_{s\in\mathcal{N}_1}\frac{\Lambda_{\text{4d}}^{2N}}
{\Lambda_3^{N+n_2+n_3}
\prod_{r\in\mathcal{N}_4\cup\widehat{\mathcal{N}}_1}(a_s-a_r)\phantom{\Big|}}
~. \nonumber
\end{eqnarray}
This agrees, term by term, with the localization result \eqref{C23} if the following
$q$ vs $\Lambda$ map is used
\begin{equation}
q_1=(-1)^{n_3+1}\frac{\Lambda_2^{n_1 + 2 n_2 + n_3}}{\Lambda_1^{n_2+n_3}}~,~~
q_2=(-1)^{n_2}\Lambda_1^{n_2+n_3}~,~~
q_3=(-1)^{n_1+1}\frac{\Lambda_3^{N+n_2+n_3}}{\Lambda_2^{n_1+2n_2+n_3}}
\end{equation}
with $q_1\,q_2\,q_3\,q_4=(-1)^N\Lambda_{\text{4d}}^{2N}$.

\subsection*{Quiver $Q_4$}
The quiver $Q_4$ is 
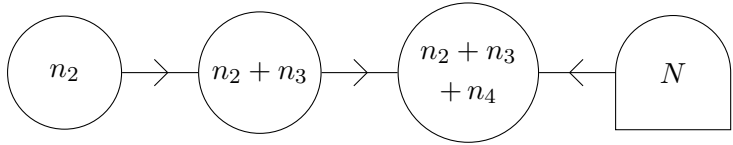
\begin{figure}[H]
\begin{center}
\begin{tikzpicture}[decoration={
markings,
mark=at position 0.6 with {\draw (-5pt,-5pt) -- (0pt,0pt);
                \draw (-5pt,5pt) -- (0pt,0pt);}}]
  \matrix[row sep=10mm,column sep=5mm] {
      \node(g1)[gauge] {$n_2$};  & & \node(g2)[gauge] {$n_2+n_3$}; & & \node(g3)[gauge,align=center] {$n_2+n_3$\\$+\, n_4$}; 
      & &\node(gfN)[gaugedflavor]{$N$};\\
  };
\graph{(g1) --[postaction={decorate}](g2)--[postaction={decorate}](g3)  (gfN)-- [postaction={decorate}](g3);};
\end{tikzpicture}
\label{Q4-4nodes}
\end{center}
\end{figure}
\vspace{-0.5cm}
\noindent
and the corresponding twisted chiral ring equations are
\begin{equation}
\begin{aligned}
\label{4nodeQ4TCR}
\cQ_2(\sigma^{(1)}_s)&=\Lambda_1^{n_2+n_3}~,\\
\cQ_3(\sigma^{(2)}_t)&=(-1)^{n_2}\Lambda_2^{n_3+n_4}
\cQ_1(\sigma^{(2)}_t)~,\\
\cP_N(\sigma^{(3)}_u)&=(-1)^{n_1+n_4}\left(\frac{\Lambda_3^{N+n_2+n_3}}
{\cQ_2(\sigma^{(3)}_u)}+\frac{\Lambda_{\text{4d}}^{2N}\,
\cQ_2(\sigma^{(3)}_u)}{\Lambda_3^{N+n_2+n_3}}\right)
\end{aligned}
\end{equation}
for $s\in\mathcal{N}_2$, $t\in\mathcal{N}_2\cup\mathcal{N}_3$ and 
$u\in\mathcal{N}_2\cup\mathcal{N}_3\cup\mathcal{N}_4$. Again we have
denoted the low-energy scales $\Lambda_I^{Q_4}$ simply as $\Lambda_I$.
Proceeding as discussed above, in this case we find
\begin{eqnarray}
\mathcal{W}\big|_{\sigma_\star}\!&=&\!\sum_{t\in\mathcal{N}_2}
\frac{(-1)^{n_1+n_2+n_4+1}\Lambda_3^{N+n_2+n_3}}{
\Lambda_1^{n_2+n_3}\Lambda_2^{n_3+n_4}
\prod_{r\in\mathcal{N}_1\cup\widehat{\mathcal{N}}_2}(a_t-a_r)\phantom{\Big|}}+
\sum_{t\in\mathcal{N}_2}\frac{\Lambda_1^{n_2+n_3}}{
\prod_{r\in\widehat{\mathcal{N}}_2\cup\mathcal{N}_3}(a_t-a_r)\phantom{\Big|}}
\label{Wquiver4app} \\
&&+\!\!
\sum_{u\in\mathcal{N}_3}\frac{(-1)^{n_2}\Lambda_2^{n_3+n_4}}{
\prod_{r\in\widehat{\mathcal{N}}_3\cup\mathcal{N}_4}
(a_u-a_r)\phantom{\Big|}}
+\!\!\sum_{v\in\mathcal{N}_4}\frac{(-1)^{n_1+n_4}\Lambda_{\text{4d}}^{2N}}
{\Lambda_3^{N+n_2+n_3}
\prod_{r\in\widehat{\mathcal{N}}_4\cup\mathcal{N}_1}(a_v-a_r)\phantom{\Big|}}
~. \nonumber
\end{eqnarray}
This expression exactly matches, term by term, with the localization result \eqref{C4}
provided the following $q$ vs $\Lambda$ map is used: 
\begin{equation}
q_1=(-1)^{n_2+n_4}\frac{\Lambda_3^{N+ n_2 +n_3}}{
\Lambda_1^{n_2+n_3}\Lambda_2^{n_3+n_4}}~,~~
q_2=(-1)^{n_2}\Lambda_1^{n_2+n_3}~,~~
q_3=(-1)^{n_2+n_3}\Lambda_2^{n_3+n_4}
\end{equation}
with $q_1\,q_2\,q_3\,q_4=(-1)^N\Lambda_{\text{4d}}^{2N}$.

\subsection*{Quiver $Q_5$}
The quiver $Q_5$ is
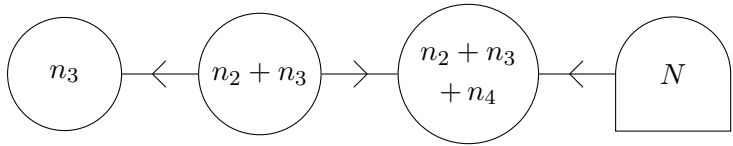
\begin{figure}[H]
\begin{center}
\begin{tikzpicture}[decoration={
markings,
mark=at position 0.6 with {\draw (-5pt,-5pt) -- (0pt,0pt);
                \draw (-5pt,5pt) -- (0pt,0pt);}}]
  \matrix[row sep=10mm,column sep=5mm] {
      \node(g1)[gauge] {$n_3$};  & & \node(g2)[gauge] {$n_2+n_3$}; & & \node(g3)[gauge,align=center] {$n_2+n_3$\\$+\, n_4$}; 
      & &\node(gfN)[gaugedflavor]{$N$};\\
  };
\graph{(g2) --[postaction={decorate}](g1) (g2)--[postaction={decorate}](g3) (gfN)--[postaction={decorate}](g3);};
\end{tikzpicture}
\label{Q5-4nodes}
\end{center}
\end{figure}
\vspace{-0.5cm}
\noindent
and the corresponding twisted chiral ring equations are
\begin{equation}
\label{4nodeQ5TCR}
\begin{aligned}
\cQ_2(\sigma^{(1)}_s)&=(-1)^{n_2+n_3}\Lambda_1^{n_2+n_3}~,\\
\cQ_1(\sigma^{(2)}_t) \,\cQ_3(\sigma^{(2)}_t) &= \Lambda_2^{n_2+2n_3+n_4} ~,\\
\cP_N(\sigma^{(3)}_u) &= 
(-1)^{n_1+n_4}\left(\frac{\Lambda_3^{N+n_2+n_3}}{\cQ_2(\sigma^{(3)}_u)} + 
\frac{\Lambda_{\text{4d}}^{2N}\,\cQ_2(\sigma^{(3)}_u)}{\Lambda_3^{N+n_2+n_3}}\right) 
\end{aligned}
\end{equation}
for $s\in\mathcal{N}_3$, $t\in\mathcal{N}_2\cup\mathcal{N}_3$ and
$u\in\mathcal{N}_2\cup\mathcal{N}_3\cup\mathcal{N}_4$ respectively. As before, to avoid
clutter we have denoted $\Lambda_I^{Q_5}$ simply as $\Lambda_I$. 
Solving these equations around the appropriate vacuum (see Tab.~\ref{vevs1}),
using \eqref{logLider} and integrating in, we find
\begin{eqnarray}
\mathcal{W}\big|_{\sigma_\star}\!&=&\!\sum_{t\in\mathcal{N}_2}
\frac{(-1)^{n_1+n_4}\Lambda_3^{N+n_2+n_3}}{
\Lambda_2^{n_2+2n_3+n_4}
\prod_{r\in\mathcal{N}_1\cup\widehat{\mathcal{N}}_2}(a_t-a_r)\phantom{\Big|}}\!+\!
\sum_{u\in\mathcal{N}_3}\frac{(-1)^{n_2+n_3+1}\Lambda_1^{n_2+n_3}}{
\prod_{r\in\mathcal{N}_2\cup\widehat{\mathcal{N}}_3}(a_u-a_r)\phantom{\Big|}}
\label{Wquiver5app} \\
&&+\!\!
\sum_{u\in\mathcal{N}_3}\frac{(-1)^{n_2+n_3+1}\Lambda_3^{N+n_2+n_3}}{
\Lambda_1^{n_1+n_3}\prod_{r\in\widehat{\mathcal{N}}_3\cup\mathcal{N}_4}
(a_u-a_r)\phantom{\Big|}}
+\!\!\sum_{v\in\mathcal{N}_4}\frac{(-1)^{n_1+n_4}\Lambda_{\text{4d}}^{2N}}
{\Lambda_3^{N+n_2+n_3}
\prod_{r\in\widehat{\mathcal{N}}_4\cup\mathcal{N}_1}(a_v-a_r)\phantom{\Big|}}
~. \nonumber
\end{eqnarray}
This expression agrees, term by term, with the localization result \eqref{C56} if the following
$q$ vs $\Lambda$ map is used:
\begin{equation}
q_1=(-1)^{n_4+1}\frac{\Lambda_3^{N+n_2 + n_3}}{\Lambda_2^{n_2+2n_3+n_4}}~,~~
q_2=(-1)^{n_3}\Lambda_1^{n_2+n_3}~,~~
q_3=(-1)^{n_2+1}\frac{\Lambda_2^{n_2+2n_3+n_4}}{\Lambda_1^{n_2+n_3}}
\end{equation}
with $q_1\,q_2\,q_3\,q_4=(-1)^N\Lambda_{\text{4d}}^{2N}$.

\subsection*{Quiver $Q_6$}
The quiver $Q_6$ is
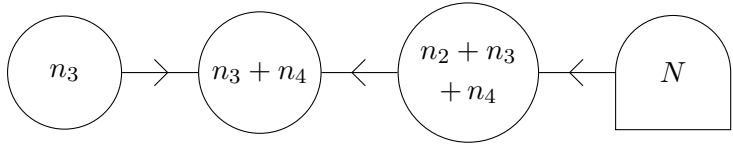
\begin{figure}[H]
\begin{center}
\begin{tikzpicture}[decoration={
markings,
mark=at position 0.6 with {\draw (-5pt,-5pt) -- (0pt,0pt);
                \draw (-5pt,5pt) -- (0pt,0pt);}}]
  \matrix[row sep=10mm,column sep=5mm] {
      \node(g1)[gauge] {$n_3$};  & & \node(g2)[gauge] {$n_3+n_4$}; & & \node(g3)[gauge,align=center] {$n_2+n_3$\\$+\, n_4$}; 
      & &\node(gfN)[gaugedflavor]{$N$};\\
  };
\graph{(g1) --[postaction={decorate}](g2) (g3)--[postaction={decorate}](g2) (gfN)--[postaction={decorate}](g3);};
\end{tikzpicture}
\label{Q6-4node}
\end{center}
\end{figure}
\vspace{-0.5cm}
\noindent
and the corresponding twisted chiral ring equations are
\begin{equation}
\label{4nodeQ6TCR}
\begin{aligned}
\cQ_2(\sigma^{(1)}_s) &=\Lambda_1^{n_3+n_4}~,\\
\cQ_1(\sigma^{(2)}_t) \, \cQ_3(\sigma^{(2)}_t) &= 
(-1)^{n_2+n_4}\, \Lambda_2^{n_2+2n_3+n_4} ~,\\
\cP_N(\sigma^{(3)}_u) &= (-1)^{N}\left(\Lambda_3^{n_1+n_2}\, \cQ_2(\sigma^{(3)}_u) + 
\frac{\Lambda_{\text{4d}}^{2N}}{\Lambda_3^{n_1+n_2}\cQ_2(\sigma^{(3)}_u)}\right)
\end{aligned}
\end{equation}
with $s\in\mathcal{N}_3$, $t\in\mathcal{N}_3\cup\mathcal{N}_4$
and $u\in\mathcal{N}_2\cup\mathcal{N}_3\cup\mathcal{N}_4$ respectively.
Again the low-energy scales of this quiver have been denoted simply as $\Lambda_I$
instead of $\Lambda_I^{Q_6}$.
We solve these equations about the classical vacuum given in Tab.~\ref{vevs1};
after inserting the solution in \eqref{logLider} and integrating in, we obtain
\begin{eqnarray}
\mathcal{W}\big|_{\sigma_\star}\!&=&\!\sum_{t\in\mathcal{N}_2}
\frac{(-1)^{N+1}\Lambda_3^{n_1+n_2}}{
\prod_{r\in\mathcal{N}_1\cup\widehat{\mathcal{N}}_2}(a_t-a_r)\phantom{\Big|}}\!+\!
\sum_{u\in\mathcal{N}_3}\frac{(-1)^{n_2+n_4}\Lambda_2^{n_2+2n_3+n_4}}{
\Lambda_1^{n_3+n_4}
\prod_{r\in\mathcal{N}_2\cup\widehat{\mathcal{N}}_3}(a_u-a_r)\phantom{\Big|}}
\label{Wquiver6app} \\
&&+\!\!
\sum_{u\in\mathcal{N}_3}\frac{\Lambda_1^{n_3+n_4}}{
\prod_{r\in\widehat{\mathcal{N}}_3\cup\mathcal{N}_4}
(a_u-a_r)\phantom{\Big|}}
+\sum_{v\in\mathcal{N}_4}\frac{(-1)^{n_1+n_3+1}\Lambda_{\text{4d}}^{2N}}
{\Lambda_2^{n_2+2n_3+n_4}\Lambda_3^{n_1+n_2}
\prod_{r\in\widehat{\mathcal{N}}_4\cup\mathcal{N}_1}(a_v-a_r)\phantom{\Big|}}
~. \nonumber
\end{eqnarray}
This expression perfectly matches, term by term, the localization result \eqref{C56}
if the $q$ vs $\Lambda$ map is
\begin{equation}
q_1=(-1)^{n_2+n_3+n_4}\Lambda_3^{n_1 + n_2}~,~~
q_2=(-1)^{n_4+1}\frac{\Lambda_2^{n_2+2n_3+n_4}}{\Lambda_1^{n_3+n_4}}~,\quad
q_3=(-1)^{n_3}\Lambda_1^{n_3+n_4}
\end{equation}
with $q_1\,q_2\,q_3\,q_4=(-1)^N\Lambda_{\text{4d}}^{2N}$.

\subsection*{Quiver $Q_7$}
The last quiver of the duality chain of  in Fig.~\ref{[n1,n2,n3,n4]chain} is
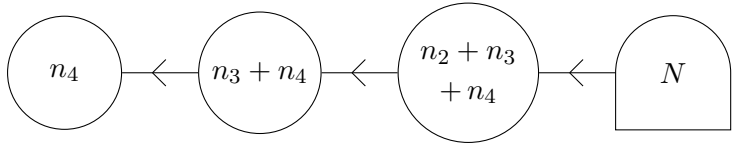
\begin{figure}[H]
\begin{center}
\begin{tikzpicture}[decoration={
markings,
mark=at position 0.6 with {\draw (-5pt,-5pt) -- (0pt,0pt);
                \draw (-5pt,5pt) -- (0pt,0pt);}}]
  \matrix[row sep=10mm,column sep=5mm] {
      \node(g1)[gauge] {$n_4$};  & & \node(g2)[gauge] {$n_3+n_4$}; & & \node(g3)[gauge,align=center] {$n_2+n_3$\\$+\, n_4$}; 
      & &\node(gfN)[gaugedflavor]{$N$};\\
  };
\graph{(g2) --[postaction={decorate}](g1) (g3)--[postaction={decorate}](g2) (gfN)--[postaction={decorate}](g3);};
\end{tikzpicture}
\label{Q7-4 nodes}
\end{center}
\end{figure}
\vspace{-0.5cm}
\noindent
and the associated twisted chiral ring equations are
\begin{subequations}
\begin{align}
\cQ_2(\sigma^{(1)}_s)&=(-1)^{n_3+n_4}\Lambda_1^{n_3+n_4}~,\\
\cQ_3(\sigma^{(2)}_t)&=(-1)^{n_2+n_3+n_4}\Lambda_2^{n_2+n_3}\cQ_1(\sigma^{(2)}_t)~,\\
\cP_N(\sigma^{(3)}_u)&=
(-1)^N\left(\Lambda_3^{n_1+n_2}\cQ_2(\sigma^{(3)}_u)
+\frac{\Lambda_{\text{4d}}^{2N}}{\Lambda_3^{n_1+n_2}
\cQ_2(\sigma^{(3)}_u)}\right)
\end{align}
\end{subequations}
with $s\in\mathcal{N}_4$, $t\in\mathcal{N}_3\cup\mathcal{N}_4$ and
$u\in\mathcal{N}_2\cup\mathcal{N}_3\cup\mathcal{N}_4$. Here, $\Lambda_I$ denote the scales
of this quiver. Solving these
equation around the vacuum reported in the last row of Tab.~\ref{vevs1}, and proceeding
as in the previous cases, we obtain
\begin{eqnarray}
\mathcal{W}\big|_{\sigma_\star}\!&=&\!\sum_{t\in\mathcal{N}_2}
\frac{(-1)^{N+1}\Lambda_3^{n_1+n_2}}{
\prod_{r\in\mathcal{N}_1\cup\widehat{\mathcal{N}}_2}(a_t-a_r)\phantom{\Big|}}\!+\!
\sum_{u\in\mathcal{N}_3}\frac{(-1)^{n_2+n_3+n_4+1}\Lambda_2^{n_2+n_3}}{
\prod_{r\in\mathcal{N}_2\cup\widehat{\mathcal{N}}_3}(a_u-a_r)\phantom{\Big|}}
\label{Wquiver7app} \\
&&+\!\!
\sum_{v\in\mathcal{N}_4}\frac{(-1)^{n_3+n_4+1}\Lambda_1^{n_3+n_4}}{
\prod_{r\in\mathcal{N}_3\cup\widehat{\mathcal{N}}_4}
(a_u-a_r)\phantom{\Big|}}
+\!\!\sum_{v\in\mathcal{N}_4}\frac{(-1)^{n_1+n_3+n_4}\Lambda_{\text{4d}}^{2N}}
{\Lambda_1^{n_3+n_4}\Lambda_2^{n_2+n_3}\Lambda_3^{n_1+n_2}
\prod_{r\in\widehat{\mathcal{N}}_4\cup\mathcal{N}_1}(a_v-a_r)\phantom{\Big|}}
~. \nonumber
\end{eqnarray}
This agrees, term by term, with the localization result \eqref{C7} using the following
$q$ vs $\Lambda$ map
\begin{equation}
q_1=(-1)^{n_2+n_3+n_4}\big(\Lambda_3\big)^{n_1+n_2}~,~~
q_2=(-1)^{n_3+n_4}\big(\Lambda_2\big)^{n_2+n_3}~,~~
q_3=(-1)^{n_4}\big(\Lambda_1\big)^{n_3+n_4}
\end{equation}
with $q_1\,q_2\,q_3\,q_4=(-1)^N\Lambda_{\text{4d}}^{2N}$.

\subsection*{Quiver $\widehat{Q}_{1}$}
The non-linear quiver $\widehat{Q}_1$ appearing in the duality chain represented in Fig.~\ref{AnotherSequence}
is
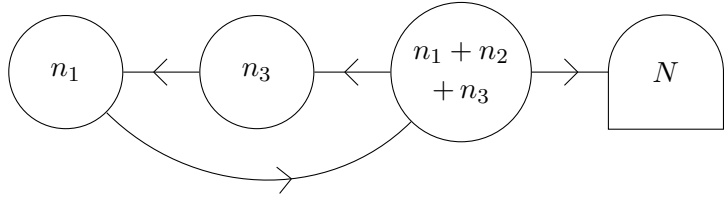
\begin{figure}[H]
\begin{center}
\begin{tikzpicture}[decoration={
markings,
mark=at position 0.6 with {\draw (-5pt,-5pt) -- (0pt,0pt);
                \draw (-5pt,5pt) -- (0pt,0pt);}}]
  \matrix[row sep=10mm,column sep=5mm] {
     \node(g1)[gauge] {$n_1$};  & & \node(g2)[gauge] {$n_3$}; & &  \node(g3)[gauge,align=center] {$n_1+n_2$\\$+\, n_3$}; 
      & &\node(gfN)[gaugedflavor]{$N$};\\
  };
\graph{(g2) --[postaction={decorate}](g1) (g3)--[postaction={decorate}](g2) (g3)--[postaction={decorate}](gfN);};
\graph{(g1) --[out=-45,in=-135,postaction={decorate}](g3)};
\end{tikzpicture}
\label{Q2'}
\end{center}
\end{figure}
\vspace{-0.5cm}
\noindent
and its corresponding twisted chiral ring equations are
\begin{equation}
\label{4nodeTriangularQuiverTCR}
\begin{aligned}
\cQ_3(\sigma^{(1)}_s) &=(-1)^{n_3}\Lambda_1^{n_1+n_2}\cQ_2(\sigma^{(1)}_s)~,\\
\cQ_3(\sigma^{(2)}_t)&= (-1)^{n_1+n_2+n_3}\, \Lambda_2^{n_2+n_3}\, \cQ_1(\sigma^{(2)}_t)
~,\\
\cP_N(\sigma^{(3)}_u) &= (-1)^{n_1}\left(\frac{\Lambda_3^{N+n_3-n_1}\, \cQ_1(\sigma^{(3)}_u)}{\cQ_2(\sigma^{(3)}_u)} + 
\frac{\Lambda_{\text{4d}}^{2N}\,\cQ_2(\sigma^{(3)}_u)}{\Lambda_3^{N+n_3-n_1}\, \cQ_1(\sigma^{(3)}_u)}\right) 
\end{aligned}
\end{equation}
for $s\in\mathcal{N}_1$, $t\in\mathcal{N}_3$ and 
$u\in\mathcal{N}_1\cup\mathcal{N}_2\cup\mathcal{N}_3$. Again we have denoted the low-energy scales $\Lambda_I^{\widehat{Q}_1}$ simply as $\Lambda_I$.
Solving these equations around the vacuum given in the first row of Tab.~\ref{tablevevs21},
using \eqref{logLider} and integrating in, we obtain
\begin{eqnarray}
\mathcal{W}\big|_{\sigma_\star}\!&=&\!\sum_{s\in\mathcal{N}_1}
\frac{(-1)^{n_3}\Lambda_1^{n_1+n_2}}{
\prod_{r\in\widehat{\mathcal{N}}_1\cup\mathcal{N}_2}(a_s-a_r)\phantom{\Big|}}\!+\!
\sum_{u\in\mathcal{N}_3}\frac{(-1)^{n_1+n_2+n_3+1}\Lambda_2^{n_2+n_3}}{
\prod_{r\in\mathcal{N}_2\cup\widehat{\mathcal{N}}_3}(a_u-a_r)\phantom{\Big|}}
\label{Wquiver2Aapp} \\
&&\!\!\!\!\!\!+
\sum_{u\in\mathcal{N}_3}\frac{(-1)^{n_2+n_3+1}\Lambda_3^{n_2+2n_3+n_4}}{
\Lambda_2^{n_2+n_3}\prod_{r\in\widehat{\mathcal{N}}_3\cup\mathcal{N}_4}
(a_u-a_r)\phantom{\Big|}}
+\sum_{s\in\mathcal{N}_1}\frac{(-1)^{n_1+n_3}\Lambda_{\text{4d}}^{2N}}
{\Lambda_1^{n_1+n_2}\Lambda_3^{n_2+2n_3+n_4}
\prod_{r\in\mathcal{N}_4\cup\widehat{\mathcal{N}}_1}(a_s-a_r)\phantom{\Big|}}
~. \nonumber
\end{eqnarray}
This matches, term by term, the localization result \eqref{C2A} using the following
$q$ vs $\Lambda$ map:
\begin{equation}
q_1=(-1)^{n_1+n_3}\Lambda_1^{n_1 + n_2}~,\quad
q_2=(-1)^{n_1+n_3}\Lambda_2^{n_2+n_3}~,\quad
q_3=(-1)^{n_2+1}\frac{\Lambda_3^{n_2+2n_3+n_4}}{\Lambda_2^{n_2+n_3}}
\end{equation}
with $q_1\,q_2\,q_3\,q_4=(-1)^N\Lambda_{\text{4d}}^{2N}$.

\subsection*{Quiver $Q_{3}$}
The quiver $Q_{3}$ is
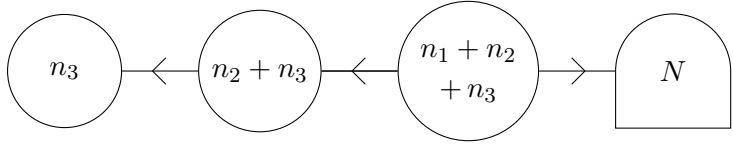
\begin{figure}[H]
\begin{center}
\begin{tikzpicture}[decoration={
markings,
mark=at position 0.6 with {\draw (-5pt,-5pt) -- (0pt,0pt);
                \draw (-5pt,5pt) -- (0pt,0pt);}}]
  \matrix[row sep=10mm,column sep=5mm] {
      \node(g1)[gauge] {$n_3$};  & & \node(g2)[gauge] {$n_2+n_3$}; & &\node(g3)[gauge,align=center] {$n_1+n_2$\\$+\, n_3$}; 
      & &\node(gfN)[gaugedflavor]{$N$};\\
  };
\graph{(g2) --[postaction={decorate}](g1) (g3)--[postaction={decorate}](g2)--(g3) -- [postaction={decorate}](gfN);};
\end{tikzpicture}
\label{Q2''}
\end{center}
\end{figure}
\vspace{-0.5cm}
\noindent
and its twisted chiral ring equations are
\begin{equation}
\label{4nodeQ2''TCR}
\begin{aligned}
\cQ_2(\sigma^{(1)}_s) &=(-1)^{n_2+n_3}\Lambda_1^{n_2+n_3}~,\\
\cQ_3(\sigma^{(2)}_t)&= (-1)^{n_1+n_2+n_3} 
\Lambda_2^{n_1+n_2}\, \cQ_1(\sigma^{(2)}_t)~,\\
\cP_N(\sigma^{(3)}_u) &= \frac{\Lambda_3^{N+n_2+n_3}}{\cQ_2(\sigma^{(3)}_u)} + 
\frac{\Lambda^{2N}_{\text{4d}}\,\cQ_2(\sigma^{(3)}_u)}{\Lambda_3^{N+n_2+n_3}}
\end{aligned}
\end{equation}
for $s\in\mathcal{N}_3$, $t\in\mathcal{N}_2\cup\mathcal{N}_3$ and
$u\in\mathcal{N}_1\cup\mathcal{N}_2\cup\mathcal{N}_3$, respectively. Again, $\Lambda_I$ denote
the low-energy scales of this quiver.
Solving these equations around the vacuum displayed in the middle row of 
Tab.~\ref{tablevevs21} and proceeding in the usual way, we get
\begin{eqnarray}
\mathcal{W}\big|_{\sigma_\star}\!&=&\!\sum_{t\in\mathcal{N}_2}
\frac{(-1)^{n_1+n_2+n_3+1}\Lambda_2^{n_1+n_2}}{
\prod_{r\in\mathcal{N}_1\cup\widehat{\mathcal{N}}_2}(a_t-a_r)\phantom{\Big|}}\!+\!
\sum_{u\in\mathcal{N}_3}\frac{(-1)^{n_2+n_3+1}\Lambda_1^{n_2+n_3}}{
\prod_{r\in\mathcal{N}_2\cup\widehat{\mathcal{N}}_3}(a_u-a_r)\phantom{\Big|}}
\label{Wquiver2Bapp} \\
&&\!\!\!\!\!\!+
\sum_{u\in\mathcal{N}_3}\frac{(-1)^{n_1}\Lambda_3^{n_2+n_3+N}}{
\Lambda_1^{n_2+n_3}\Lambda_2^{n_1+n_2}
\prod_{r\in\widehat{\mathcal{N}}_3\cup\mathcal{N}_4}
(a_u-a_r)\phantom{\Big|}}
-\sum_{s\in\mathcal{N}_1}\frac{\Lambda_{\text{4d}}^{2N}}
{\Lambda_3^{n_2+n_3+N}
\prod_{r\in\mathcal{N}_4\cup\widehat{\mathcal{N}}_1}(a_s-a_r)\phantom{\Big|}}
~. \nonumber
\end{eqnarray}
This matches, term by term, the localization result \eqref{C2B} using the following
$q$ vs $\Lambda$ map:
\begin{equation}
q_1=(-1)^{n_2+n_3}\Lambda_2^{n_1 + n_2}~,\quad
q_2=(-1)^{n_3}\Lambda_1^{n_2+n_3}~,\quad
q_3=(-1)^{n_1+n_3}\frac{\Lambda_3^{n_2+n_3+N}}{
\Lambda_1^{n_2+n_3}\Lambda_2^{n_1+n_2}}
\end{equation}
with $q_1\,q_2\,q_3\,q_4=(-1)^N\Lambda_{\text{4d}}^{2N}$.

\subsection*{Quiver $\widehat{Q}_{5}$}
The second non-linear quiver in the duality chain of Fig.~\ref{AnotherSequence} is
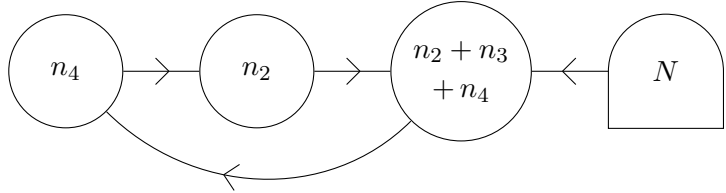
\begin{figure}[H]
\begin{center}
\begin{tikzpicture}[decoration={
markings,
mark=at position 0.6 with {\draw (-5pt,-5pt) -- (0pt,0pt);
                \draw (-5pt,5pt) -- (0pt,0pt);}}]
  \matrix[row sep=10mm,column sep=5mm] {
      \node(g1)[gauge] {$n_4$};  & & \node(g2)[gauge] {$n_2$}; & & \node(g3)[gauge,align=center] {$n_2+n_3$\\$+\, n_4$}; 
      & &\node(gfN)[gaugedflavor]{$N$};\\
  };
\graph{(g1) --[postaction={decorate}](g2) (g2)--[postaction={decorate}](g3) (gfN)--[postaction={decorate}](g3);};
\graph{(g3) --[out=-135,in=-45,postaction={decorate}](g1)};
\end{tikzpicture}
\label{Q6'}
\end{center}
\end{figure}
\vspace{-0.5cm}
\noindent
and the corresponding twisted chiral ring equations are
\begin{equation}
\label{4nodeTriangularQuiver2TCR}
\begin{aligned}
\cQ_3(\sigma^{(1)}_s) &=(-1)^{n_2+n_3+n_4}\Lambda_1^{n_3+n_4}\cQ_2(\sigma^{(1)}_s)~,\\
\cQ_3(\sigma^{(2)}_t)&= (-1)^{n_4}\, \Lambda_2^{n_2+n_3}\, \cQ_1(\sigma^{(2)}_t)~,\\
\cP_N(\sigma^{(3)}_u) &= (-1)^{n_1+n_3+n_4}\left(\frac{\Lambda_3^{n_1+2n_2+n_3}\, \cQ_1(\sigma^{(3)}_u)}{\cQ_2(\sigma^{(3)}_u)} + 
\frac{\Lambda_{\text{4d}}^{2N}\,\cQ_2(\sigma^{(3)}_u)}{\Lambda_3^{n_1+2n_2+n_3}\, \cQ_1(\sigma^{(3)}_u)}\right)
\end{aligned}
\end{equation}
for $s\in\mathcal{N}_4$, $t\in\mathcal{N}_2$ and 
$u\in\mathcal{N}_2\cup\mathcal{N}_3\cup\mathcal{N}_4$, respectively.
As usual, we have denoted the scales $\Lambda_I^{\widehat{Q}_5}$ simply as $\Lambda_I$.
Solving these equations around the vacuum indicated in the last row of
Tab.~\ref{tablevevs21}, using \eqref{logLider} and integrating in, we find
\begin{eqnarray}
\mathcal{W}\big|_{\sigma_\star}\!&=&\!\sum_{t\in\mathcal{N}_2}
\frac{(-1)^{n_1+n_3}\Lambda_3^{n_1+2n_2+n_3}}{
\Lambda_2^{n_2+n_3}
\prod_{r\in\mathcal{N}_1\cup\widehat{\mathcal{N}}_2}(a_t-a_r)\phantom{\Big|}}\!+\!
\sum_{t\in\mathcal{N}_2}\frac{(-1)^{n_4}\Lambda_2^{n_2+n_3}}{
\prod_{r\in\widehat{\mathcal{N}}_2\cup\mathcal{N}_3}(a_t-a_r)\phantom{\Big|}}
\label{Wquiver5Aapp} \\
&&\!\!\!\!\!\!+
\sum_{v\in\mathcal{N}_4}\frac{(-1)^{n_2+n_3+n_4+1}\Lambda_1^{n_3+n_4}}{
\prod_{r\in\mathcal{N}_3\cup\widehat{\mathcal{N}}_4}
(a_v-a_r)\phantom{\Big|}}
+\sum_{v\in\mathcal{N}_4}\frac{(-1)^{n_1+n_2+1}\Lambda_{\text{4d}}^{2N}}
{\Lambda_1^{n_3+n_4}\Lambda_3^{n_1+2n_2+n_3}
\prod_{r\in\widehat{\mathcal{N}}_4\cup\mathcal{N}_1}(a_v-a_r)\phantom{\Big|}}
~. \nonumber
\end{eqnarray}
This superpotential agrees, term by term, with the localization result \eqref{C5A} if
the $q$ vs $\Lambda$ map is
\begin{equation}
q_1=(-1)^{n_3+1}\frac{\Lambda_3^{n_1 +2 n_2+n_3}}{\Lambda_2^{n_2+n_3}}~,\quad
q_2=(-1)^{n_2+n_4}\Lambda_2^{n_2+n_3}~,\quad
q_3=(-1)^{n_2+n_4}\Lambda_1^{n_3+n_4}
\end{equation}
with $q_1\,q_2\,q_3\,q_4=(-1)^N\Lambda_{\text{4d}}^{2N}$.

\section{Some two-instanton results}
\label{app2instantons}

In this appendix we illustrate how the JK prescription works at two-instantons.
In order to keep things as simple as possible, we just focus on the term in the superpotential
that is proportional to $q_1q_2$. 
After using \eqref{Zso4d5d} and \eqref{zexplicit4d}, it is not difficult to realize
that this term takes the following form
\begin{eqnarray}
q_1q_2&& \!\!\!\lim_{\epsilon_1,\hat\epsilon_2\to 0}
\int \!\frac{d\chi_1}{2\pi \ii}\,\frac{d\chi_2}{2\pi \ii}\,
\frac{1}{ \big(\chi_1-\chi_2+\hat\epsilon_2\big)}\,
\prod_{s\in{\mathcal N}_1}
\frac{1}{\left(a_s-\chi_1+\frac{1}{2}(\epsilon_1+\hat\epsilon_2)\right)}
\\
&&\!\!\!\!\!\!\prod_{t\in{\mathcal N}_2} \frac{1}{\left(-a_t+\chi_1+\frac{1}{2}(\epsilon_1+\hat\epsilon_2)\right)
\left(a_t-\chi_2+\frac{1}{2}(\epsilon_1+\hat\epsilon_2)\right)}\,
\prod_{u\in{\mathcal N}_3}\frac{1}{\left(-a_u+\chi_2+\frac{1}{2}
(\epsilon_1+\hat\epsilon_2)\right)}~.
\nonumber
\end{eqnarray}
We now apply the JK prescription to compute the double integral over $\chi_1$ and $\chi_2$.
This amounts to choose two linear factors from the denominator, one containing $\chi_1$ 
and one containing $\chi_2$, such that the reference JK vector belongs to the cone defined by 
the chosen factors. Notice that this way of selecting the residues
does not use any information on 
the $\Omega$-deformation parameters. In Tab.~\ref{wpoles} we list the poles 
that are selected by this JK prescription for the quivers
$Q_1$ and $Q_2$ of the duality chain of Fig.~\ref{[n1,n2,n3,n4]chain}.

An important point that we emphasized in the main body of the paper is that the JK 
vectors $-\zeta_1^{Q_1}\,\chi_1-\zeta_2^{Q_1}\,\chi_2$ 
and $-\zeta_1^{Q_2}\,\chi_2-\zeta_2^{Q_2}\,
\chi_1$ pick up different sets of poles from the localization integrand, as a consequence of the different
signs and magnitudes of the FI parameters. 
One can see this explicitly from the entries in the third column of Tab.~\ref{wpoles}.

\begin{table}[H]
   \begin{center}
   \small{
   \begin{tabular}{|c|c|cc|}
   \hline
\,\,Quiver \phantom{\Big|} &  JK vector & ~~~~poles & \\ 
\hline \hline
\multirow{8}{*}{$\,\,Q_1 \phantom{\Big|}$}&   \multirow{8}{*}{$-\zeta_1^{Q_1}\,\chi_1-\zeta_2^{Q_1}\,\chi_2$} &
$\phantom{\Big|}\chi_1 = a_t -\frac{1}{2}(\epsilon_1+\hat\epsilon_2) $ &
 $\phantom{\Big|}t\in{\mathcal N}_2$ \\
  &    & $\phantom{\Big|}\chi_2 = a_t + \frac{1}{2}(\epsilon_1+\hat\epsilon_2)$ & 
  $\phantom{\Big|}t \in {\mathcal N}_2$ \\
  \cline{3-4}
  &    & $\phantom{\Big|}\chi_1 = a_s + \frac{1}{2}(\epsilon_1+\hat\epsilon_2)$ & 
  $\phantom{\Big|}s \in {\mathcal N}_1$ \\
  &    & $\phantom{\Big|}\chi_2=\chi_1+\hat{\epsilon}_2$ & $\phantom{\Big|}$\\
  \cline{3-4}
      &    &  $\phantom{\Big|}\chi_1 = \chi_2 - \hat\epsilon_2$ & $\phantom{\Big|}$ \\
      &    &  $\phantom{\Big|}\chi_2 = a_t +\frac{1}{2}(\epsilon_1+\hat\epsilon_2) $  
    & $\phantom{\Big|}t\in{\mathcal N}_2$ \\
 \hline
\multirow{8}{*}{$\,\,Q_2 \phantom{\Big|}$}&   \multirow{8}{*}{$-\zeta_1^{Q_2}\,\chi_2-\zeta_2^{Q_2}\,
\chi_1$} 
& $\phantom{\Big|}\chi_2 = a_t +\frac{1}{2}(\epsilon_1+\hat\epsilon_2) $ &
 $\phantom{\Big|}t\in{\mathcal N}_2$ \\
  &    & $\phantom{\Big|}\chi_1 = a_t - \frac{1}{2}(\epsilon_1+\hat\epsilon_2)$ & 
  $\phantom{\Big|}t \in {\mathcal N}_2$ \\
  \cline{3-4}
  &    & $\phantom{\Big|}\chi_2=\chi_1+\hat{\epsilon}_2$ & $\phantom{\Big|}$\\
  &    & $\phantom{\Big|}\chi_1 = a_t - \frac{1}{2}(\epsilon_1+\hat\epsilon_2)$ & 
  $\phantom{\Big|}t \in {\mathcal N}_2$ \\
  \cline{3-4}
      &    &  $\phantom{\Big|}\chi_2 = 
      a_u - \frac{1}{2}(\epsilon_1+\hat\epsilon_2)$ & $\phantom{\Big|}u\in{\mathcal N}_3$\\
      &    &  $\phantom{\Big|}\chi_1 = \chi_2-\hat\epsilon_2$  
    & $\phantom{\Big|}$ \\
 \hline
   \end{tabular}
   }
   \end{center}
\caption{We list the poles that contribute to the $q_1q_2$ term of the superpotential for the 
quivers $Q_1$ and $Q_2$ of the duality 
chain in Fig.~\ref{[n1,n2,n3,n4]chain}. 
In the second column we have shown only the parts of the JK vector that are
relevant for this two-instanton contribution.}  
\label{wpoles}
\end{table}
 
By calculating the residues over the poles selected by the 
JK vector $\eta_{Q_1}$ associated to the quiver $Q_1$, we find that the corresponding contribution 
to the superpotential proportional to $q_1q_2$ is
\begin{equation}
\begin{aligned}
w_{q_1q_2}\Big|_{\eta_{Q_1}}=& - \sum_{s \in \mathcal{N}_1} 
  \frac{( -1)^{n_1 + n_2}}{\prod_{r \in \widehat{\mathcal{N}}_1\cup \mathcal{N}_3}( a_s- a_r)
  \phantom{\Big|}
  \prod_{t \in \mathcal{N}_2 } ( a_{s} -a_t)^2
  \phantom{\Big|}} \\ 
   &+\sum_{ \substack{ t_1,t_2 \in \mathcal{N}_2 \\ t_1\not= t_2} } 
    \frac{( -1)^{ n_1+n_2} }{(a_{t_1} - a_{t_2})\phantom{\Big|}
    \prod_{r \in \mathcal{N}_1\cup\mathcal{N}_3} ( a_{t_1}-a_r)\phantom{\Big|}
   \prod_{s \in \widehat{\mathcal{N}}_2 }( a_{t_1} -a_{s})^2\phantom{\Big|}}\\
   &+\sum_{s\in \mathcal{N}_1}\sum_{t\in \mathcal{N}_2}
    \frac{( -1)^{ n_1+n_2} }{(a_{t} - a_{s})\phantom{\Big|}
    \prod_{r\in \mathcal{N}_1\cup\mathcal{N}_3} (a_{t}-a_r)\phantom{\Big|}
    \prod_{s \in  \widehat{\mathcal{N}}_2 }( a_{t} -a_{s})^2\phantom{\Big|}}\\
   &+\sum_{ \substack{ t_1,t_2 \in \mathcal{N}_2 \\ t_1\not= t_2 }}
    \frac{( -1)^{ n_1+n_2} }{(a_{t_1} - a_{t_2})\phantom{\Big|}
   \prod_{r \in \widehat{\mathcal{N}}_2 \cup \mathcal{N}_3 }( a_{t_1} -a_r)\phantom{\Big|}
   \prod_{s \in \mathcal{N}_1\cup \widehat{\mathcal{N}}_2} ( a_{t_2}-a_s)\phantom{\Big|} }~.
\end{aligned}
\label{quiverQ2Winstq1q2}
\end{equation}
Similarly, for the quiver $Q_2$ we find:
\begin{equation}
\begin{aligned}
w_{q_1q_2}\Big|_{\eta_{Q_2}} =& + \sum_{u \in \mathcal{N}_3}
\frac{( -1)^{n_1 + n_2}}{ \prod_{r \in\mathcal{N}_1 \cup\widehat{\mathcal{N}}_3  }
( a_u- a_r)  \phantom{\Big|} \prod_{t \in \mathcal{N}_2 } ( a_{u} -a_t)^2 \phantom{\Big|}}\\ 
&-\sum_{ \substack{ t_1,t_2 \in \mathcal{N}_2 \\ t_1\not= t_2} }
    \frac{( -1)^{ n_1+n_2} }{(a_{t_1} - a_{t_2})\phantom{\Big|}
    \prod_{r \in \mathcal{N}_1\cup \mathcal{N}_3} ( a_{t_1}-a_r)\phantom{\Big|}
   \prod_{s \in \widehat{\mathcal{N}}_2 }( a_{t_1} -a_s)^2\phantom{\Big|}}\\
&-\sum_{t\in \mathcal{N}_2}\sum_{u\in \mathcal{N}_3}
    \frac{( -1)^{ n_1+n_2} }{(a_{t} - a_{u})\phantom{\Big|}
    \prod_{r\in \mathcal{N}_1\cup\mathcal{N}_3} (a_{t}-a_r)\phantom{\Big|}
    \prod_{s \in  \widehat{\mathcal{N}}_2 }( a_{t} -a_{s})^2\phantom{\Big|}}\\
&-\sum_{ \substack{ t_1,t_2 \in \mathcal{N}_2 \\ t_1\not= t_2 }}
    \frac{( -1)^{ n_1+n_2} }{(a_{t_1} - a_{t_2})\phantom{\Big|}
    \prod_{r \in \mathcal{N}_1\cup \widehat{\mathcal{N}}_2} ( a_{t_1}-a_r)\phantom{\Big|}
   \prod_{s \in \widehat{\mathcal{N}}_2 \cup \mathcal{N}_3}( a_{t_2} -a_s)\phantom{\Big|}}~.
\end{aligned}
\label{quiverQ3Winstq1q2}
\end{equation}
Once again, we have found perfect agreement, term by term, between 
these results and those
obtained by solving the twisted chiral ring equations for the quivers $Q_1$ and $Q_2$.

\end{appendix}
\providecommand{\href}[2]{#2}\begingroup\raggedright\endgroup

\end{document}